\documentclass[10pt]{article}
\usepackage{graphicx}
\usepackage{amsmath}
\usepackage{amssymb}
\usepackage{mathrsfs}
\usepackage{multirow}
\usepackage{caption2}
\usepackage{authblk}
\usepackage[figuresright]{rotating}
\begin{document}
\bibliographystyle{prsty}
\begin{center}
{\large {\bf \sc{The $\Lambda_c\Sigma_c$ dibaryon and $\bar{\Lambda}_c\Sigma_c\pm \bar{\Sigma}_c\Lambda_c$  baryonium states via the QCD sum rules}}} \\[2mm]
Xiu-Wu Wang\footnote{E-mail: wangxiuwu2020@163.com.}, Zhi-Gang  Wang\footnote{E-mail: zgwang@aliyun.com.} and Guo-Liang Yu\\
 Department of Physics, North China Electric Power University, Baoding 071003, P. R. China\\
\end{center}
\begin{abstract}
In the present work, we study the $\Lambda_c\Sigma_c$ dibaryon and $\bar{\Lambda}_c\Sigma_c\pm \bar{\Sigma}_c\Lambda_c$ baryonium states via the QCD sum rules. We construct four (eight) currents with definite $J^P$ ($J^{PC}$) to interpolate the dibaryon (baryonium) states and obtain twelve QCD sum rules.   For the dibaryon states, the state with the $J^P=1^+$ lies below the $\Lambda_c\Sigma_c$ threshold, and  is a  molecule candidate.  For the   baryonium states,  the states with the $J^P=0^{-\pm}$ and $1^{-\pm}$ lie below the $\bar{\Lambda}_c\Sigma_c$ threshold, and are four   molecule  candidates. The present predictions can be confronted to the experimental data in the future and make contributions to the exotic spectroscopy.
\end{abstract}



\section{Introduction}
In the traditional quark model, the meson and baryon are made of a quark-antiquark pair  and three quarks, respectively. Since discovery of the $X(3872)$ \cite{Choi}, many  exotic  states which are possible  tetraquark, pentaquark and hexaquark candidates were observed \cite{ChenReview,GuoReview,Review-XYZ-Brambilla-PRT-2019,Review-XYZ-Wangbo-PRT-2023,
Review-XYZ-GengLS-PRT-2025,wangzg-zs}. Since the masses of the exotic states are close to the known two-particle thresholds, many of these exotic states are the possible hadronic molecules, but their inner structures are still under debates.
Several QCD-inspired approaches have been applied to study dibaryon and baryonium states, the typical exotic states,
such as the quark delocalization color screening model \cite{Xia},  the Bethe-Salpeter equation \cite{Lu,Zhu1,DongLASI}, constituent quark models \cite{Huang2,Carames}, Lattice QCD \cite{Morita1,Morita2}, QCD sum rules \cite{Kodama,Chen,Wang2,Wan,SWZG1,SWZG2}, and so on.

In the present work, we would like to focus on the QCD sum rules, which have been extensively and successfully applied to study the hidden-charm (bottom)
tetraquark (molecular) states \cite{xzAgaev,xzChen1,xzChen2,xzOzdem,xzLee1,xzZhang1,xzZhang2,xzMatheus,xzMihara,xzAlbuquerque,xzDi,Wang3,Wang4,Wang5}, pentaquark (molecular) states \cite{Wang6,Wang7} and hexaquark (molecular) states \cite{WZG,HX-Chen1,Qiao22}.

In 2023, the BESIII Collaboration observed a narrow structure $X(2085)$ in the $p\bar{\Lambda}$ system near the mass threshold  in the process $e^+e^- \to p K^-
\bar{\Lambda}$ with a statistical significance greater than $20\sigma$.
Its spin and parity were determined  to be $J^P=1^+$
in an amplitude analysis with statistical significance greater than
$5\sigma$ over other quantum numbers \cite{AblikimX2085}.

In Ref.\cite{X2085-Qiao-EPJC-2025}, Zhang, Zhang and Qiao  studied the  $p\bar{\Lambda}$ and $p\bar{\Sigma}$ states with  the QCD sum rules, and obtained the conclusion that there might exist six $p\bar{\Lambda}$ and $p\bar{\Sigma}$ states with quantum numbers $J^{P}=0^{-}, 0^{+}, 1^{-}$, which do not support $X(2085)$ being a $p\bar{\Lambda}$ or $p\bar{\Sigma}$ molecular state, or at least not its main component.

In Ref.\cite{Qiao22}, Zhang and Qiao studied the mass spectrum of  the $\Lambda_Q\bar{\Sigma}_Q$ baryonium  states with the $J^P=0^+$, $0^-$, $1^+$ and $1^-$  via the QCD sum rules by  considering the vacuum condensates up to  dimension 12, and observed that  the $\Lambda_c\bar{\Sigma}_c$ baryonium  states lie around the $5.8\,\,{\rm GeV}$ region, which is about $1\,{\rm GeV}$ above  the $\Lambda_c\bar{\Sigma}_c$ threshold, it is impossible to form the bound $\Lambda_c\bar{\Sigma}_c$  state.

In 2025, the BESIII Collaboration explored  the processes $e^+e^-\rightarrow \Sigma_c\bar{\Sigma}_c$ and  $\Lambda_c^+\bar{\Sigma}_c^-$  at $\sqrt{s}= 4.750\,, 4.781\,, 4.843\,, 4.918$ and $4.951\,\,{\rm GeV}$ for the first time, but observed no signals  for all the channels \cite{BESIII}. Furthermore, the BESIII Collaboration explored the process  $e^{+}e^{-} \to \pi^{+} \pi^{-} \Lambda_{c}^{+}\bar{\Lambda}_{c}^{-}$ for the first time, but observed no statistically significant signal  for a possible $\Lambda_{c} \bar{\Sigma}_{c}$
 \cite{BESIII2}. We still expect to observe the $\Lambda_{c} \bar{\Sigma}_{c}$ bound states when huge experimental data are accumulated.

In Refs.\cite{wangxiuwu1,wangxiuwu2}, we  studied the $\Lambda_c\Lambda_c$, $\Sigma_c\Sigma_c$ dibaryon and $\Lambda_c\bar{\Lambda}_c$, $\Sigma_c\bar{\Sigma}_c$ baryonium states via the QCD sum rules by carrying the operator product expansion (OPE) to the vacuum condensates of dimension 16, and adopt the energy scale formula to determine the suitable QCD spectral densities. And we observed that the $\bar{\Lambda}_c\Lambda_c$ state with the $J^{PC}=1^{--}$  and $\bar{\Sigma}_c\Sigma_c$ states with the $J^{PC}=0^{-+}$, $1^{--}$ lie below the $\bar{\Lambda}_c\Lambda_c$ and $\bar{\Sigma}_c\Sigma_c$ thresholds, respectively, the $\Sigma_c\Sigma_c$ state with the $J^P=0^+$ lies below the $\Sigma_c\Sigma_c$ threshold, and they are possible bound states.

In Ref.\cite{Wan}, Wan, Tang and Qiao studied the $\bar{\Lambda}_Q\Lambda_Q$ states with the $J^{PC}=0^{++}$, $0^{-+}$, $1^{++}$ and $1^{--}$ via the QCD sum rules by carrying the OPE up to the vacuum condensates of dimension 12, and observed that no bound states could be formed.

Compared with Refs.\cite{Wan,Qiao22}, we adopt the energy scale formula choose the pertinent energy scales of the QCD sum rules to enhance the pole contributions and improve the convergent behavior, and we expect to improve the predictive ability. In this work, we study the $\Lambda_c\Sigma_c$ dibaryon and $\bar{\Lambda}_c\Sigma_c\pm \bar{\Sigma}_c\Lambda_c$ baryonium states via the QCD sum rules in a systematic way.

The article is arranged as follows: we derive the QCD sum rules of the dibaryon and baryonium states in Sect.2; we present the numerical results and discussions in Sect.3; Sect.4 is reserved for our conclusions.

\section{The QCD sum rules for the dibaryon and baryonium  states}
 We adopt  two basic components,
\begin{eqnarray}
\notag \mathcal{J}_{\Lambda}(x)&=&\varepsilon^{ijk}u^{\texttt{T}}_i(x)C\gamma_5d_j(x)c_k(x)\, ,\\
  \mathcal{J}_{\Sigma}(x)&=&\varepsilon^{ijk}
u^{\texttt{T}}_i(x)C\gamma_{\mu}d_j(x)\gamma^{\mu}\gamma_5c_k(x) \, ,
\end{eqnarray}
with the same quantum numbers as the $\Lambda_c$ and $\Sigma_c$ to construct
  twelve  currents,
\begin{eqnarray}\label{di-baryon-current}
 \mathcal{J}_{1}&=&\mathcal{J}_{\Lambda}^\texttt{T}C\mathcal{J}_{\Sigma} \, ,\nonumber \\
 \mathcal{J}_{2}&=&\mathcal{J}_{\Lambda}^\texttt{T}C\gamma_5\mathcal{J}_{\Sigma} \, ,\nonumber\\
\mathcal{J}_{1\alpha}&=&\mathcal{J}_{\Lambda}^\texttt{T}C\gamma_{\alpha}\mathcal{J}_{\Sigma} \, ,\nonumber\\
\mathcal{J}_{2\alpha}&=&\mathcal{J}_{\Lambda}^\texttt{T}C\gamma_5\gamma_{\alpha}\mathcal{J}_{\Sigma} \, ,
\end{eqnarray}
\begin{eqnarray}\label{baryonium-current}
\notag \mathcal{J}_{1}^{\pm}&=& \frac{1}{\sqrt{2}} \left(\bar{\mathcal{J}}_{\Lambda}\mathcal{J}_{\Sigma}\pm \bar{\mathcal{J}}_{\Sigma}\mathcal{J}_{\Lambda}\right)\, ,\\
\notag \mathcal{J}_{2}^{\pm}&=& \frac{1}{\sqrt{2}} \left(\bar{\mathcal{J}}_{\Lambda}\gamma_5\mathcal{J}_{\Sigma}\pm \bar{\mathcal{J}}_{\Sigma}\gamma_5\mathcal{J}_{\Lambda}\right)\,,\\
\notag \mathcal{J}_{1\alpha}^{\pm}&=& \frac{1}{\sqrt{2}} \left(\bar{\mathcal{J}}_{\Lambda}\gamma_\alpha\mathcal{J}_{\Sigma}\mp \bar{\mathcal{J}}_{\Sigma}\gamma_\alpha\mathcal{J}_{\Lambda}\right)\,,\\
 \mathcal{J}_{2\alpha}^{\pm}&=& \frac{1}{\sqrt{2}} \left(\bar{\mathcal{J}}_{\Lambda}\gamma_\alpha\gamma_5\mathcal{J}_{\Sigma}\pm \bar{\mathcal{J}}_{\Sigma}\gamma_\alpha\gamma_5\mathcal{J}_{\Lambda}\right)\,,
\end{eqnarray}
where the  $i, j, k$ are color indexes, the $C$ represents the charge conjugation matrix, and the superscript \texttt{T} denotes  transpose of the Dirac spinors. Considering the transpose properties  $\mathcal{J}_{1/2}^{\texttt{T}}=\mathcal{J}_{1/2}$,  $\mathcal{J}_{1\mu}^{\texttt{T}}=-\mathcal{J}_{1\mu}$ and $\mathcal{J}_{2\mu}^{\texttt{T}}=\mathcal{J}_{2\mu}$, the currents in Eq.\eqref{di-baryon-current} are complete to interpolate the $\Lambda_c\Sigma_c$ dibaryon states.

According to properties   of the parity and charge conjugation operators $\widehat{P}$ and $\widehat{\mathcal{C}}$,  $\widehat{P}\psi\widehat{P}^{-1}=\gamma^0\psi$  and $\widehat{\mathcal{C}}\psi\widehat{\mathcal{C}}^{-1}=C\bar{\psi}^{\texttt{T}}$, it is direct to obtain the transformation properties   of the interpolating  currents,
\begin{eqnarray}
\notag \widehat{P}\mathcal{J}_{1}\widehat{P}^{-1}&=&-\mathcal{J}_1 \, ,\\
\notag \widehat{P}\mathcal{J}_{2}\widehat{P}^{-1}&=&+\mathcal{J}_2 \, ,\\
\notag \widehat{P}\mathcal{J}_{1\alpha}\widehat{P}^{-1}&=&-\mathcal{J}_1^{\alpha} \, , \\
\widehat{P}\mathcal{J}_{2\alpha}\widehat{P}^{-1}&=&+\mathcal{J}_{2}^{\alpha} \, ,
\end{eqnarray}
\begin{eqnarray}
\notag \widehat{P}\mathcal{J}_{1}^{\pm}\widehat{P}^{-1}&=&+\mathcal{J}_{1}^{\pm}\,,\\
\notag \widehat{P}\mathcal{J}_{2}^{\pm}\widehat{P}^{-1}&=&-\mathcal{J}_{2}^{\pm}\,,\\
\notag \widehat{P}\mathcal{J}_{1\alpha}^{\pm}\widehat{P}^{-1}&=&+\mathcal{J}_{1}^{\alpha\pm}\,,\\
\widehat{P}\mathcal{J}_{2\alpha}^{\pm}\widehat{P}^{-1}&=&-\mathcal{J}_{2}^{\alpha\pm}\, ,
\end{eqnarray}
\begin{eqnarray}
\notag \widehat{\mathcal{C}}\mathcal{J}_{n}^{\pm}\widehat{\mathcal{C}}^{-1}
&=&\pm\mathcal{J}_{n}^{\pm}\, , \nonumber\\
\widehat{\mathcal{C}}\mathcal{J}_{n\alpha}^{\pm}\widehat{\mathcal{C}}^{-1}&=&
\pm\mathcal{J}_{n\alpha}^{\pm}\, ,
\end{eqnarray}
with $n=1$, $2$. The currents in Eq.\eqref{di-baryon-current} are not eigenstates of the charge-conjugation operators.

Now we write down the  two-point correlation functions in the QCD sum rules,
\begin{eqnarray}
\notag &&\Pi(p^2)=\texttt{i}\int d^4x e^{\texttt{i}p\cdot x}\langle 0 |\mathcal{T}\big\{ \mathcal{J} (x) \mathcal{J}^\dag(0) \big\}| 0\rangle \, ,\\
\notag &&\Pi_{\alpha\beta}(p^2)=\texttt{i}\int d^4x e^{\texttt{i}p\cdot x}\langle 0 | \mathcal{T}\big\{ \mathcal{J}_{\alpha} (x) \mathcal{J}_{\beta}^\dag(0)\big\} | 0\rangle \, ,
\end{eqnarray}
 where
 \begin{eqnarray}
 \mathcal{J} (x)&=& \mathcal{J}_1(x)\, , \, \mathcal{J}_2(x)\, , \, \mathcal{J}_{1}^{\pm}(x)\, , \, \mathcal{J}_{2}^{\pm}(x)\, , \, \nonumber\\
 \mathcal{J}_\alpha (x)&=& \mathcal{J}_{1\alpha}(x)\, , \, \mathcal{J}_{2\alpha}(x)\, , \, \mathcal{J}_{1\alpha}^{\pm}(x)\, , \, \mathcal{J}_{2\alpha}^{\pm}(x)\, , \,
 \end{eqnarray}
 $\texttt{i}^2=-1$, and the $\mathcal{T}$ is the time order operator.

On the hadron side, we insert a complete set of intermediate hadronic states with the same quantum numbers as the interpolating currents into the correlation functions, and obtain the results,
\begin{eqnarray}
\notag
 \Pi(p^2)&=&\frac{\lambda_{\mathcal{Z}}^2}{M_{\mathcal{Z}}^2-p^2}+\cdots \, ,\\
\Pi_{\alpha\beta}(p^2)&=&\frac{\lambda_{\mathcal{Z}}^2}{M_{\mathcal{Z}}^2-p^2}\left(-g_{\alpha\beta}+\frac{p_\alpha p_\beta}{p^2}\right)+\cdots \, ,
\end{eqnarray}
 where
\begin{eqnarray}
 \langle0 | \mathcal{J}(0)|\mathcal{Z}(p) \rangle &=&\lambda_{\mathcal{Z}} \, ,\nonumber\\
 \langle0 | \mathcal{J}_{\alpha}(0)|\mathcal{Z}(p) \rangle &=&\lambda_{\mathcal{Z}}\, \varepsilon_\alpha \, ,
\end{eqnarray}
 the $\mathcal{Z}$ are the ground states, the $\varepsilon_\alpha$ are the polarization vectors, and the $\lambda_{\mathcal{Z}}$ are the pole residues.

The components (or currents) $\mathcal{J}_\Lambda(x)$ and $\mathcal{J}_\Sigma(x)$ have  positive parity, and they
  couple potentially to the charmed baryons ($\mathcal{B}$) with both positive ($+$) and negative ($-$) parities, as multiplying
  $\texttt{i}\gamma^5$  to the currents changes their parities \cite{wangzg-zs},
    \begin{eqnarray}
\langle 0| \mathcal{J}_{\Lambda/\Sigma} (0)|\mathcal{B}^{+}(q)\rangle &=&\lambda^{+}_{\Lambda/\Sigma}\, U^{+}(q,s) \, , \nonumber \\
\langle 0|  \mathcal{J}_{\Lambda/\Sigma}(0) |\mathcal{B}^{-}(q)\rangle &=&\lambda^{-}_{\Lambda/\Sigma}\, i\gamma_5 U^{-}(q,s) \, ,
\end{eqnarray}
 where the $U^{\pm}(q,s)$ are Dirac spinors. There exists an additional P-wave in the  $\mathcal{B}^{-}$,  we usually expect that the   $\mathcal{B}^{-}$ lies above the corresponding  $\mathcal{B}^{+}$  about $0.5\,\rm{GeV}$.
The currents $\mathcal{J}(x)$ and $\mathcal{J}_{\alpha}(x)$ have definite parities, and couple  potentially to the dibaryon  or baryonium states $\mathcal{Z}$ with definite parities. The $\mathcal{Z}$ have two color-neutral clusters, which have the same quantum numbers as the S-wave $\Lambda_c\Sigma_c$, $\Lambda_c^\prime\Sigma_c^\prime$,  $\bar{\Lambda}_c\Sigma_c\pm\bar{\Sigma}_c\Lambda_c$ and $\bar{\Lambda}^\prime_c\Sigma_c^\prime\pm\bar{\Sigma}^\prime_c\Lambda_c^\prime$ pairs, where we take the notations $\Lambda_c^\prime$ and $\Sigma_c^\prime$ to represent  the  $\mathcal{B}^{-}$. If we take the $\mathcal{Z}$ to present the ground states, the excited states $\mathcal{Z}^\prime$ would have two additional  P-waves and would lie above the ground states $\mathcal{Z}$ about $1\,\rm{GeV}$. The $\mathcal{Z}^\prime$ have no contamination in the present work due to  the continuum threshold parameters, see Eq.\eqref{s0-threshold}.

On the QCD side, we carry out the OPE routinely with the help of the full  quark propagators \cite{wangzg-zs,Wang3},  it is accurate enough for us to choose the terms $\langle\overline{q}q\rangle$, $\langle\frac{\alpha_s}{\pi}GG\rangle$, $\langle\overline{q}g_s\sigma Gq\rangle$,  $\langle\overline{q}q\rangle^2$, $\langle\overline{q}g_s\sigma Gq\rangle\langle\overline{q}q\rangle$, $\langle\overline{q}g_s\sigma Gq\rangle^2$, $\langle\frac{\alpha_s}{\pi}GG\rangle\langle\overline{q}q\rangle^2$, $\langle\overline{q}q\rangle^4$, $\langle\overline{q}g_s\sigma Gq\rangle\langle\overline{q}q\rangle^3$, $\langle\overline{q}g_s\sigma Gq\rangle^2\langle\overline{q}q\rangle^2$ and $\langle\frac{\alpha_s}{\pi}GG\rangle\langle\overline{q}q\rangle^4$ with the truncation of the order $\mathcal{O}(\alpha_s^k )$ with $k\leq 1$ \cite{wangxiuwu1}.

The perturbative contributions are embodied in the Wilson's coefficients, while the nonperturbative contributions are absorbed in the vacuum condensates. For the quark condensate $\langle\bar{q}q\rangle$, it symbolizes spontaneous breaking of the chiral symmetry through the Gell-Mann-Oakes-Renner relation $f_{\pi}^2m_{\pi}^2=-2(m_d+u_d)\langle\bar{q}q\rangle$, where the $f_\pi$ is the decay constant of the pion. As for the other vacuum condensates, such as $\langle\overline{q}g_s\sigma Gq\rangle$, $\langle\overline{q}q\rangle^2$, etc, are parameters introduced by hand to describe the nonperturbative QCD vacuum. These vacuum condensates are nonzero vacuum expectations for
the normal-ordered quark-gluon operators to describe the hadron properties. We take the four-quark vacuum condensate $\langle0|:\bar {q}_{\alpha}^{i}q_{\beta}^{j}\bar{q}_{\lambda}^{m}q_{\tau}^{n}:|0\rangle$ as an example, and  insert a complete set of
intermediate states into the normal-ordered operator,
\begin{eqnarray}
\langle0|:\bar{q}_{\alpha}^{i}q_{\beta}^{j}\bar{q}_{\lambda}^{m}q_{\tau}^{n}:|0\rangle &=& \frac{1}{16N_{c}^{2}}\langle0|:\bar{q}q:|0 \rangle^{2}\left(\delta_{ij}\delta_{m n}\delta_{\alpha\beta}\delta_{\lambda\tau}-\delta_{i n}\delta_{j m}\delta_{\alpha\tau}\delta_{\beta\lambda}\right)+\zeta\,,
\end{eqnarray}
where the $i$, $j$, $m$ and $n$ are color indexes, the $\alpha$, $\beta$, $\lambda$ and $\tau$ are Dirac spinor indexes, the $N_c$ is the number of color, the $\zeta$ represents the contribution besides the vacuum state.  Shifman, Vainshtein and Zakharov took the hypothesis of vacuum saturation  and argued that the vacuum contribution dominates the sum of all the intermediate states \cite{Shifman1,Shifman2}, thus the vacuum saturation  holds if one sets $\zeta=0$. Furthermore, the accuracy of  factorization of the higher dimensional vacuum condensates is of the order of $\frac{1}{N_c^2}$ \cite{Shifman1,Shifman2,Ioffe1}, the vacuum saturation holds well in the
large-$N_c$ limit. In reality, the deviation is around 0.11 for $N_c=3$. In the present work,  we would like to study  violation of the vacuum saturation by setting  $\langle\overline{q}q\rangle^2 \to \xi\langle\overline{q}q\rangle^2$, $\langle\overline{q}g_s\sigma Gq\rangle\langle\overline{q}q\rangle \to \xi \langle\overline{q}g_s\sigma Gq\rangle\langle\overline{q}q\rangle$, etc, and $\xi=1.11$.

We obtain the QCD spectral densities through dispersion relation, match the QCD side with the hadron side, and   perform the Borel transformation to get the QCD sum rules,
\begin{eqnarray}\label{QCDSR-main}
\lambda_{\mathcal{Z}}^2\exp\left(-\frac{M_{\mathcal{Z}}^2}{T^2}\right)&=&
\int_{\Delta^2}^{s_0}ds \rho_{QCD}(s)\exp\left(-\frac{s}{T^2}\right)\,,
\end{eqnarray}
\begin{eqnarray}\label{QCDSR-mass}
M_{\mathcal{Z}}^2&=&\frac{-\frac{d}{d\tau}\int_{\Delta^2}^{s_0}ds \rho_{QCD}(s)\exp\left(-\frac{s}{T^2}\right)}{\int_{\Delta^2}^{s_0}ds \rho_{QCD}(s)\exp \left(-\frac{s}{T^2}\right)}\mid_{\tau=\frac{1}{T^2}}\, ,
\end{eqnarray}
where $\Delta^2=4m_c^2$, the $\rho_{QCD}(s)$ are the QCD spectral densities, the $T^2$ is the Borel parameter, and the $s_0$ are the continuum threshold parameters.

\section{Numerical results and discussions}
We take the standard values of the vacuum condensates,  $\langle\overline{q}q\rangle=-(0.24\pm0.01\,{\rm GeV})^3$, $\langle\overline{q}g_s\sigma Gq\rangle=m_0^2\langle\overline{q}q\rangle\,$GeV$^2$, $m_0^2=(0.8\pm0.1)\,{\rm GeV}^2$, $\langle \frac{\alpha_s
GG}{\pi}\rangle=0.012\pm0.004\,\rm{GeV}^4$  at the energy scale $\mu=1\;{\rm GeV}$ \cite{Reinders,ColangeloReview}, and the $
\overline{MS}$ mass $m_c(m_c)=1.275\pm0.025\,{\rm GeV}$ from the Particle Data Group \cite{PDG}. These parameters rely on the energy-scale, the dependence  is given by,
\begin{eqnarray}
\notag \langle\overline{q}q\rangle(\mu)&=&\langle\overline{q}q\rangle(1\;{\rm GeV})\left[\frac{\alpha_s(1\;{\rm GeV})}{\alpha_s(\mu)}\right]^{\frac{12}{33-2n_f}}\, ,\\
\notag \langle\overline{q}g_s\sigma Gq\rangle(\mu)& =&\langle\overline{q}g_s\sigma Gq\rangle(1\;{\rm GeV})\left[\frac{\alpha_s(1\;{\rm GeV})}{\alpha_s(\mu)}\right]^{\frac{2}{33-2n_f}}\, ,\\
\notag  m_c(\mu)&=&m_c(m_c)\left[\frac{\alpha_s(\mu)}{\alpha_s(m_c)}\right]^{\frac{12}{33-2n_f}}\, ,\\
\notag \alpha_s(\mu)&=&\frac{1}{b_0\texttt{t}}\left[1-\frac{b_1}{b_0^2}\frac{\rm{log}\texttt{t}}{\texttt{t}}+\frac{b_1^2(\rm{log}^2\texttt{t}-\rm{log}\texttt{t}-1)+\emph{b}_0\emph{b}_2}{b_0^4\texttt{t}^2}\right]\, ,
\end{eqnarray}
where $\texttt{t}=\rm{log}\frac{\mu^2}{\Lambda_{\emph{QCD}}^2}$, $\emph b_0=\frac{33-2\emph{n}_\emph{f}}{12\pi}$, $b_1=\frac{153-19n_f}{24\pi^2}$, $b_2=\frac{2857-\frac{5033}{9}n_f+\frac{325}{27}n_f^2}{128\pi^3}$
and $\Lambda_{QCD}=213$ MeV, $296$ MeV, $339$ MeV for the flavors $n_f=5,4,3$, respectively \cite{Shifman1,Shifman2,PDG,Narison}, in this work, we choose the flavor number $n_f=4$ for all the exotic  states.

As for the continuum threshold parameters $s_0$, we take the experiential values \cite{WZG,wangxiuwu1,wangxiuwu2,XZWZG1,XZWZG2},
\begin{eqnarray}\label{s0-threshold}
\sqrt{s_0}=M_{\mathcal{Z}}+(0.50\sim0.70)\pm0.10\,\,{\rm GeV}\, ,
\end{eqnarray}
where the $M_{\mathcal{Z}}$ are the ground state masses, $0.5\sim0.7\,{\rm GeV}$ is the energy gap between the ground states and first radial excited states, and $\pm0.1\,{\rm GeV}$ stands for uncertainty. Such a constraint has been applied to study   the $\Lambda_c\Lambda_c$, $\Sigma_c\Sigma_c$ dibaryon states and $\bar{\Lambda}_c\Lambda_c$, $\bar{\Sigma}_c\Sigma_c$ baryonium states  \cite{wangxiuwu1,wangxiuwu2}.

Till now, the available experimental data  are extremely scarce for the energy gaps between the ground states and  first radial excited states for the hidden-charm tetraquark or pentaquark  states, not to mention the dibaryon and baryonium states.

In Ref.\cite{WZGs06}, we assign the $\Lambda_c(2765)$ as the first radial excitation of the $\Lambda_c$ baryon,  the energy gap is around $479\,{\rm MeV}$. The first radial excitation of the $\Sigma_c$ baryon has not been observed yet experimentally,  the theoretical value of the mass is around $2900\,{\rm MeV}$ \cite{SigmaS01,SigmaS02,SigmaS03,SigmaS04}, thus the energy gap  is around $450\,{\rm MeV}$. In the present work, we focus on the $\Lambda_c\Sigma_c$ dibaryon and $\bar{\Lambda}_c\Sigma_c\pm \bar{\Sigma}_c\Lambda_c$  baryonium states, it is safe and reasonable for us to adopt the phenomenological constraint $\sqrt{s_0}=M_{\mathcal{Z}}+(0.50\sim0.70)\pm0.10\,\,{\rm GeV}$.

 And we chose analogous constraints in other works, i.e. we chose $\sqrt{s_0}=M_{B}+0.50\sim0.55\pm0.10\,{\rm GeV}$ for the triply-heavy baryon states $B$ \cite{WZGs01}, $\sqrt{s_0}=M_{X}+0.50\pm0.10\,{\rm GeV}$ for the fully-heavy tetraquark states $X$ \cite{WZGs02a,WZGs02b}, $\sqrt{s_0}=M_{X/Z}+0.55\pm0.10\,{\rm GeV}$ for the hidden-charm and hidden-bottom tetraquark states $X_Q$ and $Z_Q$ \cite{WZGs03a,WZGs03b}, $\sqrt{s_0}=M_{P}+0.65\pm0.10\,{\rm GeV}$
for the hidden-charm pentaquark states $P_c$ and $P_{cs}$ states \cite{WZGs04a,WZGs04b}, $\sqrt{s_0}=M_{P}+0.60\pm0.10\,{\rm GeV}$ for the fully-heavy pentaquark states $P_Q$ \cite{WZGs05a}, which work well.

Need to point out that, the QCD sum rules in Eqs.\eqref{QCDSR-main}-\eqref{QCDSR-mass} depend on the energy scale $\mu$ \cite{Chen,Wang4,Wang5}, we apply the energy scale formula to determine the best energy scales of the QCD spectral densities \cite{Wang3,SWZG3},
\begin{eqnarray}
\mu=\sqrt{M_{X/Y/Z}^2-4\mathbb{M}_c^2}\, ,
\end{eqnarray}
where the $\mathbb{M}_c$ is the effective charm quark mass, we adopt the value $\mathbb{M}_c=1.84\pm0.01$ {\rm GeV} \cite{wangzg}.

The numerical values of the masses $M_{\mathcal{Z}}$ and pole residues $\lambda_{\mathcal{Z}}$ are extracted from the Borel windows under two basic principles:
 pole dominance  and convergence of the OPE. Quantitative descriptions of the two principles are realized by the parameters PC and $D(n)$, namely, the pole contributions (PC) and contributions of the vacuum condensates of dimension $n$, where
\begin{eqnarray}
{\rm PC}=\frac{\int_{4m_c^2}^{s_0}ds\rho_{QCD}(s)\exp\left(-\frac{s}{T^2}\right)}{\int_{4m_c^2}^{\infty}ds\rho_{QCD}(s)\exp\left(-\frac{s}{T^2}\right)}\, ,
\end{eqnarray}
\begin{eqnarray}
D(n)=\frac{\int_{4m_c^2}^{s_0}ds\rho_{QCD;n}(s)\exp\left(-\frac{s}{T^2}\right)}{\int_{4m_c^2}^{s_0}ds\rho_{QCD}(s)\exp\left(-\frac{s}{T^2}\right)}\, ,
\end{eqnarray}
 the $\rho_{QCD;n}(s)$ are the spectral densities involving  the vacuum condensates with dimension $n$.

Firstly, we adopt  vacuum saturation via setting $\xi=1$ for the higher dimensional vacuum condensates, and take the hadron  masses  $M_{\Sigma_c}=M_{\overline\Sigma_c}=2.455$ {\rm GeV} and $M_{\Lambda_c}=M_{\overline\Lambda_c}=2.285$ GeV  from the Particle Data Group \cite{PDG}.

We obtain the Borel windows for all the currents via trial and error, and extract the masses and pole residues from them directly, the numerical results are displayed in Table \ref{BorelP-mass}, where the uncertainties  are estimated  via the formula,
\begin{eqnarray}
\Delta f=\sqrt{\sum\limits_\textsf{i}\left(\frac{\partial f}{\partial x_\textsf{i}}\right)^2|_{x_\textsf{i}=\bar x_\textsf{i}}(x_\textsf{i}-\bar x_\textsf{i})^2}\approx\sqrt{\sum\limits_\textsf{i} \left[f(\bar x_\textsf{i}\pm \Delta x_\textsf{i})-f(\bar x_\textsf{i})\right]^2}\, ,
\end{eqnarray}
where the $f$ represent the masses and  pole residues, the $\Delta x_\textsf{i}$ refer to the uncertainties of the input parameters $\langle\bar qq\rangle$, $\langle\overline{q}g_s\sigma Gq\rangle$, $m_c(m_c)$, $s_0$ and $\mathbb{M}_c$, the $\bar x_\textsf{i}$ denote the central values of these parameters.

\begin{figure}
 \centering
 \includegraphics[totalheight=5cm,width=7cm]{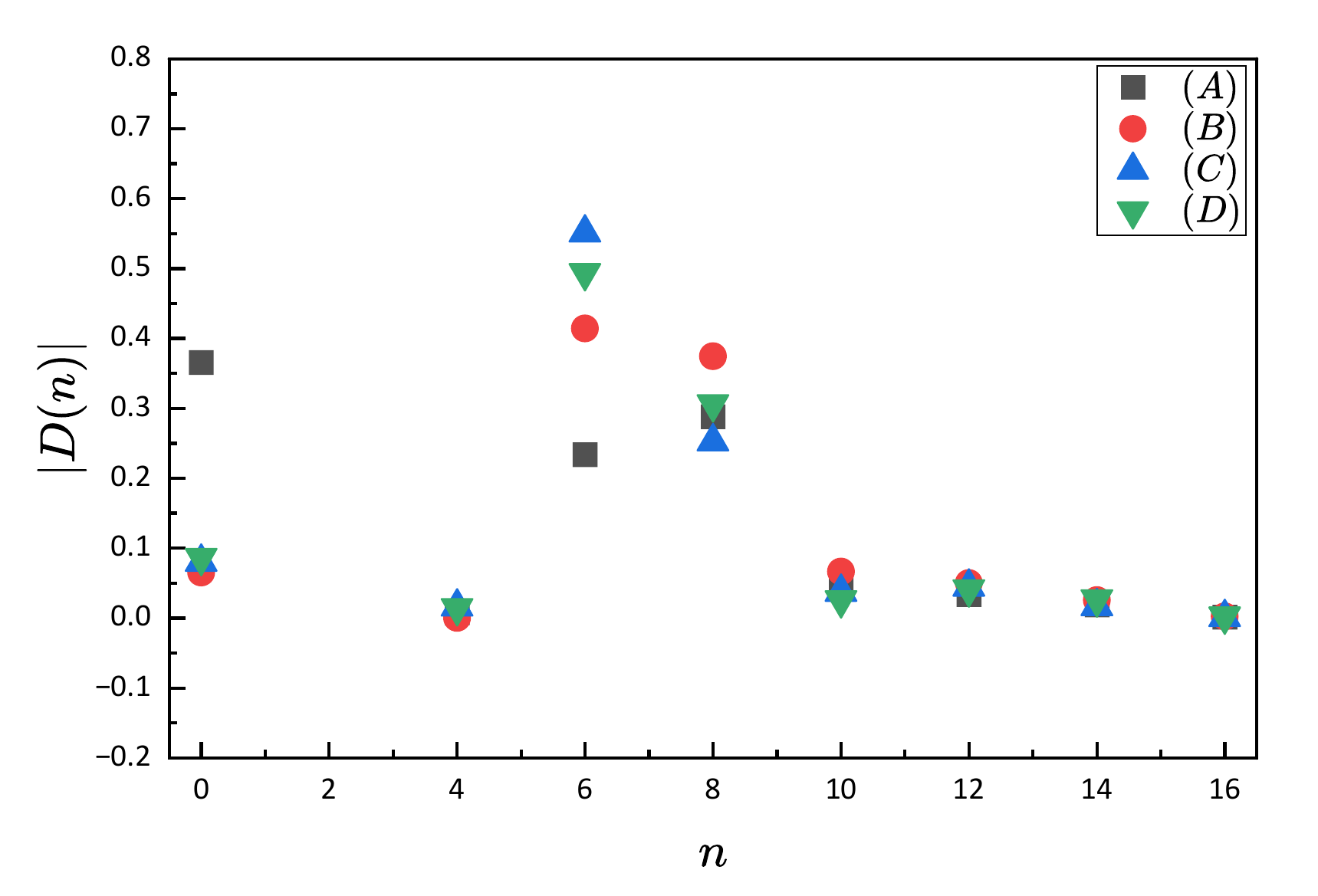}
 \includegraphics[totalheight=5cm,width=7cm]{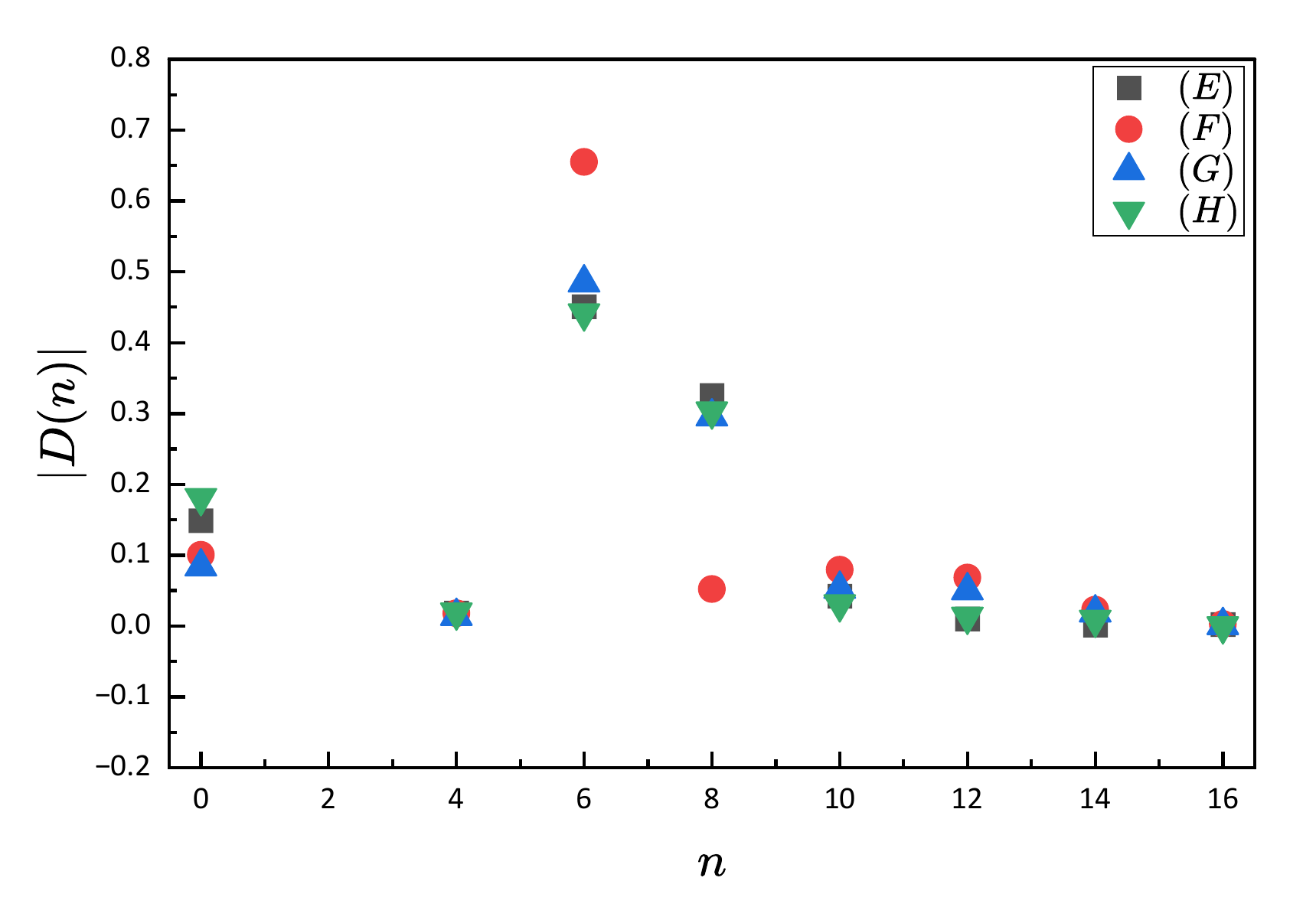}
 \caption{The $|D(n)|$ with variations of  $n$ for the central values of the input parameters, where $A$, $B$, $C$, $D$, $E$, $F$, $G$ and $H$ correspond to the currents $\mathcal{J}_1$, $\mathcal{J}_2$, $\mathcal{J}_{1\mu}$, $\mathcal{J}_{2\mu}$, $\mathcal{J}_1^{\pm}$, $\mathcal{J}_2^{\pm}$, $\mathcal{J}_{1\mu}^{\pm}$ and $\mathcal{J}_{2\mu}^{\pm}$, respectively.}\label{DC-fig}
\end{figure}

\begin{figure}
 \centering
 \includegraphics[totalheight=5cm,width=7cm]{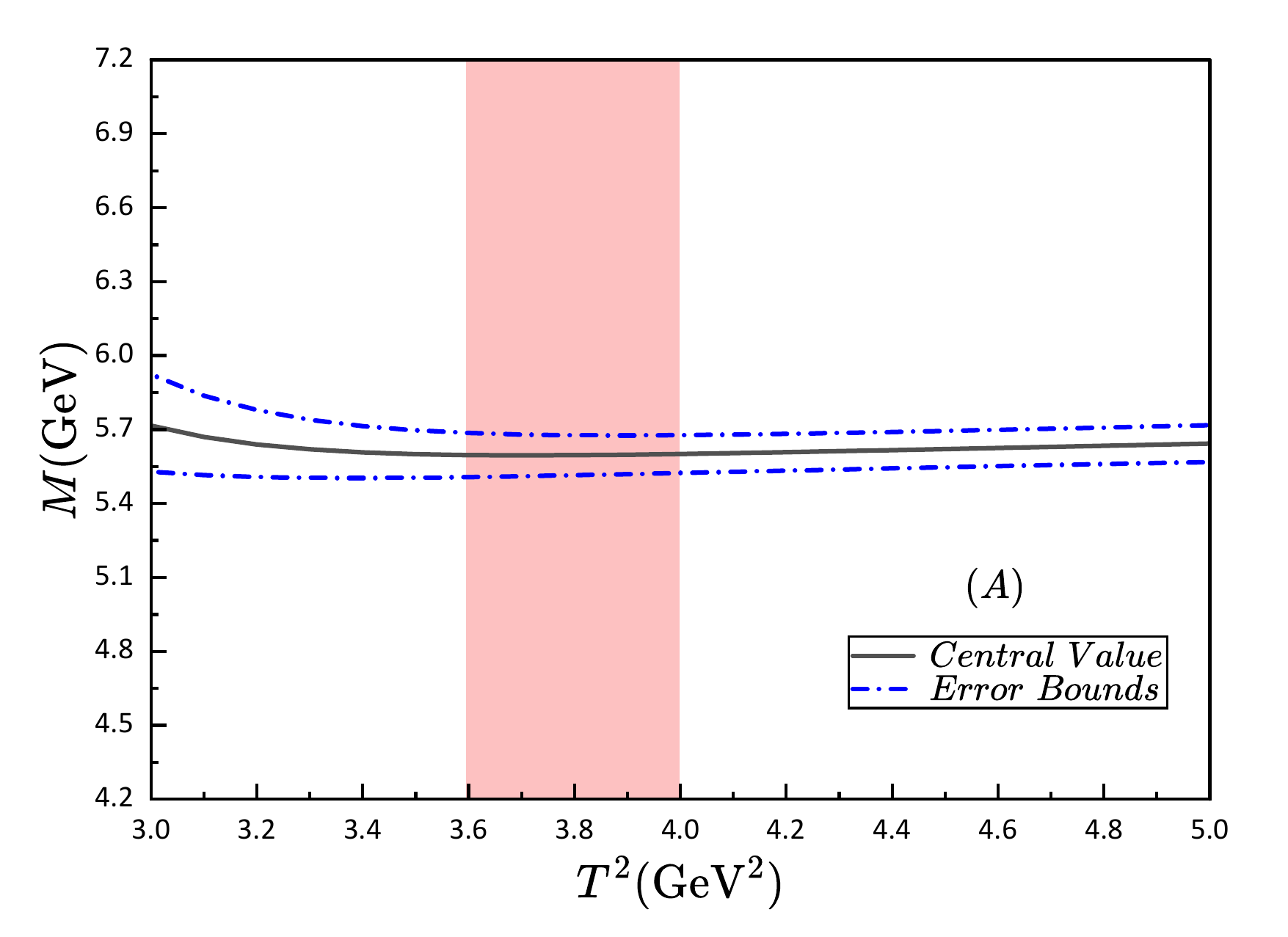}
 \includegraphics[totalheight=5cm,width=7cm]{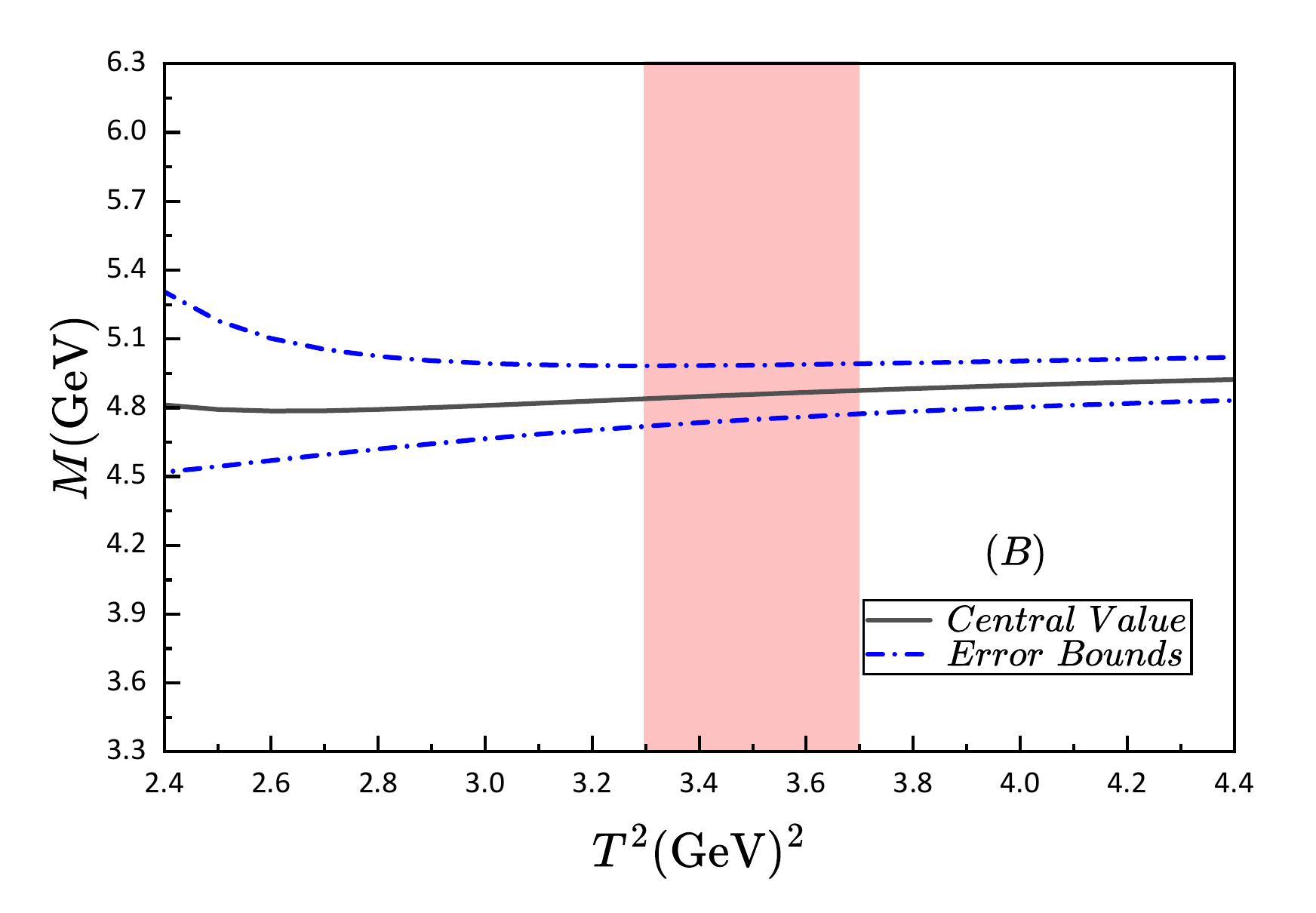}
 \includegraphics[totalheight=5cm,width=7cm]{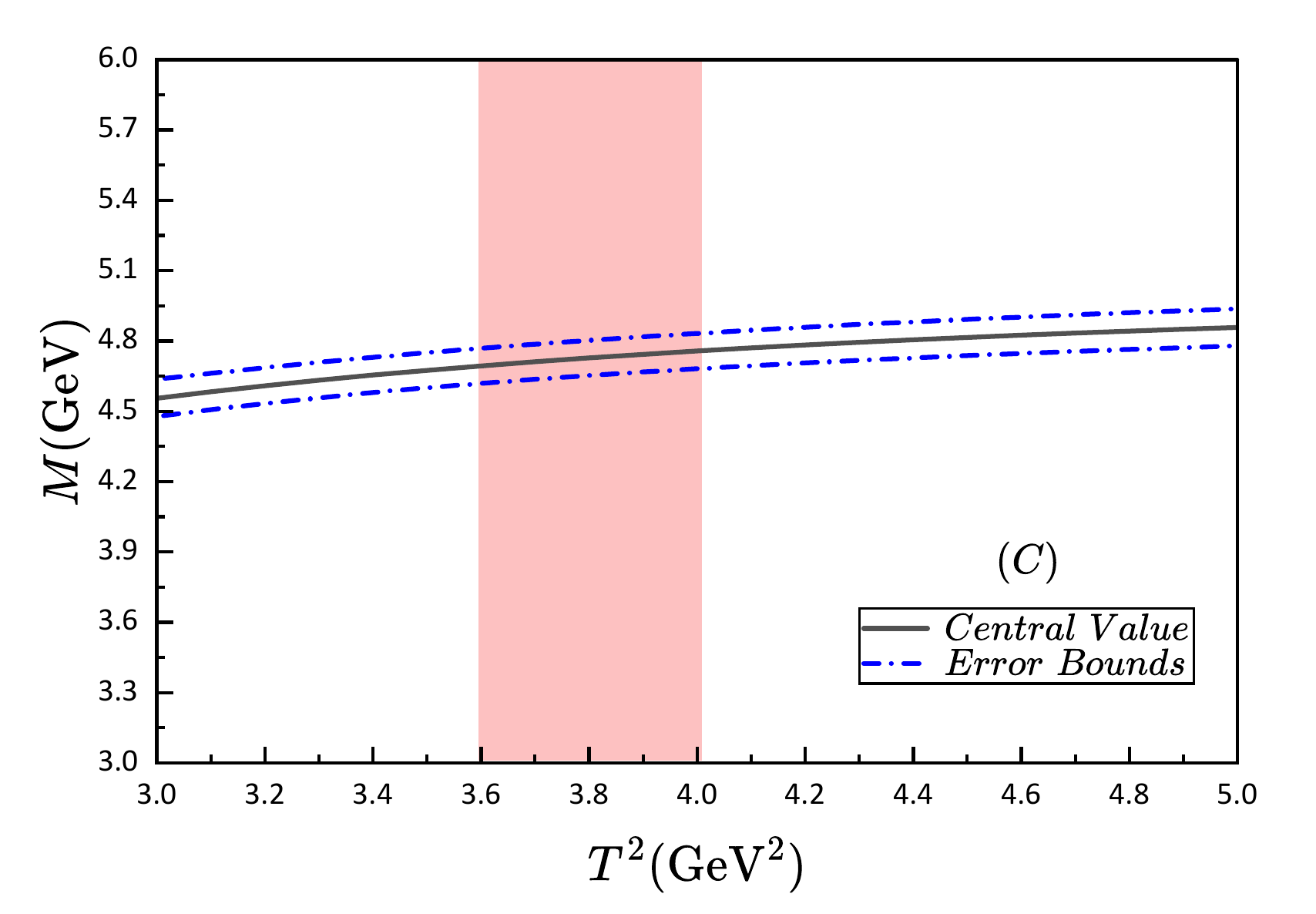}
 \includegraphics[totalheight=5cm,width=7cm]{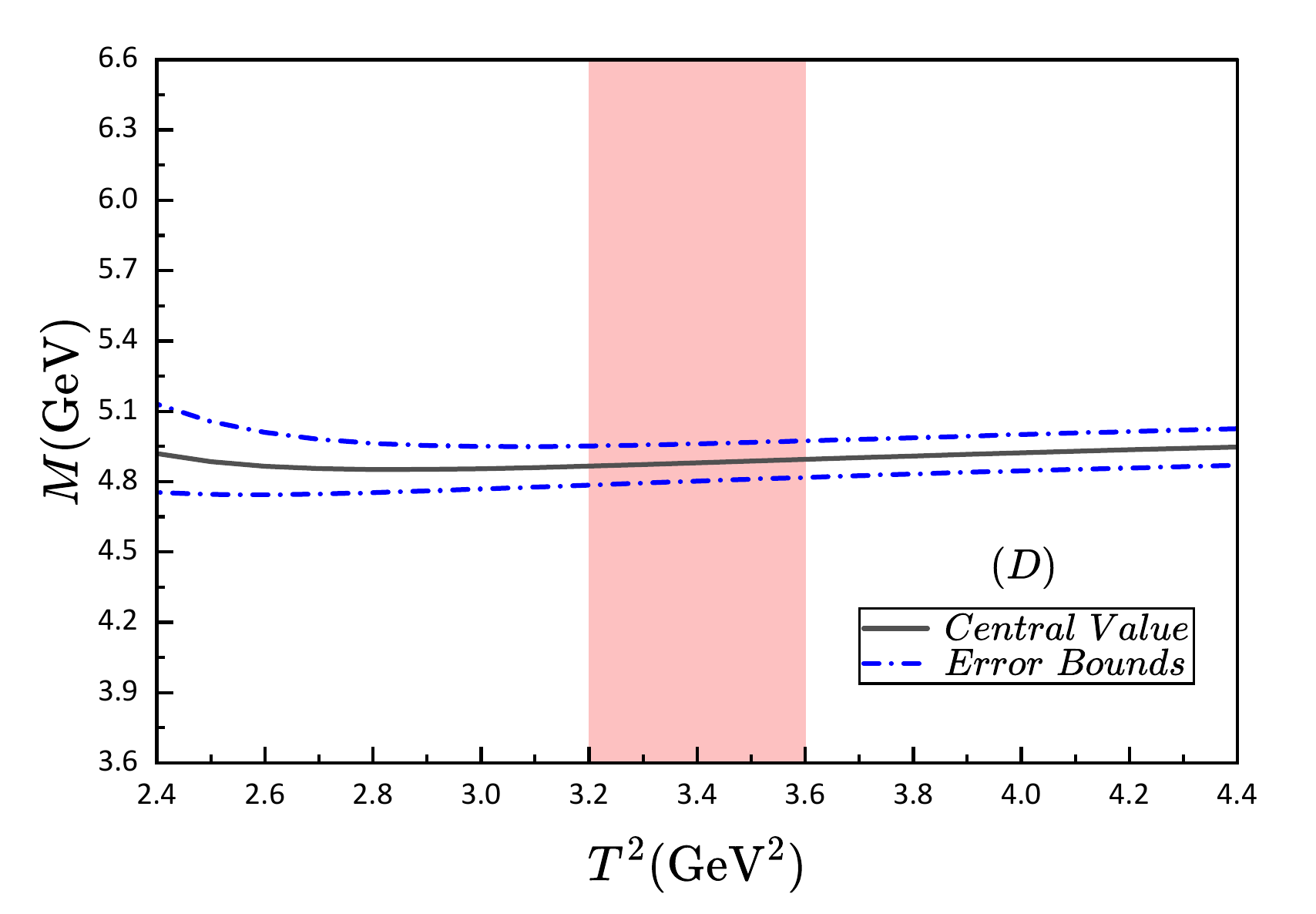}
 \includegraphics[totalheight=5cm,width=7cm]{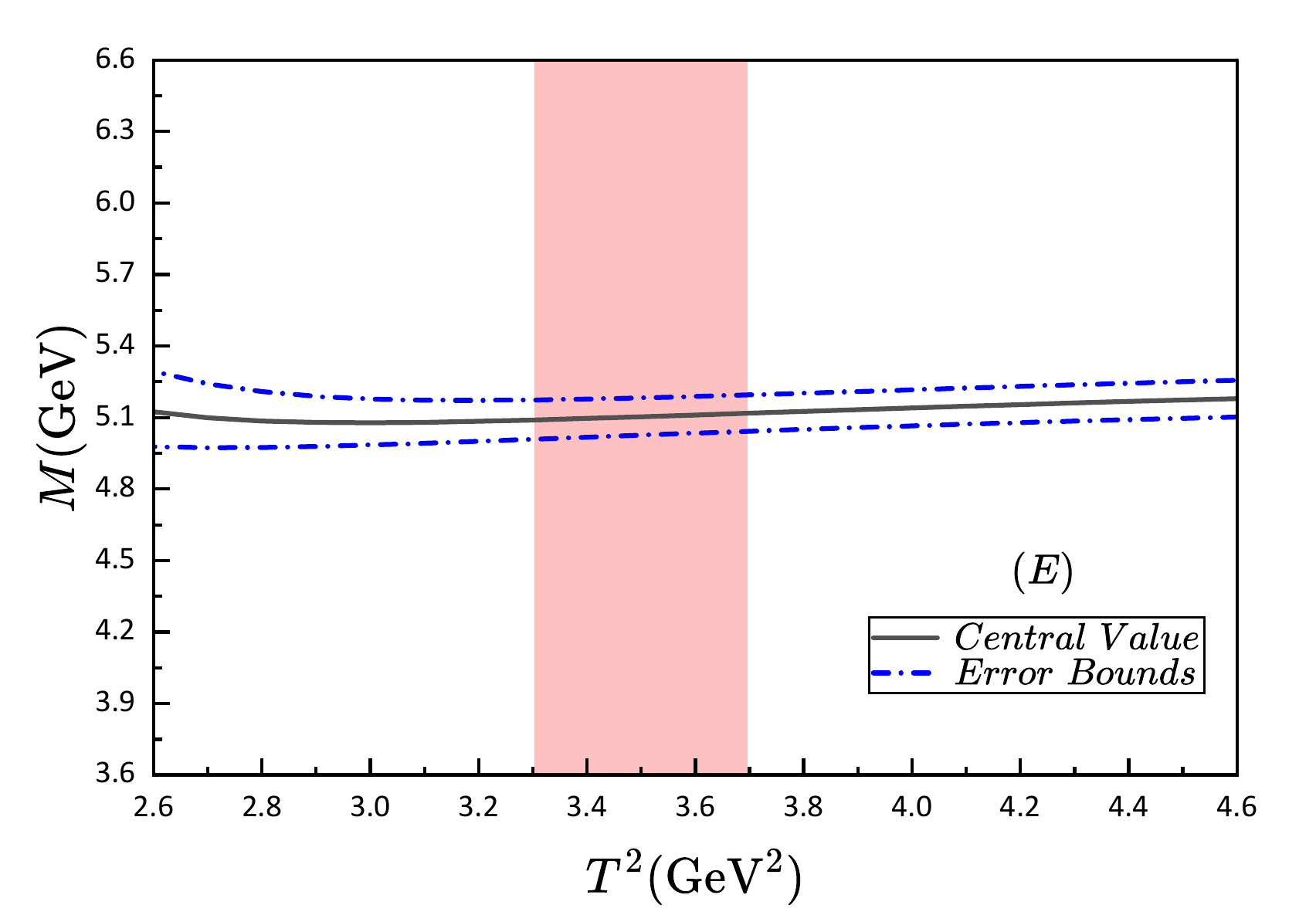}
 \includegraphics[totalheight=5cm,width=7cm]{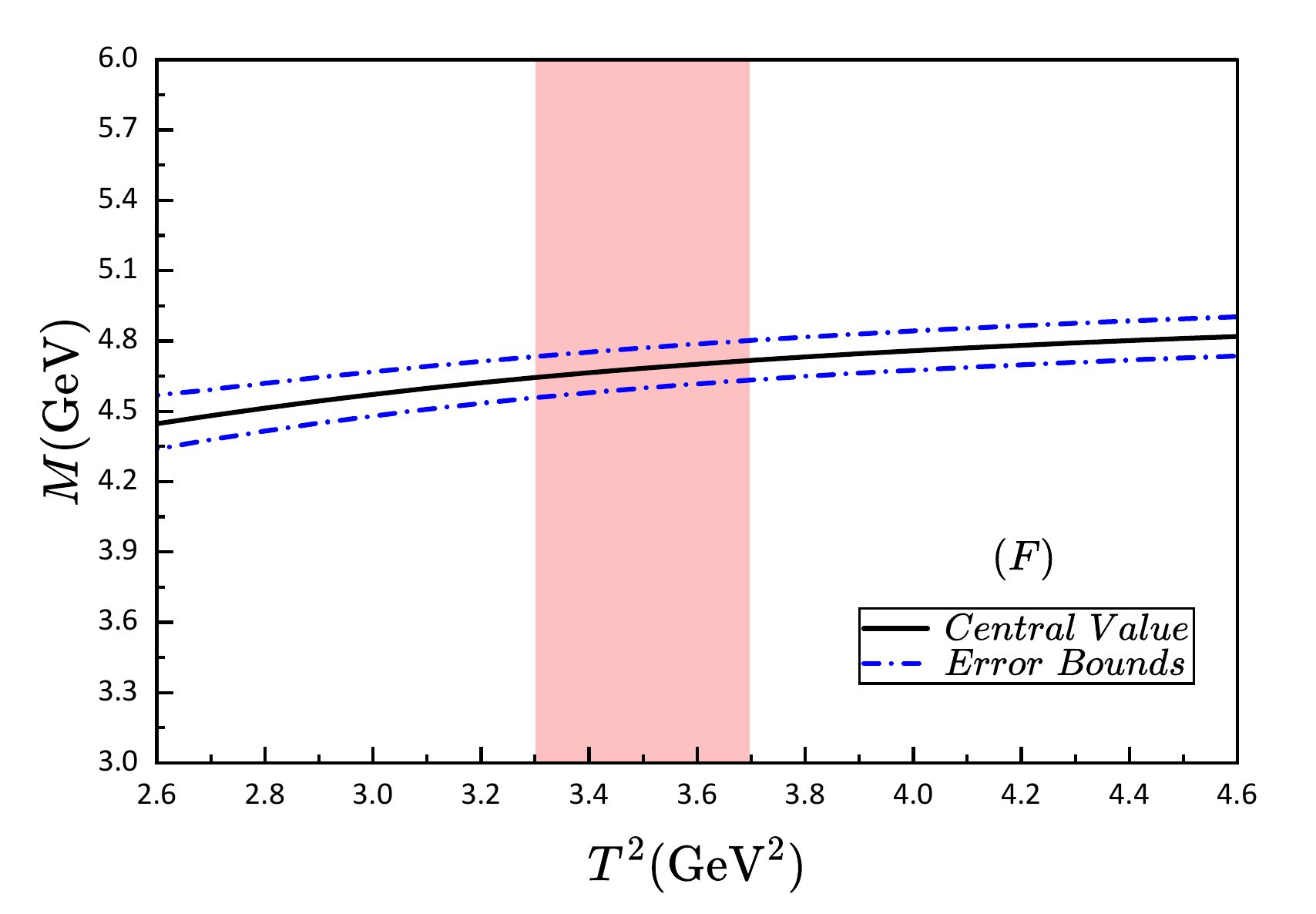}
 \includegraphics[totalheight=5cm,width=7cm]{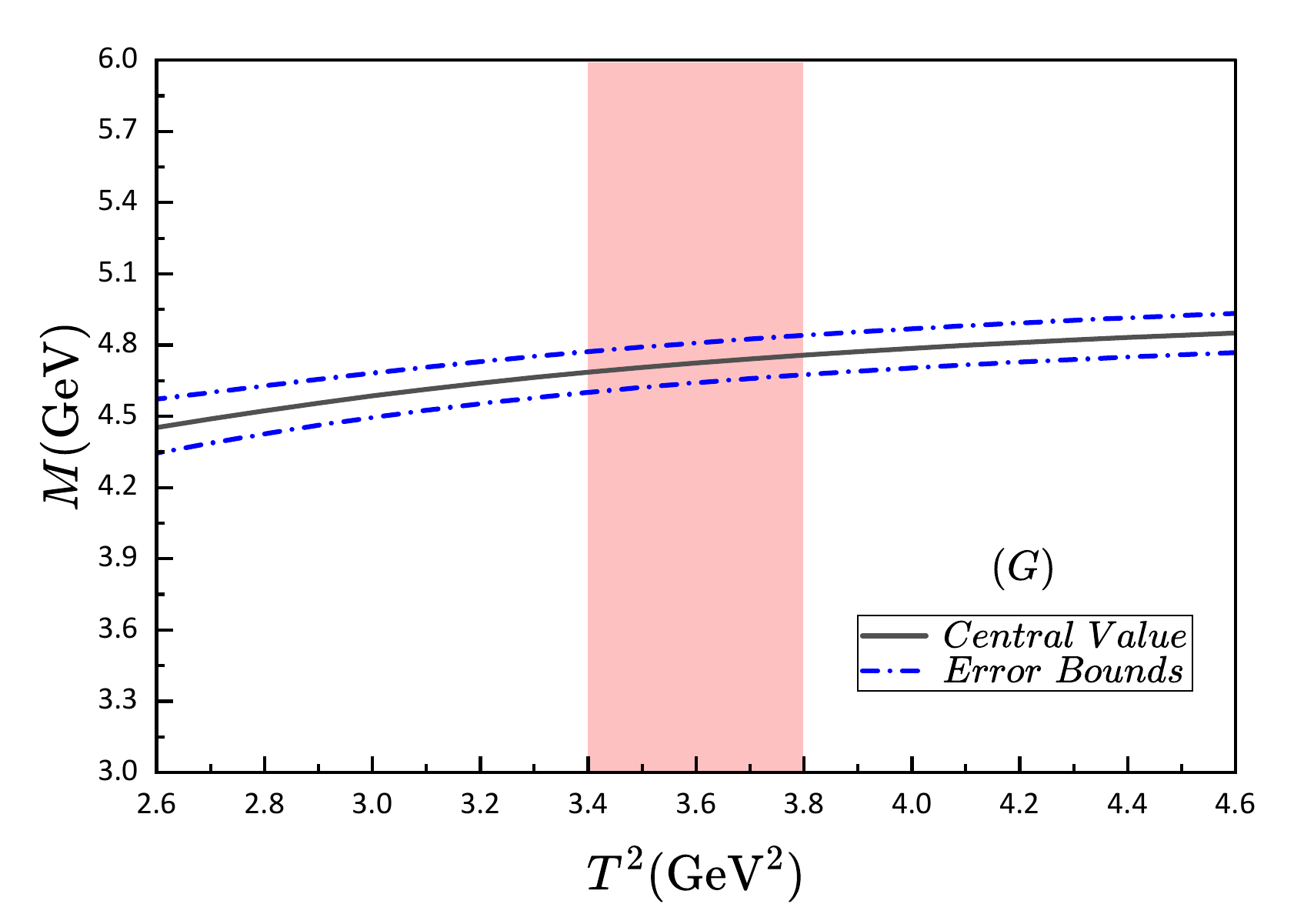}
 \includegraphics[totalheight=5cm,width=7cm]{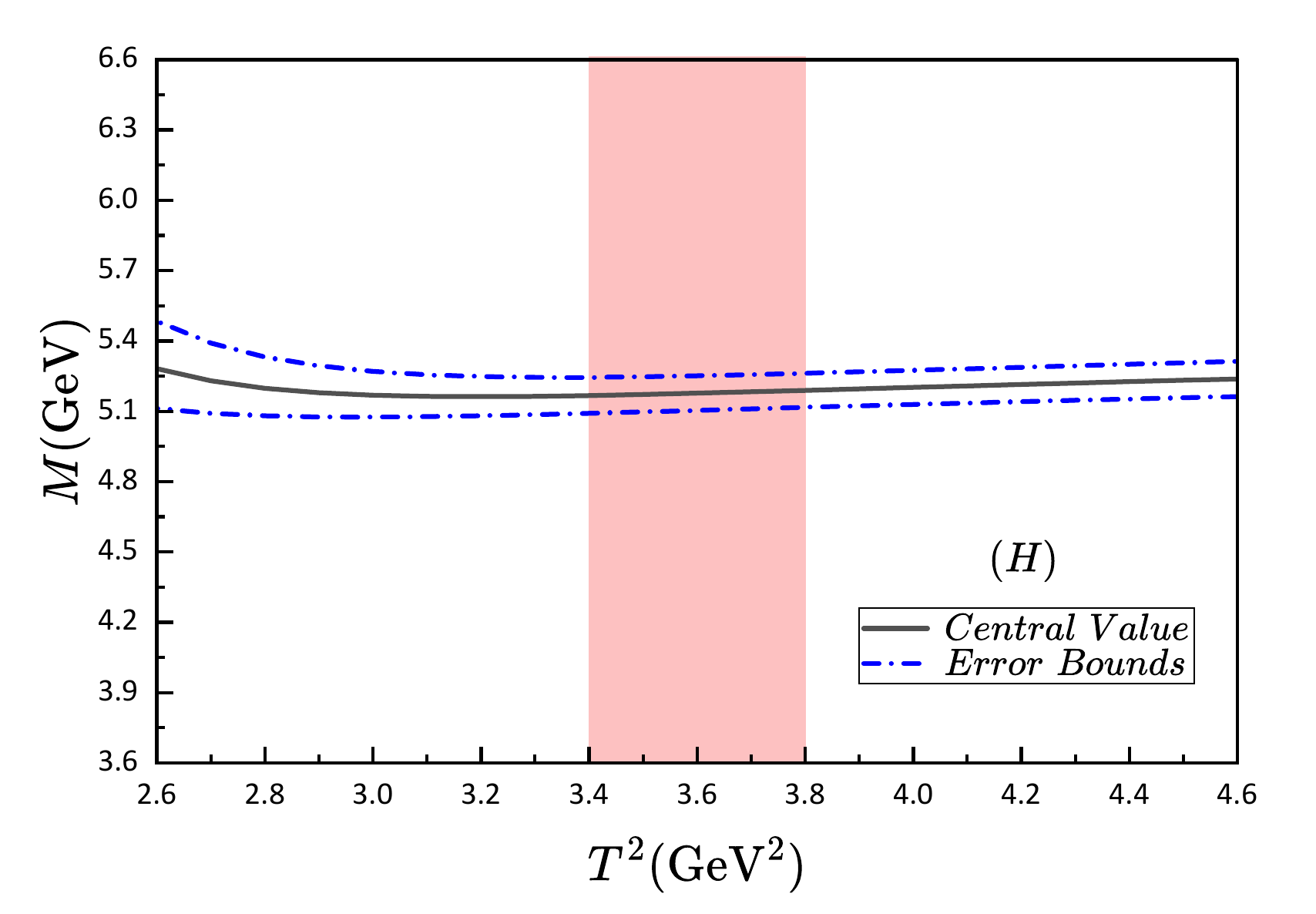}
 \caption{The $M-T^2$ curves, where $A$, $B$, $C$, $D$, $E$, $F$, $G$ and $H$ correspond to the currents $\mathcal{J}_1$, $\mathcal{J}_2$, $\mathcal{J}_{1\mu}$, $\mathcal{J}_{2\mu}$, $\mathcal{J}_1^{\pm}$, $\mathcal{J}_2^{\pm}$, $\mathcal{J}_{1\mu}^{\pm}$ and $\mathcal{J}_{2\mu}^{\pm}$, respectively.}\label{Masses-fig}
\end{figure}

We show contributions of the vacuum condensates dimensionally   for all the exotic states  in the Fig.\ref{DC-fig}, and show the $M-T^2$ curves in the Fig.\ref{Masses-fig}. Need to point out that the  threshold parameters $\sqrt{s_0}$ should be carefully considered, we should choose suitable  $\sqrt{s_0}$ to satisfy the two basic requirements of the QCD sum rules, what's more, the energy scale formula should be hold. One can consult  the similar  steps for choosing  this parameter in Ref.\cite{wangxiuwu1}.

The mass extracted from the Borel window of the pseudoscalar $\mathcal{J}_1(x)$ is $5.60$ GeV, which is $860$ MeV above the $\Lambda_c\Sigma_c$ threshold, the $\Lambda_c\Sigma_c$ dibaryon resonance  with the $J^P=0^-$ is hard to be observed  experimentally. For the scalar current $\mathcal{J}_2(x)$, the predicted mass  is $4.86^{+0.13}_{-0.11}$ GeV, which lies above the $\Lambda_c\Sigma_c$ threshold  even if one take the lower bound into consideration, so the $\Lambda_c\Sigma_c$ state with the $J^P=0^+$ is a  resonance. Similarly, the state  with the $J^P=1^-$ from the vector current $\mathcal{J}_{2\mu}(x) $ is also a dibaryon resonance. The central value of the predicted mass of the  state with the $J^P=1^+$ from the axialvector current $\mathcal{J}_{1\mu}(x)$  lies about $10$ MeV below the $\Lambda_c\Sigma_c$ threshold,   there might exist an axialvector  molecular state. In the same way, we conclude that the $\bar{\Lambda}_c\Sigma_c\pm\bar{\Sigma}_c\Lambda_c$ states with the $J^P=0^{-\pm}$ and $1^{-\pm}$ are also possible molecular states. For the baryonium states, the states with positive and negative charge-conjugations have degenerated masses and pole residues due to the particular structures, see Eq.\eqref{baryonium-current}. We take the correlation functions $\Pi_{1}^{\pm}(p^2)$ of the currents $\mathcal{J}_{1}^{\pm}(x)$ for example to interpret the  argument. After Wick contraction, the correlation functions $\Pi_{1}^{\pm}(p^2)$ on the QCD sides are derived as,
\begin{eqnarray}
\notag &&\Pi_{1}^+(p^2)=-\varepsilon^{ijk}\varepsilon^{lmn}\varepsilon^{i'j'k'}\varepsilon^{l'm'n'}\texttt{i}\int d^4x e^{\texttt{i}p\cdot x}  \\
\notag && \Big\{Tr\left[\gamma^{\mu}\gamma_5B^{nn'}(x)\gamma_5\gamma^{\mu'}B^{k'k}(-x)\right]\\
\notag &&\cdot Tr\left[\gamma_5 Q^{j'j}(-x)\gamma_{5}CQ^{i'i\texttt{T}}(-x)C\right]Tr\left[\gamma_\mu Q^{mm'}(x)\gamma_{\mu'}CQ^{ll'\texttt{T}}(x)C\right]\\
\notag &&+\frac{1}{2}Tr\left[\gamma^{\mu}\gamma_5B^{nn'}(x)\gamma_5\gamma^{\mu'}B^{k'k}(-x)\right]\\
\notag &&\cdot Tr\left[\gamma_{\mu'} Q^{j'j}(-x)\gamma_{5}CQ^{i'i\texttt{T}}(-x)C\right]Tr\left[\gamma_\mu Q^{mm'}(x)\gamma_{5}CQ^{ll'\texttt{T}}(x)C\right]\\
\notag && +\frac{1}{2}Tr\left[\gamma^{\mu}\gamma_5B^{nn'}(x)\gamma_5\gamma^{\mu'}B^{k'k}(-x)\right]\\
\notag &&\cdot Tr\left[\gamma_5 Q^{j'j}(-x)\gamma_{\mu}CQ^{i'i\texttt{T}}(-x)C\right]Tr\left[\gamma_5 Q^{mm'}(x)\gamma_{\mu'}CQ^{ll'\texttt{T}}(x)C\right]\Big\} \, ,
\end{eqnarray}
\begin{eqnarray}
\notag &&\Pi_{1}^-(p^2)=-\varepsilon^{ijk}\varepsilon^{lmn}\varepsilon^{i'j'k'}\varepsilon^{l'm'n'}\texttt{i}\int d^4x e^{\texttt{i}p\cdot x}  \\
\notag && \Big\{Tr\left[\gamma^{\mu}\gamma_5B^{nn'}(x)\gamma_5\gamma^{\mu'}B^{k'k}(-x)\right]\\
\notag &&\cdot Tr\left[\gamma_5 Q^{j'j}(-x)\gamma_{5}CQ^{i'i\texttt{T}}(-x)C\right]Tr\left[\gamma_\mu Q^{mm'}(x)\gamma_{\mu'}CQ^{ll'\texttt{T}}(x)C\right]\\
\notag &&-\frac{1}{2}Tr\left[\gamma^{\mu}\gamma_5B^{nn'}(x)\gamma_5\gamma^{\mu'}B^{k'k}(-x)\right]\\
\notag &&\cdot Tr\left[\gamma_{\mu'} Q^{j'j}(-x)\gamma_{5}CQ^{i'i\texttt{T}}(-x)C\right]Tr\left[\gamma_\mu Q^{mm'}(x)\gamma_{5}CQ^{ll'\texttt{T}}(x)C\right]\\
\notag && -\frac{1}{2}Tr\left[\gamma^{\mu}\gamma_5B^{nn'}(x)\gamma_5\gamma^{\mu'}B^{k'k}(-x)\right]\\
\notag &&\cdot Tr\left[\gamma_5 Q^{j'j}(-x)\gamma_{\mu}CQ^{i'i\texttt{T}}(-x)C\right]Tr\left[\gamma_5 Q^{mm'}(x)\gamma_{\mu'}CQ^{ll'\texttt{T}}(x)C\right]\Big\} \, ,
\end{eqnarray}
\noindent where, $Q^{ab}(x)$ and $B^{ab}(x)$ ($a$ and $b$ are the color indices) are the full light and heavy quark propagators, respectively. Note that, the difference between $u-quark$ and $d-quark$ full propagators has been neglected since the masses of light quarks are set as $m_u\approx m_d\approx0$ in the present work, the masses of the light quarks have little contribution to the final results. One can show that the terms $\varepsilon^{ijk}\varepsilon^{i'j'k'}Tr\left[\gamma_{\mu'} Q^{j'j}(-x)\gamma_{5}CQ^{i'i\texttt{T}}(-x)C\right]\equiv0$ and $\varepsilon^{ijk}\varepsilon^{i'j'k'}Tr\left[\gamma_5 Q^{j'j}(-x)\gamma_{\mu}CQ^{i'i\texttt{T}}(-x)C\right]\equiv0$, accordingly, $\Pi_{1}^-(p^2)=\Pi_{1}^+(p^2)$. Similar situations are there for the other correlation functions: $\Pi_{2}^-(p^2)=\Pi_{2}^+(p^2)$, $\Pi_{1\alpha\beta}^-(p^2)=\Pi_{1\alpha\beta}^+(p^2)$ and  $\Pi_{2\alpha\beta}^-(p^2)=\Pi_{2\alpha\beta}^+(p^2)$, where $\Pi_{2}^\pm(p^2)$ and $\Pi_{1/2\alpha\beta}^\pm(p^2)$ are the correlation functions of the currents $\mathcal{J}_2^\pm(x)$ and $\mathcal{J}_{1/2\alpha}^\pm(x)$, respectively. The detailed expressions of all the spectral densities are documented in detail in the attached appendix for reference.

In Ref.\cite{Qiao22}, Zhang and Qiao studied the mass spectrum of  the $\Lambda_Q\bar{\Sigma}_Q$ baryonium  states with the $J^P=0^+$, $0^-$, $1^+$ and $1^-$  via the QCD sum rules, and obtained the conclusion that  no $\Lambda_c\bar{\Sigma}_c$ bound state could be formed. Compared to Ref.\cite{Qiao22}, in the present work, we take account of more vacuum condensates, and resort to the energy scale formula to improve the convergent behaviors of the OPE and to enhance the pole contributions, which lead to quite different Borel windows and quite different predictions.

In the QCD sum rules, we usually choose the local currents to interpolate the tetraquark, pentaquark and hexaquark  molecular states  having two color-neutral clusters, which are not necessary to be the physical mesons and baryons. The local currents require that the molecular states have the average spatial sizes $\sqrt{\langle r^2\rangle}$ of the same magnitudes as the conventional mesons and baryons, and they are also compact objects, just like the diquark-antidiquark type
tetraquark states or diquark-diquark-antiquark type pentaquark states, and these molecular states are not necessary to be loosely bound. In the local limit, the meson-meson, meson-baryon,  baryon-baryon and baryon-antibaryon pairs lose themselves and merge into color-singlet-color-singlet type tetraquark,  pentaquark and hexaquark states \cite{WZG-color-neutral}. Thus, the so-called ``molecular states" in the QCD sum rules are different from the commonly  called  ``molecular states" completely, we call them ``molecular states" just because they have two color-neutral clusters, which have the same quantum numbers as the physical hadrons except for the masses. In fact, they are color-singlet-color-singlet type  multiquark states \cite{wangzg-zs,WangZG-Landau-PRD}. Although we usually compare the mass of the $\mathcal{Z}$ with the two-particle threshold to judge whether it is a bound state or a resonance state, the bound energy has no explicit physical meaning and has no relation to the $\mathcal{Z}$'s decay width. If we want to know the decay width, we have to study  it with the three-point QCD sum rules directly.

From Fig.\ref{DC-fig}, we observe  that the higher  dimensional contributions with $n\geq10$ play less important roles, thus the convergence of the OPE is well satisfied, which is  similar to the case of the $\Lambda_c\Lambda_c$, $\bar{\Lambda}_c\Lambda_c$, $\Sigma_c\Sigma_c$ and $\bar{\Sigma}_c\Sigma_c$ states \cite{wangxiuwu1,wangxiuwu2}.

\begin{table}
\begin{center}
\begin{tabular}{|c|c|c|c|c|c|c|c|c|}\hline\hline
  &$J^{P(C)}$ &$T^2 ({\rm GeV}^2)$    &$\sqrt{s_0}({\rm GeV})$ &$\mu ({\rm GeV})$ &$\rm PC$ &$M({\rm GeV})$ &$\lambda(10^{-3}{\rm GeV}^8)$ \\ \hline

$\mathcal{J}_1$            &$0^-$      &$3.6\sim4.0$  &$6.10\pm0.1$  &$4.2$  & $(42-53)\% $        &$5.60^{+0.08}_{-0.08}$      &$3.18^{+0.48}_{-0.45}$ \\

 $\mathcal{J}_2$           &$0^+$      &$3.3\sim3.7$  &$5.40\pm0.1$  &$3.2$  &$(42-55)\% $      &$4.86^{+0.13}_{-0.11}$      &$1.10^{+0.19}_{-0.17}$ \\

$\mathcal{J}_{1\mu}$       &$1^+$      &$3.6\sim4.0$  &$5.40\pm0.1$  &$3.0$  & $(40-52)\% $        &$4.73^{+0.08}_{-0.08}$      &$1.52^{+0.23}_{-0.22}$ \\

$\mathcal{J}_{2\mu}$       &$1^-$      &$3.2\sim3.6$  &$5.45\pm0.1$  &$3.2$  & $(40-53)\% $        &$4.88^{+0.08}_{-0.08}$      &$1.06^{+0.18}_{-0.16}$ \\
\hline

$\mathcal{J}^{\pm}_1$      &$0^{+\pm}$ &$3.3\sim3.7$  &$5.67\pm0.1$  &$3.5$  & $(43-56)\% $        &$5.10^{+0.08}_{-0.08}$      &$1.68^{+0.08}_{-0.26}$ \\

$\mathcal{J}_2^{\pm}$      &$0^{-\pm}$ &$3.3\sim3.7$  &$5.35\pm0.1$  &$2.9$  & $(40-54)\% $        &$4.69^{+0.09}_{-0.09}$      &$1.24^{+0.19}_{-0.18}$ \\

$\mathcal{J}_{1\mu}^{\pm}$ &$1^{-\pm}$ &$3.4\sim3.8$  &$5.40\pm0.1$  &$3.6$  & $(40-53)\% $        &$4.72^{+0.09}_{-0.08}$      &$1.31^{+0.20}_{-0.19}$ \\

$\mathcal{J}_{2\mu}^{\pm}$ &$1^{+\pm}$ &$3.4\sim3.8$  &$5.75\pm0.1$  &$3.6$  & $(42-54)\% $        &$5.18^{+0.08}_{-0.07}$      &$1.76^{+0.28}_{-0.25}$ \\
\hline\hline

\end{tabular}
\end{center}
\caption{ The Borel parameters, continuum threshold parameters, best energy scales of the QCD spectral densities, pole contributions, hadron masses and pole residues. }\label{BorelP-mass}
\end{table}

\begin{figure}
 \centering
 \includegraphics[totalheight=5cm,width=7cm]{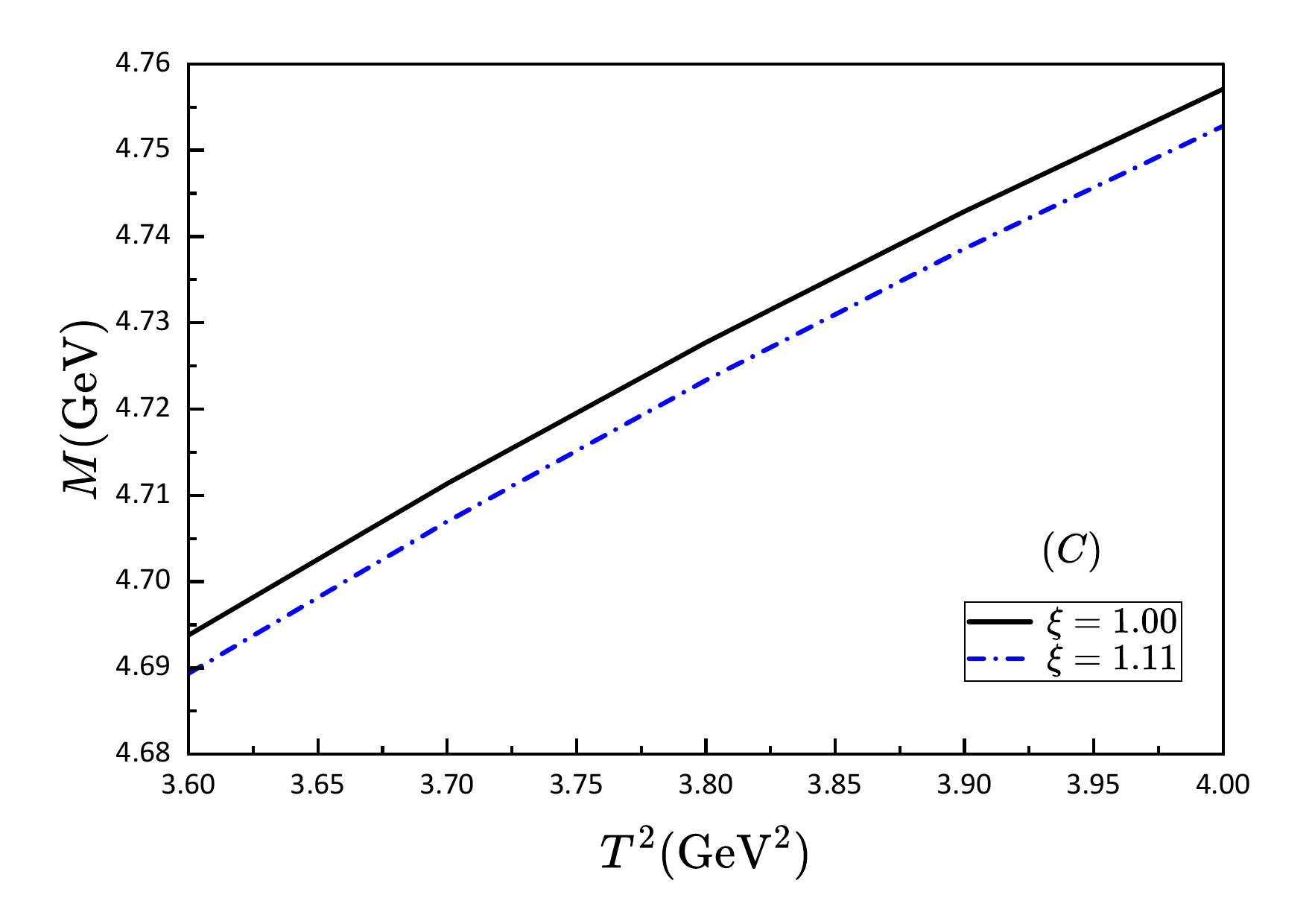}
 \caption{The $M-T^2$ curves among the Borel platforms with $\xi=1.11$ and $\xi=1.00$.}\label{Xi-fig}
\end{figure}

Now, we take the  $\Lambda_c\Sigma_c$ state with the $J^P=1^+$ as an example to illustrate  reliability of the vacuum saturation  via setting $\xi=1.11$ for all the higher dimensional vacuum condensates. As shown in  Fig.\ref{Xi-fig}, the $M-T^2$ curves only have slight difference for $\xi=1.00$ and $1.11$ under the same parameters $\sqrt{s_0}=5.40\,{\rm GeV}$, $T^2=3.6-4.0\,{\rm GeV^2}$ and $\mu=3.0\,{\rm GeV}$. The extracted mass is $4.72_{-0.07}^{+0.07}\,{\rm GeV}$, which lies  about $5\,{\rm MeV}$ below the ones  from the vacuum saturation. Considering  uncertainties of the predicted masses, the violation of  vacuum saturation is tiny. In the Borel window, the pole contribution is about $(41-53)\%$, which coincides with that from the vacuum saturation. Furthermore, the OPE converges very well. All in all,  the factorization hypothesis is reliable.

\section{Conclusions}
In the present work, we study the $\Lambda_c\Sigma_c$ dibaryon and $\bar{\Lambda}_c\Sigma_c\pm \bar{\Sigma}_c\Lambda_c$ baryonium states via the QCD sum rules. We construct four (eight) currents with definite $J^P$ ($J^{PC}$) to interpolate the dibaryon (baryonium) states without introducing orbit angular momentum, and obtain twelve QCD sum rules for these exotic states.  Then we routinely obtain the Borel platforms fulfilling requirements of the QCD sum rules, and extract the masses and pole residues. For the $\Lambda_c\Sigma_c$ dibaryon states with the $J^P=0^-$, $0^+$, $1^+$ and $1^-$, the predicted masses are $5.60^{+0.08}_{-0.08}$ GeV, $4.86^{+0.13}_{-0.11}$ GeV, $4.73^{+0.08}_{-0.08}$ GeV and $4.88^{+0.08}_{-0.08}$ GeV, respectively, the state with the $J^P=1^+$ lies below the $\Lambda_c\Sigma_c$ threshold, and  is a possible molecular state.  For the  $\bar{\Lambda}_c\Sigma_c\pm \bar{\Sigma}_c\Lambda_c$ baryonium states with the $J^{PC}=0^{+\pm}$, $0^{-\pm}$, $1^{-\pm}$ and $1^{+\pm}$, the predicted masses  are $5.10^{+0.08}_{-0.08}$ GeV, $4.69^{+0.09}_{-0.09}$ GeV, $4.72^{+0.09}_{-0.08}$ GeV and $5.18^{+0.08}_{-0.07}$ GeV, respectively. The $\bar{\Lambda}_c\Sigma_c\pm \bar{\Sigma}_c\Lambda_c$ states with the $J^P=0^{-\pm}$ and $1^{-\pm}$ lie below the $\bar{\Lambda}_c\Sigma_c$ threshold and are four  possible molecular states. The present predictions can be confronted to the experimental data in the future and make contributions to the exotic spectroscopy.

\section*{Acknowledgements}
This work is supported by the National Natural Science Foundation with Grant Number 12575083 and the Natural Science Foundation of HeBei Province with Grant Number A2024502002.

\section*{Appendix}
The spectral density $\rho(s)$ of each current is expressed in four or three types of integrals,
\begin{eqnarray}
\notag \rho(s)=\sum\limits_{\mathcal{A},\mathcal{B},\mathcal{C},\mathcal{D},n,i}\rho_{\mathcal{A},\mathcal{B},\mathcal{C},\mathcal{D}}(s,n)_i\,,
\end{eqnarray}
where, $\mathcal{A}$, $\mathcal{B}$, $\mathcal{C}$ and $\mathcal{D}$ refer to the four types of integrals, $n$ is the dimension of the condensate, $i$ represents different kinds of vacuum condensates sharing the same dimension $n$. For the following expressions, $\tilde{m}^2=\frac{m_c^2}{(1-y)y}$ and $\bar{m}^2=\frac{(y+z)m_c^2}{y z}$, $y$ and $z$ are the parameters introduced during the calculation of the integrals in the momentum space of the two heavy quarks. For simplicity, set $\int_{y_i}^{y_f}dy\int_{z_i}^{1-y}dz\rightarrow\int dydz$
and $\int_{y_i}^{y_f}dy\rightarrow\int dy\,$. For type $\mathcal{A}$ and $\mathcal{B}$, $y_i=\frac{1}{2}(1-\sqrt{1-4m_c^2/s})$, $y_f=\frac{1}{2}(1+\sqrt{1-4m_c^2/s})$ and $z_i=\frac{y m_c^2}{y s-m_c^2}$. For type $\mathcal{C}$ and $\mathcal{D}$, $y_i=0$, $y_f=1$ and $z_i=0$.

\vspace{0.5cm}
The spectral density $\rho_1(s)$ of the current $\mathcal{J}_1$:
\begin{eqnarray}
\notag  \rho_{\mathcal{A}}(s,12)_1 &=& - \frac{m_c^2 \langle\bar{q}q\rangle^4}{36 \pi^2} \int dy\, \,{\rm exp}\left(-\frac{s}{T^2}\right) \\
\notag && + \frac{\langle\bar{q}q\rangle^4}{72 \pi^2} \int dy\, (1-y) y (3s-2\tilde{m}^2)\,{\rm exp}\left(-\frac{s}{T^2}\right) \,,
\end{eqnarray}
\begin{eqnarray}
\notag \rho_{\mathcal{A}}(s,12)_2  &=& - \frac{g_s^2 m_c^2 \langle\bar{q}q\rangle^4}{3888 \pi^4} \int dy\, \,{\rm exp}\left(-\frac{s}{T^2}\right) \\
\notag && + \frac{g_s^2 \langle\bar{q}q\rangle^4}{11664 \pi^4} \int dy\, (1-y) y (3s-2\tilde{m}^2)\,{\rm exp}\left(-\frac{s}{T^2}\right) \,,
\end{eqnarray}
\begin{eqnarray}
\notag \rho_{\mathcal{A}}(s,14)  &=& -\frac{\langle\bar{q}q\rangle^3 \langle\bar{q}g_s\sigma Gq\rangle}{12 \pi^2} \int dy\,  (1-y) y \,{\rm exp}\left(-\frac{s}{T^2}\right) \\
\notag &&  + \frac{5\langle\bar{q}q\rangle^3 \langle\bar{q}g_s\sigma Gq\rangle}{576 \pi^2} \int dy\, (1-y)\,{\rm exp}\left(-\frac{s}{T^2}\right) \,,
\end{eqnarray}
\begin{eqnarray}
\notag \rho_{\mathcal{B}}(s,0) &=& -\frac{m_c^2}{19660800 \pi^{10}} \int dydz\,(1-y-z)^5 (s-\bar{m}^2)^6 \,{\rm exp}\left(-\frac{s}{T^2}\right) \\
\notag && + \frac{1}{68812800 \pi^{10}} \int dydz\,  y z (1-y-z)^6(s-\bar{m}^2)^5 (15s^2 -9s\bar{m}^2 + \bar{m}^4) \,{\rm exp}\left(-\frac{s}{T^2}\right) \,,
\end{eqnarray}
\begin{eqnarray}
\notag \rho_{\mathcal{B}}(s,4)  &=& \frac{m_c^4 \langle g_s^2GG\rangle}{5898240 \pi^{10}} \int dydz\,  \frac{(1-y-z)^5}{y^3}  (s-\bar{m}^2)^3 \,{\rm exp}\left(-\frac{s}{T^2}\right) \\
\notag && - \frac{m_c^2 \langle g_s^2GG\rangle}{7864320 \pi^{10}} \int dydz\, \frac{(1-y-z)^5}{y^2} (s-\bar{m}^2)^4 \,{\rm exp}\left(-\frac{s}{T^2}\right) \\
\notag &&  - \frac{5 m_c^2 \langle g_s^2GG\rangle}{12582912 \pi^{10}} \int dydz\, \frac{(1-y-z)^4}{y} (s-\bar{m}^2)^4 \,{\rm exp}\left(-\frac{s}{T^2}\right) \\
\notag &&   -  \frac{m_c^2 \langle g_s^2GG\rangle}{11796480 \pi^{10}} \int dydz\, \frac{z (1-y-z)^6}{y^2} (s-\bar{m}^2)^2 (7s^2 -6s\bar{m}^2 + \bar{m}^4)\,{\rm exp}\left(-\frac{s}{T^2}\right) \\
\notag && + \frac{\langle g_s^2GG\rangle}{37748736 \pi^{10}} \int dydz\,  z (1-y-z)^4  (s-\bar{m}^2)^4 (7s-5\bar{m}^2) \,{\rm exp}\left(-\frac{s}{T^2}\right) \\
\notag && - \frac{\langle g_s^2GG\rangle}{47185920 \pi^{10}} \int dydz\,  y z(1-y-z)^3 (s-\bar{m}^2)^4 (7s-2\bar{m}^2) \,{\rm exp}\left(-\frac{s}{T^2}\right) \\
\notag &&   + \frac{\langle g_s^2GG\rangle}{471859200 \pi^{10}} \int dydz\,(1-y-z)^5 (s-\bar{m}^2)^4 (7s-2\bar{m}^2) \,{\rm exp}\left(-\frac{s}{T^2}\right) \\
\notag &&  +  \frac{\langle g_s^2GG\rangle}{23592960 \pi^{10}} \int dydz\,  y z (1-y-z)^4  (s-\bar{m}^2)^3 (18s^2 -11s\bar{m}^2 + 3\bar{m}^4) \,{\rm exp}\left(-\frac{s}{T^2}\right) \\
\notag &&  +  \frac{\langle g_s^2GG\rangle}{47185920 \pi^{10}} \int dydz\, z(1-y-z)^5 (s-\bar{m}^2)^3 (28s^2 -21s\bar{m}^2 + 3\bar{m}^4)\,{\rm exp}\left(-\frac{s}{T^2}\right) \\
\notag &&   -  \frac{\langle g_s^2GG\rangle}{1415577600 \pi^{10}} \int dydz\,  (1-y-z)^6  (s-\bar{m}^2)^3 (28s^2 -21s\bar{m}^2 + 3\bar{m}^4)\,{\rm exp}\left(-\frac{s}{T^2}\right) \,,
\end{eqnarray}
\begin{eqnarray}
\notag \rho_{\mathcal{B}}(s,6)_1  &=& - \frac{m_c^2 \langle\bar{q}q\rangle^2}{1536 \pi^6} \int dydz\,  (1-y-z)^2 (s-\bar{m}^2)^3 \,{\rm exp}\left(-\frac{s}{T^2}\right) \\
\notag && + \frac{\langle\bar{q}q\rangle^2}{9216 \pi^6} \int dydz\,  y z (1-y-z)^2  (s-\bar{m}^2)^3 (3s-\bar{m}^2) \,{\rm exp}\left(-\frac{s}{T^2}\right) \\
\notag &&  +  \frac{\langle\bar{q}q\rangle^2}{4608 \pi^6} \int dydz\,  y z(1-y-z)^3 (s-\bar{m}^2)^2 ( 7s^2 - 6s\bar{m}^2 + \bar{m}^4)\,{\rm exp}\left(-\frac{s}{T^2}\right) \,,
\end{eqnarray}
\begin{eqnarray}
\notag \rho_{\mathcal{B}}(s,6)_2  &=& -\frac{g_s^2 m_c^2 \langle\bar{q}q\rangle^2}{41472 \pi^8} \int dydz\, (1-y-z)^2 (s-\bar{m}^2)^3 \,{\rm exp}\left(-\frac{s}{T^2}\right) \\
\notag && + \frac{g_s^2 \langle\bar{q}q\rangle^2}{41472 \pi^8} \int dydz\, y z(1-y-z)^3 (s-\bar{m}^2)^2 ( 7s^2 - 6s\bar{m}^2 + \bar{m}^4) \,{\rm exp}\left(-\frac{s}{T^2}\right) \,,
\end{eqnarray}
\begin{eqnarray}
\notag \rho_{\mathcal{B}}(s,8)  &=&  \frac{25 m_c^2 \langle\bar{q}q\rangle \langle\bar{q}g_s\sigma Gq\rangle}{6144 \pi^6} \int dydz\,(1-y-z) (s-\bar{m}^2)^2 \,{\rm exp}\left(-\frac{s}{T^2}\right) \\
\notag &&  - \frac{5 m_c^2 \langle\bar{q}q\rangle \langle\bar{q}g_s\sigma Gq\rangle}{12288 \pi^6} \int dydz\,  \frac{(1-y-z)^2}{y} (s-\bar{m}^2)^2 \,{\rm exp}\left(-\frac{s}{T^2}\right) \\
\notag &&  - \frac{25 \langle\bar{q}q\rangle \langle\bar{q}g_s\sigma Gq\rangle}{55296 \pi^6} \int dydz\, y z(1-y-z) (s-\bar{m}^2)^2 (5s-2\bar{m}^2) \,{\rm exp}\left(-\frac{s}{T^2}\right) \\
\notag && + \frac{5 \langle\bar{q}q\rangle \langle\bar{q}g_s\sigma Gq\rangle}{73728 \pi^6} \int dydz\, z (1-y-z)^2 (s-\bar{m}^2)^2 (5s-2\bar{m}^2) \,{\rm exp}\left(-\frac{s}{T^2}\right) \\
\notag &&  -  \frac{25\langle\bar{q}q\rangle \langle\bar{q}g_s\sigma Gq\rangle}{9216 \pi^6} \int dydz\, y z (1-y-z)^2 (s-\bar{m}^2) ( 5s^2 - 5s\bar{m}^2 + \bar{m}^4) \,{\rm exp}\left(-\frac{s}{T^2}\right) \,,
\end{eqnarray}
\begin{eqnarray}
\notag \rho_{\mathcal{B}}(s,10)_1 &=& - \frac{491 m_c^2 \langle\bar{q}g_s\sigma Gq\rangle^2}{294912 \pi^6} \int dydz\,(s-\bar{m}^2)\,{\rm exp}\left(-\frac{s}{T^2}\right) \\
\notag && + \frac{85 m_c^2 \langle\bar{q}g_s\sigma Gq\rangle^2}{73728 \pi^6} \int dydz\,  \frac{(1-y-z)}{y}(s-\bar{m}^2)\,{\rm exp}\left(-\frac{s}{T^2}\right) \\
\notag &&  + \frac{m_c^2 \langle\bar{q}g_s\sigma Gq\rangle^2}{131072 \pi^6} \int dydz\, \frac{(1-y-z)^2}{y z}(s-\bar{m}^2) \,{\rm exp}\left(-\frac{s}{T^2}\right) \\
\notag && + \frac{491 \langle\bar{q}g_s\sigma Gq\rangle^2}{884736 \pi^6} \int dydz\,  y z (s-\bar{m}^2) (2s-\bar{m}^2) \,{\rm exp}\left(-\frac{s}{T^2}\right) \\
\notag && - \frac{155 \langle\bar{q}g_s\sigma Gq\rangle^2}{294912 \pi^6} \int dydz\,  z(1-y-z) (s-\bar{m}^2)(2s-\bar{m}^2) \,{\rm exp}\left(-\frac{s}{T^2}\right) \\
\notag &&  - \frac{203 \langle\bar{q}g_s\sigma Gq\rangle^2}{7077888 \pi^6} \int dydz\, (1-y-z)^2 (s-\bar{m}^2) (2s-\bar{m}^2) \,{\rm exp}\left(-\frac{s}{T^2}\right) \\
\notag &&  -  \frac{5 \langle\bar{q}g_s\sigma Gq\rangle^2}{49152 \pi^6} \int dydz\, z (1-y-z)^2 (3s^2-3s\bar{m}^2+\bar{m}^4)\,{\rm exp}\left(-\frac{s}{T^2}\right) \\
\notag &&  + \frac{491 \langle\bar{q}g_s\sigma Gq\rangle^2}{442368 \pi^6} \int dydz\, y z(1-y-z) (10s^2-12s\bar{m}^2+3\bar{m}^4)\,{\rm exp}\left(-\frac{s}{T^2}\right) \\
\notag &&  +  \frac{331 \langle\bar{q}g_s\sigma Gq\rangle^2}{10616832 \pi^6} \int dydz\,(1-y-z)^3 (10s^2-12s\bar{m}^2+3\bar{m}^4) \,{\rm exp}\left(-\frac{s}{T^2}\right) \,,
\end{eqnarray}
\begin{eqnarray}
\notag \rho_{\mathcal{B}}(s,10)_2 &=& \frac{7 m_c^4 \langle g_s^2GG\rangle \langle\bar{q}q\rangle^2 }{62208 \pi^6} \int dydz\,  \frac{(1-y-z)^2}{y^3} \,{\rm exp}\left(-\frac{s}{T^2}\right) \\
\notag && - \frac{11 m_c^2 \langle g_s^2GG\rangle \langle\bar{q}q\rangle^2}{73728 \pi^6} \int dydz\, (s-\bar{m}^2)\,{\rm exp}\left(-\frac{s}{T^2}\right) \\
\notag && - \frac{7 m_c^2 \langle g_s^2GG\rangle \langle\bar{q}q\rangle^2}{20736 \pi^6} \int dydz\, \frac{(1-y-z)^2}{y^2}(s-\bar{m}^2) \,{\rm exp}\left(-\frac{s}{T^2}\right) \\
\notag && - \frac{5 m_c^2 \langle g_s^2GG\rangle \langle\bar{q}q\rangle^2}{36864 \pi^6} \int dydz\,\frac{(1-y-z)}{y}(s-\bar{m}^2) \,{\rm exp}\left(-\frac{s}{T^2}\right) \\
\notag && - \frac{m_c^2 \langle g_s^2GG\rangle \langle\bar{q}q\rangle^2}{27648 \pi^6} \int dydz\, \frac{z (1-y-z)^2}{y^2}  (3s - 2\bar{m}^2) \,{\rm exp}\left(-\frac{s}{T^2}\right) \\
\notag && - \frac{5 m_c^2 \langle g_s^2GG\rangle \langle\bar{q}q\rangle^2}{31104 \pi^6} \int dydz\, \frac{z(1-y-z)^3}{y^2} (2s-\bar{m}^2) \,{\rm exp}\left(-\frac{s}{T^2}\right) \\
\notag && + \frac{11 \langle g_s^2GG\rangle \langle\bar{q}q\rangle^2}{221184 \pi^6} \int dydz\, y z (s-\bar{m}^2) (2s-\bar{m}^2) \,{\rm exp}\left(-\frac{s}{T^2}\right) \\
\notag &&  + \frac{5\langle g_s^2GG\rangle \langle\bar{q}q\rangle^2}{110592 \pi^6} \int dydz\,z(1-y-z) (s-\bar{m}^2) (2s-\bar{m}^2) \,{\rm exp}\left(-\frac{s}{T^2}\right) \\
\notag && + \frac{5 \langle g_s^2GG\rangle \langle\bar{q}q\rangle^2}{221184 \pi^6} \int dydz\,  (1-y-z)^2 (s-\bar{m}^2) (2s-\bar{m}^2) \,{\rm exp}\left(-\frac{s}{T^2}\right) \\
\notag && +  \frac{5 \langle g_s^2GG\rangle \langle\bar{q}q\rangle^2}{110592 \pi^6} \int dydz\,  z (1-y-z)^2 (10s^2-12s\bar{m}^2+3\bar{m}^4) \,{\rm exp}\left(-\frac{s}{T^2}\right) \\
\notag && - \frac{\langle g_s^2GG\rangle \langle\bar{q}q\rangle^2}{331776 \pi^6} \int dydz\, (1-y-z)^3 (10s^2-12s\bar{m}^2+3\bar{m}^4) \,{\rm exp}\left(-\frac{s}{T^2}\right) \\
\notag && + \frac{11 \langle g_s^2GG\rangle \langle\bar{q}q\rangle^2}{110592 \pi^6} \int dydz\, y z (1-y-z) (10s^2-12s\bar{m}^2+3\bar{m}^4) \,{\rm exp}\left(-\frac{s}{T^2}\right) \,,
\end{eqnarray}
\begin{eqnarray}
\notag \rho_{\mathcal{B}}(s,12) &=& \frac{g_s^2 \langle\bar{q}q\rangle^4}{972 \pi^4} \int dydz\, y z (2s-\bar{m}^2) \,{\rm exp}\left(-\frac{s}{T^2}\right) \,,
\end{eqnarray}
\begin{eqnarray}
\notag \rho_{\mathcal{C}}(s,14) &=& \frac{m_c^2 \langle\bar{q}q\rangle^3 \langle\bar{q}g_s\sigma Gq\rangle}{36 \pi^2} \int dy\,\delta(s-\tilde{m}^2) \left(\frac{s}{T^2} + 1\right)\,{\rm exp}\left(-\frac{s}{T^2}\right) \\
\notag && - \frac{5 m_c^2 \langle\bar{q}q\rangle^3 \langle\bar{q}g_s\sigma Gq\rangle}{576 \pi^2} \int dy\,\delta(s-\tilde{m}^2) \frac{1}{y} \,{\rm exp}\left(-\frac{s}{T^2}\right) \\
\notag && - \frac{\langle\bar{q}q\rangle^3 \langle\bar{q}g_s\sigma Gq\rangle}{72 \pi^2} \int dy\,\delta(s-\tilde{m}^2)  (1-y) y  \left(\frac{s^2}{T^2} + 4s \right)\,{\rm exp}\left(-\frac{s}{T^2}\right) \\
\notag &&  + \frac{5 \langle\bar{q}q\rangle^3 \langle\bar{q}g_s\sigma Gq\rangle}{1152 \pi^2} \int dy\, \delta(s-\tilde{m}^2) (1-y) s\,{\rm exp}\left(-\frac{s}{T^2}\right) \,,
\end{eqnarray}
\begin{eqnarray}
\notag \rho_{\mathcal{C}}(s,16)_1 &=& - \frac{m_c^2 \langle\bar{q}q\rangle^2 \langle\bar{q}g_s\sigma Gq\rangle^2}{96 \pi^2 T^6} \int dy\,\delta(s-\tilde{m}^2) s^2 \,{\rm exp}\left(-\frac{s}{T^2}\right) \\
\notag && + \frac{5 m_c^2 \langle\bar{q}q\rangle^2 \langle\bar{q}g_s\sigma Gq\rangle^2}{768 \pi^2 T^4} \int dy\,\delta(s-\tilde{m}^2) \frac{s}{y} \,{\rm exp}\left(-\frac{s}{T^2}\right) \\
\notag && - \frac{17 m_c^2 \langle\bar{q}q\rangle^2 \langle\bar{q}g_s\sigma Gq\rangle^2}{18432 \pi^2 T^2} \int dy\,\delta(s-\tilde{m}^2) \frac{1}{(1-y) y} \,{\rm exp}\left(-\frac{s}{T^2}\right) \\
\notag &&  +  \frac{\langle\bar{q}q\rangle^2 \langle\bar{q}g_s\sigma Gq\rangle^2}{192 \pi^2} \int dy\,\delta(s-\tilde{m}^2) (1-y) y \left(\frac{s^3}{T^6} +\frac{3s^2}{T^4} +\frac{6s}{T^2} +6 \right)\,{\rm exp}\left(-\frac{s}{T^2}\right) \\
\notag && - \frac{5 \langle\bar{q}q\rangle^2 \langle\bar{q}g_s\sigma Gq\rangle^2}{1536 \pi^2} \int dy\,\delta(s-\tilde{m}^2) (1-y) \left(\frac{s^2}{T^4} +\frac{2s}{T^2} + 2 \right)\,{\rm exp}\left(-\frac{s}{T^2}\right) \\
\notag && + \frac{17 \langle\bar{q}q\rangle^2 \langle\bar{q}g_s\sigma Gq\rangle^2}{110592 \pi^2} \int dy\,\delta(s-\tilde{m}^2) \left(\frac{s}{T^2} +1 \right)\,{\rm exp}\left(-\frac{s}{T^2}\right) \,,
\end{eqnarray}
\begin{eqnarray}
\notag \rho_{\mathcal{C}}(s,16)_2 &=& \frac{109 m_c^4 \langle g_s^2GG\rangle \langle\bar{q}q\rangle^4}{139968 \pi^2 T^4} \int dy\,\delta(s-\tilde{m}^2) \frac{1}{y^3} \,{\rm exp}\left(-\frac{s}{T^2}\right) \\
\notag && - \frac{m_c^2 \langle g_s^2GG\rangle \langle\bar{q}q\rangle^4}{1296 \pi^2 T^6} \int dy\,\delta(s-\tilde{m}^2) s^2 \,{\rm exp}\left(-\frac{s}{T^2}\right) \\
\notag &&  +  \frac{\langle g_s^2GG\rangle \langle\bar{q}q\rangle^4}{2592 \pi^2} \int dy\,\delta(s-\tilde{m}^2)  (1-y) y \left(\frac{s^3}{T^6} +\frac{3s^2}{T^4} +\frac{6s}{T^2} + 6 \right)\,{\rm exp}\left(-\frac{s}{T^2}\right) \\
\notag && + \frac{\langle g_s^2GG\rangle \langle\bar{q}q\rangle^4}{6912 \pi^2}\int dy\,\delta(s-\tilde{m}^2)  \left(\frac{s}{T^2} +1 \right)\,{\rm exp}\left(-\frac{s}{T^2}\right) \,,
\end{eqnarray}
\begin{eqnarray}
\notag \rho_{\mathcal{D}}(s,12)  &=& \frac{g_s^2 \langle\bar{q}q\rangle^4}{5832 \pi^4} \int dydz\,\delta(s-\bar{m}^2) y z s^2\,{\rm exp}\left(-\frac{s}{T^2}\right) \,,
\end{eqnarray}
\begin{eqnarray}
\notag \rho_{\mathcal{D}}(s,10) &=& -\frac{m_c^2 \langle g_s^2GG\rangle \langle\bar{q}q\rangle^2}{41472 \pi^6} \int dydz\,\delta(s-\bar{m}^2) \frac{z(1-y-z)^3}{y^2} s^2\,{\rm exp}\left(-\frac{s}{T^2}\right) \,.
\end{eqnarray}
The spectral density $\rho_2(s)$ of the current $\mathcal{J}_2$:
\begin{eqnarray}
\notag \rho_{\mathcal{A}}(s,12)_1 &=& \frac{m_c^2 \langle\bar{q}q\rangle^4}{54 \pi^2} \int dy\, \,{\rm exp}\left(-\frac{s}{T^2}\right) \\
\notag &&  + \frac{\langle\bar{q}q\rangle^4}{108 \pi^2} \int dy\, (1-y) y (3s-2\tilde{m}^2)\,{\rm exp}\left(-\frac{s}{T^2}\right) \,,
\end{eqnarray}
\begin{eqnarray}
\notag \rho_{\mathcal{A}}(s,12)_2 &=& \frac{g_s^2 m_c^2 \langle\bar{q}q\rangle^4}{1944 \pi^4} \int dy\, \,{\rm exp}\left(-\frac{s}{T^2}\right) \\
\notag &&  + \frac{g_s^2 \langle\bar{q}q\rangle^4}{5832 \pi^4} \int dy\, (1-y) y (3s - 2\tilde{m}^2)\,{\rm exp}\left(-\frac{s}{T^2}\right) \,,
\end{eqnarray}
\begin{eqnarray}
\notag \rho_{\mathcal{A}}(s,14) &=& - \frac{\langle\bar{q}q\rangle^3 \langle\bar{q}g_s\sigma Gq\rangle}{18 \pi^2} \int dy\, (1-y) y \,{\rm exp}\left(-\frac{s}{T^2}\right) \\
\notag &&    + \frac{\langle\bar{q}q\rangle^3 \langle\bar{q}g_s\sigma Gq\rangle}{1728 \pi^2}\int dy\, (1-y)\,{\rm exp}\left(-\frac{s}{T^2}\right) \,,
\end{eqnarray}
\begin{eqnarray}
\notag \rho_{\mathcal{B}}(s,0) &=& \frac{m_c^2}{29491200 \pi^{10}} \int dydz\, (1-y-z)^5 (s-\bar{m}^2)^6  \,{\rm exp}\left(-\frac{s}{T^2}\right) \\
\notag &&   + \frac{1}{103219200 \pi^{10}} \int dydz\, y z (1-y-z)^6 (s-\bar{m}^2)^5 (15s^2-9s\bar{m}^2+\bar{m}^4) \,{\rm exp}\left(-\frac{s}{T^2}\right) \,,
\end{eqnarray}
\begin{eqnarray}
\notag \rho_{\mathcal{B}}(s,4) &=&  -\frac{m_c^4 \langle g_s^2GG\rangle}{8847360 \pi^{10}} \int dydz\, \frac{(1-y-z)^5}{y^3} (s-\bar{m}^2)^3 \,{\rm exp}\left(-\frac{s}{T^2}\right) \\
\notag && + \frac{m_c^2 \langle g_s^2GG\rangle}{11796480 \pi^{10}} \int dydz\, \frac{(1-y-z)^5}{y^2} (s-\bar{m}^2)^4 \,{\rm exp}\left(-\frac{s}{T^2}\right) \\
\notag &&  + \frac{m_c^2 \langle g_s^2GG\rangle}{37748736 \pi^{10}} \int dydz\, \frac{(1-y-z)^4}{y} (s-\bar{m}^2)^4 \,{\rm exp}\left(-\frac{s}{T^2}\right) \\
\notag && - \frac{m_c^2 \langle g_s^2GG\rangle}{3145728 \pi^{10}} \int dydz\,(1-y-z)^3 (s-\bar{m}^2)^4  \,{\rm exp}\left(-\frac{s}{T^2}\right) \\
\notag &&   - \frac{m_c^2 \langle g_s^2GG\rangle}{17694720 \pi^{10}} \int dydz\, \frac{z (1-y-z)^6}{y^2} (s-\bar{m}^2)^2 (7s^2-6s\bar{m}^2+\bar{m}^4) \,{\rm exp}\left(-\frac{s}{T^2}\right) \\
\notag &&  + \frac{\langle g_s^2GG\rangle}{566231040 \pi^{10}} \int dydz\, z(1-y-z)^4 (s-\bar{m}^2)^4 (7s-2\bar{m}^2) \,{\rm exp}\left(-\frac{s}{T^2}\right) \\
\notag &&  - \frac{11 \langle g_s^2GG\rangle}{141557760 \pi^{10}} \int dydz\, y z(1-y-z)^3 (s-\bar{m}^2)^4 (7s-2\bar{m}^2) \,{\rm exp}\left(-\frac{s}{T^2}\right) \\
\notag && - \frac{\langle g_s^2GG\rangle}{353894400 \pi^{10}} \int dydz\,(1-y-z)^5 (s-\bar{m}^2)^4 (7s-2\bar{m}^2) \,{\rm exp}\left(-\frac{s}{T^2}\right) \\
\notag &&  + \frac{\langle g_s^2GG\rangle}{707788800 \pi^{10}} \int dydz\, z(1-y-z)^5 (s-\bar{m}^2)^3 (28s^2-21s\bar{m}^2+3\bar{m}^4) \,{\rm exp}\left(-\frac{s}{T^2}\right) \\
\notag &&  + \frac{13 \langle g_s^2GG\rangle}{141557760 \pi^{10}} \int dydz\, y z (1-y-z)^4 (s-\bar{m}^2)^3 (28s^2-21s\bar{m}^2+3\bar{m}^4) \,{\rm exp}\left(-\frac{s}{T^2}\right) \\
\notag &&  + \frac{\langle g_s^2GG\rangle}{1061683200 \pi^{10}} \int dydz\, (1-y-z)^6 (s-\bar{m}^2)^3 (28s^2-21s\bar{m}^2+3\bar{m}^4) \,{\rm exp}\left(-\frac{s}{T^2}\right) \,,
\end{eqnarray}
\begin{eqnarray}
\notag \rho_{\mathcal{B}}(s,6)_1 &=& \frac{m_c^2 \langle\bar{q}q\rangle^2}{768 \pi^6} \int dydz\, (1-y-z)^2 (s-\bar{m}^2)^3\,{\rm exp}\left(-\frac{s}{T^2}\right) \\
\notag &&  + \frac{\langle\bar{q}q\rangle^2}{4608 \pi^6} \int dydz\, y z (1-y-z)^2 (s-\bar{m}^2)^3 (3s-\bar{m}^2) \,{\rm exp}\left(-\frac{s}{T^2}\right) \\
\notag &&  + \frac{\langle\bar{q}q\rangle^2}{2304 \pi^6} \int dydz\, y z(1-y-z)^3 (s-\bar{m}^2)^2 (7s^2-6s\bar{m}^2+\bar{m}^4)\,{\rm exp}\left(-\frac{s}{T^2}\right) \,,
\end{eqnarray}
\begin{eqnarray}
\notag \rho_{\mathcal{B}}(s,6)_2 &=& \frac{g_s^2 m_c^2 \langle\bar{q}q\rangle^2}{62208 \pi^8} \int dydz\, (1-y-z)^2 (s-\bar{m}^2)^3 \,{\rm exp}\left(-\frac{s}{T^2}\right) \\
\notag && + \frac{g_s^2 \langle\bar{q}q\rangle^2}{62208 \pi^8} \int dydz\, y z(1-y-z)^3 (s-\bar{m}^2)^2 (7s^2-6s\bar{m}^2+\bar{m}^4) \,{\rm exp}\left(-\frac{s}{T^2}\right) \,,
\end{eqnarray}
\begin{eqnarray}
\notag \rho_{\mathcal{B}}(s,8) &=&  -\frac{5 m_c^2 \langle\bar{q}q\rangle \langle\bar{q}g_s\sigma Gq\rangle}{1024 \pi^6} \int dydz\,(1-y-z) (s-\bar{m}^2)^2 \,{\rm exp}\left(-\frac{s}{T^2}\right) \\
\notag &&  + \frac{m_c^2 \langle\bar{q}q\rangle \langle\bar{q}g_s\sigma Gq\rangle}{12288 \pi^6} \int dydz\, \frac{(1-y-z)^2}{y} (s-\bar{m}^2)^2 \,{\rm exp}\left(-\frac{s}{T^2}\right) \\
\notag && - \frac{5 \langle\bar{q}q\rangle \langle\bar{q}g_s\sigma Gq\rangle}{9216 \pi^6} \int dydz\, y z(1-y-z) (s-\bar{m}^2)^2 (5s-2\bar{m}^2) \,{\rm exp}\left(-\frac{s}{T^2}\right) \\
\notag &&  + \frac{\langle\bar{q}q\rangle \langle\bar{q}g_s\sigma Gq\rangle}{24576 \pi^6} \int dydz\, z (1-y-z)^2 (s-\bar{m}^2)^2 (5s-2\bar{m}^2) \,{\rm exp}\left(-\frac{s}{T^2}\right) \\
\notag && - \frac{\langle\bar{q}q\rangle \langle\bar{q}g_s\sigma Gq\rangle}{4608 \pi^6} \int dydz\, z(1-y-z)^3 (s-\bar{m}^2) (5s^2-5s\bar{m}^2+\bar{m}^4) \,{\rm exp}\left(-\frac{s}{T^2}\right) \\
\notag &&  - \frac{5 \langle\bar{q}q\rangle \langle\bar{q}g_s\sigma Gq\rangle}{1536 \pi^6} \int dydz\, y z (1-y-z)^2 (s-\bar{m}^2) (5s^2-5s\bar{m}^2+\bar{m}^4) \,{\rm exp}\left(-\frac{s}{T^2}\right) \,,
\end{eqnarray}
\begin{eqnarray}
\notag \rho_{\mathcal{B}}(s,10)_1 &=& \frac{7025 m_c^2 \langle\bar{q}g_s\sigma Gq\rangle^2}{4718592 \pi^6} \int dydz\, (s-\bar{m}^2)\,{\rm exp}\left(-\frac{s}{T^2}\right) \\
\notag && + \frac{13 m_c^2 \langle\bar{q}g_s\sigma Gq\rangle^2}{294912 \pi^6} \int dydz\, \frac{(1-y-z)}{y} (s-\bar{m}^2) \,{\rm exp}\left(-\frac{s}{T^2}\right) \\
\notag &&   - \frac{61 m_c^2 \langle\bar{q}g_s\sigma Gq\rangle^2}{393216 \pi^6} \int dydz\, \frac{(1-y-z)^2}{y z} (s-\bar{m}^2) \,{\rm exp}\left(-\frac{s}{T^2}\right) \\
\notag &&   + \frac{55\langle\bar{q}g_s\sigma Gq\rangle^2}{110592 \pi^6} \int dydz\, y z (s-\bar{m}^2) (2s-\bar{m}^2) \,{\rm exp}\left(-\frac{s}{T^2}\right) \\
\notag && - \frac{53 \langle\bar{q}g_s\sigma Gq\rangle^2}{884736 \pi^6} \int dydz\, z(1-y-z) (s-\bar{m}^2) (2s-\bar{m}^2) \,{\rm exp}\left(-\frac{s}{T^2}\right) \\
\notag &&   -  \frac{215 \langle\bar{q}g_s\sigma Gq\rangle^2}{7077888 \pi^6} \int dydz\, (1-y-z)^2 (s-\bar{m}^2) (2s-\bar{m}^2) \,{\rm exp}\left(-\frac{s}{T^2}\right) \\
\notag &&  + \frac{55\langle\bar{q}g_s\sigma Gq\rangle^2}{55296 \pi^6} \int dydz\, y z(1-y-z) (10s^2-12s\bar{m}^2+3\bar{m}^4) \,{\rm exp}\left(-\frac{s}{T^2}\right) \\
\notag &&   + \frac{145 \langle\bar{q}g_s\sigma Gq\rangle^2}{884736 \pi^6} \int dydz\, z (1-y-z)^2 (10s^2-12s\bar{m}^2+3\bar{m}^4) \,{\rm exp}\left(-\frac{s}{T^2}\right) \\
\notag && - \frac{23 \langle\bar{q}g_s\sigma Gq\rangle^2}{10616832 \pi^6} \int dydz\,(1-y-z)^3 (10s^2-12s\bar{m}^2+3\bar{m}^4) \,{\rm exp}\left(-\frac{s}{T^2}\right) \,,
\end{eqnarray}
\begin{eqnarray}
\notag \rho_{\mathcal{B}}(s,10)_2 &=& -  \frac{41 m_c^4 \langle g_s^2GG\rangle \langle\bar{q}q\rangle^2}{186624 \pi^6} \int dydz\, \frac{(1-y-z)^2}{y^3} \,{\rm exp}\left(-\frac{s}{T^2}\right) \\
\notag &&  + \frac{5 m_c^2 \langle g_s^2GG\rangle \langle\bar{q}q\rangle^2}{24576 \pi^6} \int dydz\, (s-\bar{m}^2) \,{\rm exp}\left(-\frac{s}{T^2}\right) \\
\notag &&  +  \frac{41 m_c^2 \langle g_s^2GG\rangle \langle\bar{q}q\rangle^2}{62208 \pi^6} \int dydz\,   \frac{(1-y-z)^2}{y^2} (s-\bar{m}^2) \,{\rm exp}\left(-\frac{s}{T^2}\right) \\
\notag && - \frac{5 m_c^2 \langle g_s^2GG\rangle \langle\bar{q}q\rangle^2}{36864 \pi^6} \int dydz\, \frac{(1-y-z)}{y}  (s-\bar{m}^2)  \,{\rm exp}\left(-\frac{s}{T^2}\right) \\
\notag &&  - \frac{m_c^2 \langle g_s^2GG\rangle \langle\bar{q}q\rangle^2}{13824 \pi^6} \int dydz\, \frac{z (1-y-z)^2}{y^2} (3s - 2\bar{m}^2) \,{\rm exp}\left(-\frac{s}{T^2}\right) \\
\notag && - \frac{m_c^2 \langle g_s^2GG\rangle \langle\bar{q}q\rangle^2}{3456 \pi^6} \int dydz\,  \frac{z(1-y-z)^3}{y^2} (2s-\bar{m}^2)\,{\rm exp}\left(-\frac{s}{T^2}\right) \\
\notag &&  - \frac{7 \langle g_s^2GG\rangle \langle\bar{q}q\rangle^2}{221184 \pi^6} \int dydz\, y z (s-\bar{m}^2)(2s-\bar{m}^2) \,{\rm exp}\left(-\frac{s}{T^2}\right) \\
\notag &&  - \frac{5\langle g_s^2GG\rangle \langle\bar{q}q\rangle^2}{55296 \pi^6} \int dydz\, (1-y-z)^2 (s-\bar{m}^2)(2s-\bar{m}^2) \,{\rm exp}\left(-\frac{s}{T^2}\right) \\
\notag && - \frac{5 \langle g_s^2GG\rangle \langle\bar{q}q\rangle^2}{110592 \pi^6} \int dydz\, z(1-y-z) (s-\bar{m}^2)(2s-\bar{m}^2) \,{\rm exp}\left(-\frac{s}{T^2}\right) \\
\notag &&  - \frac{5\langle g_s^2GG\rangle \langle\bar{q}q\rangle^2}{110592 \pi^6} \int dydz\, z (1-y-z)^2 (10s^2-12s\bar{m}^2+3\bar{m}^4) \,{\rm exp}\left(-\frac{s}{T^2}\right) \\
\notag &&  + \frac{\langle g_s^2GG\rangle \langle\bar{q}q\rangle^2}{82944 \pi^6} \int dydz\,(1-y-z)^3  (10s^2-12s\bar{m}^2+3\bar{m}^4) \,{\rm exp}\left(-\frac{s}{T^2}\right) \\
\notag && + \frac{\langle g_s^2GG\rangle \langle\bar{q}q\rangle^2}{4096 \pi^6} \int dydz\, y z(1-y-z) (10s^2-12s\bar{m}^2+3\bar{m}^4) \,{\rm exp}\left(-\frac{s}{T^2}\right) \,,
\end{eqnarray}
\begin{eqnarray}
\notag \rho_{\mathcal{B}}(s,12) &=& \frac{g_s^2 \langle\bar{q}q\rangle^4}{486 \pi^4} \int dydz\, y z  (2s-\bar{m}^2)\,{\rm exp}\left(-\frac{s}{T^2}\right) \,,
\end{eqnarray}
\begin{eqnarray}
\notag \rho_{\mathcal{C}}(s,14) &=&  - \frac{m_c^2 \langle\bar{q}q\rangle^3 \langle\bar{q}g_s\sigma Gq\rangle}{54 \pi^2} \int dy\,\delta(s-\tilde{m}^2) \left( \frac{s}{T^2} + 1\right) \,{\rm exp}\left(-\frac{s}{T^2}\right) \\
\notag &&  -  \frac{m_c^2 \langle\bar{q}q\rangle^3 \langle\bar{q}g_s\sigma Gq\rangle}{1728 \pi^2} \int dy\,\delta(s-\tilde{m}^2) \frac{1}{y} \,{\rm exp}\left(-\frac{s}{T^2}\right) \\
\notag &&  - \frac{\langle\bar{q}q\rangle^3 \langle\bar{q}g_s\sigma Gq\rangle}{108 \pi^2} \int dy\,\delta(s-\tilde{m}^2) (1-y) y \left( \frac{s^2}{T^2} + 4s \right) \,{\rm exp}\left(-\frac{s}{T^2}\right) \\
\notag &&  + \frac{\langle\bar{q}q\rangle^3 \langle\bar{q}g_s\sigma Gq\rangle}{3456 \pi^2} \int dy\,\delta(s-\tilde{m}^2)  (1-y) s\,{\rm exp}\left(-\frac{s}{T^2}\right) \,,
\end{eqnarray}
\begin{eqnarray}
\notag \rho_{\mathcal{C}}(s,16)_1 &=&  \frac{m_c^2 \langle\bar{q}q\rangle^2 \langle\bar{q}g_s\sigma Gq\rangle^2}{144 \pi^2 T^6} \int dy\,\delta(s-\tilde{m}^2) s^2 \,{\rm exp}\left(-\frac{s}{T^2}\right) \\
\notag &&  -  \frac{m_c^2 \langle\bar{q}q\rangle^2 \langle\bar{q}g_s\sigma Gq\rangle^2}{2304 \pi^2 T^4} \int dy\,\delta(s-\tilde{m}^2) \frac{s}{y} \,{\rm exp}\left(-\frac{s}{T^2}\right) \\
\notag &&  - \frac{29 m_c^2 \langle\bar{q}q\rangle^2 \langle\bar{q}g_s\sigma Gq\rangle^2}{55296 \pi^2 T^2} \int dy\,\delta(s-\tilde{m}^2) \frac{1}{(1-y) y} \,{\rm exp}\left(-\frac{s}{T^2}\right) \\
\notag &&  +  \frac{\langle\bar{q}q\rangle^2 \langle\bar{q}g_s\sigma Gq\rangle^2}{288 \pi^2} \int dy\,\delta(s-\tilde{m}^2) (1-y) y \left( \frac{s^3}{T^6} + \frac{3s^2}{T^4} + \frac{6s}{T^2}+6\right) \,{\rm exp}\left(-\frac{s}{T^2}\right) \\
\notag &&  - \frac{\langle\bar{q}q\rangle^2 \langle\bar{q}g_s\sigma Gq\rangle^2}{4608 \pi^2} \int dy\,\delta(s-\tilde{m}^2) (1-y) \left( \frac{s^2}{T^4} + \frac{2s}{T^2} + 2\right)\,{\rm exp}\left(-\frac{s}{T^2}\right) \\
\notag &&  - \frac{13 \langle\bar{q}q\rangle^2 \langle\bar{q}g_s\sigma Gq\rangle^2}{165888 \pi^2} \int dy\,\delta(s-\tilde{m}^2) \left( \frac{s}{T^2} + 1\right) \,{\rm exp}\left(-\frac{s}{T^2}\right) \,,
\end{eqnarray}
\begin{eqnarray}
\notag \rho_{\mathcal{C}}(s,16)_2 &=& - \frac{m_c^4 \langle g_s^2GG\rangle \langle\bar{q}q\rangle^4}{1944 \pi^2 T^4} \int dy\,\delta(s-\tilde{m}^2) \frac{1}{y^3} \,{\rm exp}\left(-\frac{s}{T^2}\right) \\
\notag &&  + \frac{m_c^2 \langle g_s^2GG\rangle \langle\bar{q}q\rangle^4}{1944 \pi^2 T^6} \int dy\,\delta(s-\tilde{m}^2) s^2 \,{\rm exp}\left(-\frac{s}{T^2}\right) \\
\notag &&  - \frac{m_c^2 \langle g_s^2GG\rangle \langle\bar{q}q\rangle^4}{3888 \pi^2 T^4} \int dy\,\delta(s-\tilde{m}^2) \frac{(1-y)s}{y^2} \,{\rm exp}\left(-\frac{s}{T^2}\right) \\
\notag &&  + \frac{m_c^2 \langle g_s^2GG\rangle \langle\bar{q}q\rangle^4}{648 \pi^2 T^2} \int dy\,\delta(s-\tilde{m}^2)\frac{1}{y^2} \,{\rm exp}\left(-\frac{s}{T^2}\right) \\
\notag &&  + \frac{\langle g_s^2GG\rangle \langle\bar{q}q\rangle^4}{3888 \pi^2} \int dy\,\delta(s-\tilde{m}^2) (1-y) y \left( \frac{s^3}{T^6} + \frac{3s^2}{T^4} + \frac{6s}{T^2} + 6 \right) \,{\rm exp}\left(-\frac{s}{T^2}\right) \\
\notag &&  - \frac{\langle g_s^2GG\rangle \langle\bar{q}q\rangle^4}{5184 \pi^2} \int dy\,\delta(s-\tilde{m}^2) \left( \frac{s}{T^2} + 1 \right)\,{\rm exp}\left(-\frac{s}{T^2}\right) \,,
\end{eqnarray}
\begin{eqnarray}
\notag \rho_{\mathcal{D}}(s,10) &=&  -\frac{m_c^2 \langle g_s^2GG\rangle \langle\bar{q}q\rangle^2}{20736 \pi^6} \int dydz\,\delta(s-\bar{m}^2) \frac{z(1-y-z)^3}{y^2} s^2 \,{\rm exp}\left(-\frac{s}{T^2}\right) \,,
\end{eqnarray}
\begin{eqnarray}
\notag \rho_{\mathcal{D}}(s,12) &=&  \frac{g_s^2 \langle\bar{q}q\rangle^4}{2916 \pi^4} \int dydz\,\delta(s-\bar{m}^2) y z s^2\,{\rm exp}\left(-\frac{s}{T^2}\right) \,.
\end{eqnarray}
The spectral density $\rho_3(s)$ of the current $\mathcal{J}_{1\alpha}$:
\begin{eqnarray}
\notag \rho_{\mathcal{A}}(s,12)_1 &=& \frac{11 m_c^2  \langle\bar{q}q\rangle^4}{432 \pi^2} \int dy\, \,{\rm exp}\left(-\frac{s}{T^2}\right) \\
\notag && + \frac{\langle\bar{q}q\rangle^4}{432 \pi^2} \int dy\, (1-y) y (17s - 10\tilde{m}^2)\,{\rm exp}\left(-\frac{s}{T^2}\right) \,,
\end{eqnarray}
\begin{eqnarray}
\notag \rho_{\mathcal{A}}(s,12)_2&=& \frac{7 g_s^2 m_c^2  \langle\bar{q}q\rangle^4}{8748 \pi^4} \int dy\,  \,{\rm exp}\left(-\frac{s}{T^2}\right) \\
\notag && + \frac{g_s^2 \langle\bar{q}q\rangle^4}{23328 \pi^4} \int dy\, (1-y) y (49s - 27\tilde{m}^2) \,{\rm exp}\left(-\frac{s}{T^2}\right) \,,
\end{eqnarray}
\begin{eqnarray}
\notag \rho_{\mathcal{A}}(s,14) &=& -\frac{5\langle\bar{q}q\rangle^3 \langle\bar{q}g_s\sigma Gq\rangle}{72 \pi^2} \int dy\,  (1-y) y \,{\rm exp}\left(-\frac{s}{T^2}\right) \\
\notag && + \frac{5 \langle\bar{q}q\rangle^3 \langle\bar{q}g_s\sigma Gq\rangle}{1152 \pi^2} \int dy\, (1-y) \,{\rm exp}\left(-\frac{s}{T^2}\right) \,,
\end{eqnarray}
\begin{eqnarray}
\notag \rho_{\mathcal{B}}(s,0) &=& \frac{m_c^2}{23592960 \pi^{10}} \int dydz\,(1-y-z)^5 (s-\bar{m}^2)^6 \,{\rm exp}\left(-\frac{s}{T^2}\right) \\
\notag && + \frac{m_c^2}{707788800\pi^{10}} \int dydz\, (1-y-z)^6 (s-\bar{m}^2)^5 (13s-\bar{m}^2) \,{\rm exp}\left(-\frac{s}{T^2}\right) \\
\notag && +\frac{11}{825753600 \pi^{10}} \int dydz\, y z(1-y-z)^5 (s-\bar{m}^2)^6 (8s-\bar{m}^2) \,{\rm exp}\left(-\frac{s}{T^2}\right) \,,
\end{eqnarray}
\begin{eqnarray}
\notag \rho_{\mathcal{B}}(s,4) &=& - \frac{m_c^4 \langle g_s^2GG\rangle}{212336640 \pi^{10}} \int dydz\, \frac{(1-y-z)^6}{y^3} (s-\bar{m}^2)^2 (7s-\bar{m}^2) \,{\rm exp}\left(-\frac{s}{T^2}\right) \\
\notag && +\frac{m_c^2 \langle g_s^2GG\rangle}{14155776 \pi^{10}} \int dydz\,  \frac{(1-y-z)^5}{y^2} (s-\bar{m}^2)^3\left[ (1-y) (s-\bar{m}^2) - 2s y \right]\,{\rm exp}\left(-\frac{s}{T^2}\right) \\
\notag && + \frac{95 m_c^2  \langle g_s^2GG\rangle}{339738624 \pi^{10}} \int dydz\, \frac{(1-y-z)^4}{y} (s-\bar{m}^2)^4 \,{\rm exp}\left(-\frac{s}{T^2}\right) \\
\notag && + \frac{7 m_c^2  \langle g_s^2GG\rangle}{56623104 \pi^{10}} \int dydz\,(1-y-z)^3 (s-\bar{m}^2)^4 \,{\rm exp}\left(-\frac{s}{T^2}\right) \\
\notag && - \frac{7 m_c^2  \langle g_s^2GG\rangle}{339738624 \pi^{10}} \int dydz\, \frac{(1-y-z)^5}{y z} (s-\bar{m}^2)^4\,{\rm exp}\left(-\frac{s}{T^2}\right) \\
\notag && - \frac{11m_c^2 \langle g_s^2GG\rangle}{141557760 \pi^{10}} \int dydz\, \frac{z(1-y-z)^5}{y^2} (s-\bar{m}^2)^3 (5s-\bar{m}^2) \,{\rm exp}\left(-\frac{s}{T^2}\right) \\
\notag && + \frac{m_c^2  \langle g_s^2GG\rangle}{283115520 \pi^{10}} \int dydz\, \frac{(1-y-z)^6}{y^2} (s-\bar{m}^2)^3 (9s-\bar{m}^2) \,{\rm exp}\left(-\frac{s}{T^2}\right) \\
\notag &&  + \frac{m_c^2 \langle g_s^2GG\rangle}{212336640 \pi^{10}} \int dydz\, \frac{(1-y-z)^5}{y} (s-\bar{m}^2)^3 (9s-\bar{m}^2) \,{\rm exp}\left(-\frac{s}{T^2}\right) \\
\notag && + \frac{13 m_c^2 \langle g_s^2GG\rangle}{226492416 \pi^{10}} \int dydz\, (1-y-z)^4 (s-\bar{m}^2)^3 (9s-\bar{m}^2) \,{\rm exp}\left(-\frac{s}{T^2}\right) \\
\notag && -  \frac{11 m_c^2 \langle g_s^2GG\rangle}{10192158720 \pi^{10}} \int dydz\, \frac{(1-y-z)^6}{y z} (s-\bar{m}^2)^3 (9s-\bar{m}^2) \,{\rm exp}\left(-\frac{s}{T^2}\right) \\
\notag && - \frac{\langle g_s^2GG\rangle}{209715200 \pi^{10}} \int dydz\,(1-y-z)^5 (s-\bar{m}^2)^4 (6s-\bar{m}^2) \,{\rm exp}\left(-\frac{s}{T^2}\right) \\
\notag && + \frac{\langle g_s^2GG\rangle}{113246208 \pi^{10}} \int dydz\, z (1-y-z)^4 (s-\bar{m}^2)^4 (46s-9\bar{m}^2) \,{\rm exp}\left(-\frac{s}{T^2}\right) \\
\notag && + \frac{\langle g_s^2GG\rangle}{94371840 \pi^{10}} \int dydz\, y z(1-y-z)^3 (s-\bar{m}^2)^4 (424s-19\bar{m}^2) \,{\rm exp}\left(-\frac{s}{T^2}\right) \,,
\end{eqnarray}
\begin{eqnarray}
\notag \rho_{\mathcal{B}}(s,6)_1 &=& \frac{7 m_c^2  \langle\bar{q}q\rangle^2}{3456 \pi^6} \int dydz\, (1-y-z)^2 (s-\bar{m}^2)^3\,{\rm exp}\left(-\frac{s}{T^2}\right) \\
\notag && + \frac{5m_c^2 \langle\bar{q}q\rangle^2}{82944 \pi^6} \int dydz\,(1-y-z)^3  (s-\bar{m}^2)^2 (7s-\bar{m}^2) \,{\rm exp}\left(-\frac{s}{T^2}\right) \\
\notag && + \frac{\langle\bar{q}q\rangle^2}{36864 \pi^6} \int dydz\, y z (1-y-z)^2 (s-\bar{m}^2)^3 (115 s-27\bar{m}^2) \,{\rm exp}\left(-\frac{s}{T^2}\right) \,,
\end{eqnarray}
\begin{eqnarray}
\notag \rho_{\mathcal{B}}(s,6)_2 &=& \frac{5 g_s^2 m_c^2  \langle\bar{q}q\rangle^2}{248832 \pi^8} \int dydz\, (1-y-z)^2 (s-\bar{m}^2)^3\,{\rm exp}\left(-\frac{s}{T^2}\right) \\
\notag && + \frac{g_s^2 m_c^2  \langle\bar{q}q\rangle^2}{746496 \pi^8} \int dydz\,(1-y-z)^3 (s-\bar{m}^2)^2 (7s-\bar{m}^2) \,{\rm exp}\left(-\frac{s}{T^2}\right) \\
\notag && + \frac{11g_s^2 \langle\bar{q}q\rangle^2}{995328 \pi^8} \int dydz\, y z (1-y-z)^2 (s-\bar{m}^2)^3 (5 s-\bar{m}^2) \,{\rm exp}\left(-\frac{s}{T^2}\right) \,,
\end{eqnarray}
\begin{eqnarray}
\notag \rho_{\mathcal{B}}(s,8) &=& -\frac{67 m_c^2  \langle\bar{q}q\rangle \langle\bar{q}g_s\sigma Gq\rangle}{13824 \pi^6} \int dydz\, (1-y-z) (s-\bar{m}^2)^2\,{\rm exp}\left(-\frac{s}{T^2}\right) \\
\notag && + \frac{37 m_c^2  \langle\bar{q}q\rangle \langle\bar{q}g_s\sigma Gq\rangle}{36864 \pi^6} \int dydz\, \frac{(1-y-z)^2}{y} (s-\bar{m}^2)^2\,{\rm exp}\left(-\frac{s}{T^2}\right) \\
\notag && + \frac{m_c^2 \langle\bar{q}q\rangle \langle\bar{q}g_s\sigma Gq\rangle}{110592 \pi^6} \int dydz\, \frac{(1-y-z)^3}{y} (s-\bar{m}^2) (5 s-\bar{m}^2) \,{\rm exp}\left(-\frac{s}{T^2}\right) \\
\notag && - \frac{m_c^2  \langle\bar{q}q\rangle \langle\bar{q}g_s\sigma Gq\rangle}{221184 \pi^6} \int dydz\, (1-y-z)^2 (s-\bar{m}^2) (295s-83\bar{m}^2) \,{\rm exp}\left(-\frac{s}{T^2}\right) \\
\notag && - \frac{\langle\bar{q}q\rangle \langle\bar{q}g_s\sigma Gq\rangle}{110592 \pi^6} \int dydz\, y z(1-y-z) (s-\bar{m}^2)^2 (1123s-267\bar{m}^2) \,{\rm exp}\left(-\frac{s}{T^2}\right) \\
\notag && + \frac{\langle\bar{q}q\rangle \langle\bar{q}g_s\sigma Gq\rangle}{73728 \pi^6} \int dydz\, z (1-y-z)^2 (s-\bar{m}^2)^2 (76s-21\bar{m}^2) \,{\rm exp}\left(-\frac{s}{T^2}\right) \,,
\end{eqnarray}
\begin{eqnarray}
\notag \rho_{\mathcal{B}}(s,10)_1 &=& \frac{1165 m_c^2  \langle\bar{q}g_s\sigma Gq\rangle^2}{1327104 \pi^6} \int dydz\, (s-\bar{m}^2) \,{\rm exp}\left(-\frac{s}{T^2}\right) \\
\notag &&  - \frac{2611 m_c^2  \langle\bar{q}g_s\sigma Gq\rangle^2}{2654208 \pi^6} \int dydz\, \frac{(1-y-z)}{y} (s-\bar{m}^2)\,{\rm exp}\left(-\frac{s}{T^2}\right) \\
\notag && + \frac{137 m_c^2  \langle\bar{q}g_s\sigma Gq\rangle^2}{5308416 \pi^6} \int dydz\, \frac{(1-y-z)^2}{y z} (s-\bar{m}^2)\,{\rm exp}\left(-\frac{s}{T^2}\right) \\
\notag && -\frac{179 m_c^2 \langle\bar{q}g_s\sigma Gq\rangle^2}{10616832 \pi^6} \int dydz\, \frac{(1-y-z)^2}{y} (3s-\bar{m}^2)  \,{\rm exp}\left(-\frac{s}{T^2}\right) \\
\notag &&  + \frac{481 m_c^2 \langle\bar{q}g_s\sigma Gq\rangle^2}{5308416 \pi^6} \int dydz\, (1-y-z) (3s-\bar{m}^2) \,{\rm exp}\left(-\frac{s}{T^2}\right) \\
\notag && + \frac{61 m_c^2  \langle\bar{q}g_s\sigma Gq\rangle^2}{7962624 \pi^6} \int dydz\, \frac{(1-y-z)^3}{y z}  (3s-\bar{m}^2)\,{\rm exp}\left(-\frac{s}{T^2}\right) \\
\notag && + \frac{\langle\bar{q}g_s\sigma Gq\rangle^2}{3538944\pi^6 } \int dydz\, y z (s-\bar{m}^2) (16039s-15363\bar{m}^2) \,{\rm exp}\left(-\frac{s}{T^2}\right) \\
\notag &&  - \frac{\langle\bar{q}g_s\sigma Gq\rangle^2}{1769472 \pi^6} \int dydz\, z(1-y-z) (s-\bar{m}^2) (1591s-501\bar{m}^2) \,{\rm exp}\left(-\frac{s}{T^2}\right) \\
\notag && + \frac{\langle\bar{q}g_s\sigma Gq\rangle^2}{393216 \pi^6} \int dydz\, (1-y-z)^2 (s-\bar{m}^2) (29s-15\bar{m}^2) \,{\rm exp}\left(-\frac{s}{T^2}\right) \,,
\end{eqnarray}
\begin{eqnarray}
\notag \rho_{\mathcal{B}}(s,10)_2 &=& -\frac{5 m_c^4  \langle g_s^2GG\rangle \langle\bar{q}q\rangle^2}{497664 \pi^6} \int dydz\, \frac{(1-y-z)^3}{y^3} \,{\rm exp}\left(-\frac{s}{T^2}\right) \\
\notag && + \frac{m_c^2 \langle g_s^2GG\rangle \langle\bar{q}q\rangle^2}{20736 \pi^6} \int dydz\, \frac{(1-y-z)^2}{y^2} \left[ 14(1-y)(s-\bar{m}^2) + 7s y \right] \,{\rm exp}\left(-\frac{s}{T^2}\right) \\
\notag && + \frac{199 m_c^2  \langle g_s^2GG\rangle \langle\bar{q}q\rangle^2}{331776 \pi^6} \int dydz\, (s-\bar{m}^2)\,{\rm exp}\left(-\frac{s}{T^2}\right) \\
\notag && + \frac{149 m_c^2 \langle g_s^2GG\rangle \langle\bar{q}q\rangle^2}{331776 \pi^6} \int dydz\, \frac{(1-y-z)}{y}(s-\bar{m}^2)\,{\rm exp}\left(-\frac{s}{T^2}\right) \\
\notag && + \frac{5 m_c^2  \langle g_s^2GG\rangle \langle\bar{q}q\rangle^2}{165888 \pi^6} \int dydz\, \frac{(1-y-z)^3}{y^2} (3s-\bar{m}^2)\,{\rm exp}\left(-\frac{s}{T^2}\right) \\
\notag && + \frac{m_c^2 \langle g_s^2GG\rangle \langle\bar{q}q\rangle^2}{20736 \pi^6} \int dydz\, (1-y-z) (3s - \bar{m}^2)\,{\rm exp}\left(-\frac{s}{T^2}\right) \\
\notag && + \frac{m_c^2 \langle g_s^2GG\rangle \langle\bar{q}q\rangle^2}{82944 \pi^6} \int dydz\, \frac{(1-y-z)^2}{y} (3s-\bar{m}^2) \,{\rm exp}\left(-\frac{s}{T^2}\right) \\
\notag && - \frac{m_c^2  \langle g_s^2GG\rangle \langle\bar{q}q\rangle^2}{165888 \pi^6} \int dydz\, \frac{z (1-y-z)^2}{y^2} (164s -81\bar{m}^2)\,{\rm exp}\left(-\frac{s}{T^2}\right) \\
\notag && + \frac{m_c^2  \langle g_s^2GG\rangle \langle\bar{q}q\rangle^2}{1990656 \pi^6} \int dydz\, \frac{(1-y-z)^3}{y z} (369s - 403\bar{m}^2)\,{\rm exp}\left(-\frac{s}{T^2}\right) \\
\notag && + \frac{\langle g_s^2GG\rangle \langle\bar{q}q\rangle^2}{442368 \pi^6} \int dydz\, y z (s-\bar{m}^2) (431s-159\bar{m}^2) \,{\rm exp}\left(-\frac{s}{T^2}\right) \\
\notag && + \frac{\langle g_s^2GG\rangle \langle\bar{q}q\rangle^2}{221184 \pi^6} \int dydz\, z(1-y-z) (s-\bar{m}^2) (173s-63\bar{m}^2) \,{\rm exp}\left(-\frac{s}{T^2}\right) \\
\notag && - \frac{\langle g_s^2GG\rangle \langle\bar{q}q\rangle^2}{884736 \pi^6} \int dydz\, (1-y-z)^2 (s-\bar{m}^2) (191s-75\bar{m}^2) \,{\rm exp}\left(-\frac{s}{T^2}\right) \,,
\end{eqnarray}
\begin{eqnarray}
\notag \rho_{\mathcal{B}}(s,12) &=& \frac{5 g_s^2 m_c^2  \langle\bar{q}q\rangle^4}{69984 \pi^4} \int dydz\, \,{\rm exp}\left(-\frac{s}{T^2}\right) \,,
\end{eqnarray}
\begin{eqnarray}
\notag \rho_{\mathcal{C}}(s,14) &=& - \frac{11 m_c^2 \langle\bar{q}q\rangle^3 \langle\bar{q}g_s\sigma Gq\rangle}{432 \pi^2} \int dy\,\delta(s-\tilde{m}^2) \left( \frac{s}{ T^2} + 1 \right)\,{\rm exp}\left(-\frac{s}{T^2}\right) \\
\notag && + \frac{11 m_c^2  \langle\bar{q}q\rangle^3 \langle\bar{q}g_s\sigma Gq\rangle}{1728 \pi^2} \int dy\,\delta(s-\tilde{m}^2) \frac{1}{y} \,{\rm exp}\left(-\frac{s}{T^2}\right) \\
\notag && - \frac{\langle\bar{q}q\rangle^3 \langle\bar{q}g_s\sigma Gq\rangle}{432 \pi^2} \int dy\,\delta(s-\tilde{m}^2) (1-y) y \left( \frac{7s^2}{ T^2} + 24s \right)\,{\rm exp}\left(-\frac{s}{T^2}\right) \\
\notag && + \frac{37 \langle\bar{q}q\rangle^3 \langle\bar{q}g_s\sigma Gq\rangle}{10368 \pi^2} \int dy\,\delta(s-\tilde{m}^2) (1-y) s\,{\rm exp}\left(-\frac{s}{T^2}\right) \,,
\end{eqnarray}
\begin{eqnarray}
\notag \rho_{\mathcal{C}}(s,16)_1 &=& \frac{11 m_c^2  \langle\bar{q}q\rangle^2 \langle\bar{q}g_s\sigma Gq\rangle^2}{1152\pi^2 T^6} \int dy\,\delta(s-\tilde{m}^2) s^2 \,{\rm exp}\left(-\frac{s}{T^2}\right) \\
\notag && - \frac{11 m_c^2  \langle\bar{q}q\rangle^2 \langle\bar{q}g_s\sigma Gq\rangle^2}{2304 \pi^2 T^4} \int dy\,\delta(s-\tilde{m}^2) \frac{s}{y} \,{\rm exp}\left(-\frac{s}{T^2}\right) \\
\notag && + \frac{25 m_c^2  \langle\bar{q}q\rangle^2 \langle\bar{q}g_s\sigma Gq\rangle^2}{110592 \pi^2 T^2} \int dy\,\delta(s-\tilde{m}^2) \frac{1}{(1-y) y} \,{\rm exp}\left(-\frac{s}{T^2}\right) \\
\notag && + \frac{\langle\bar{q}q\rangle^2 \langle\bar{q}g_s\sigma Gq\rangle^2}{1728\pi^2} \int dy\,\delta(s-\tilde{m}^2) (1-y) y \left( \frac{17s^3}{ T^6} + \frac{45 s^2}{T^4} + \frac{78 s}{T^2} +66 \right)\,{\rm exp}\left(-\frac{s}{T^2}\right) \\
\notag && - \frac{\langle\bar{q}q\rangle^2 \langle\bar{q}g_s\sigma Gq\rangle^2}{13824 \pi^2} \int dy\,\delta(s-\tilde{m}^2) (1-y) \left( \frac{37 s^2}{T^4} + \frac{45 s}{T^2} + 16 \right)\,{\rm exp}\left(-\frac{s}{T^2}\right) \\
\notag && + \frac{\langle\bar{q}q\rangle^2 \langle\bar{q}g_s\sigma Gq\rangle^2}{5971968 \pi^2} \int dy\,\delta(s-\tilde{m}^2) \left( \frac{669s}{ T^2} + 739 \right)\,{\rm exp}\left(-\frac{s}{T^2}\right) \,,
\end{eqnarray}
\begin{eqnarray}
\notag \rho_{\mathcal{C}}(s,16)_2 &=& - \frac{11 m_c^4 \langle g_s^2GG\rangle \langle\bar{q}q\rangle^4}{15552 \pi^2 T^4} \int dy\,\delta(s-\tilde{m}^2) \frac{1}{y^3} \,{\rm exp}\left(-\frac{s}{T^2}\right) \\
\notag && + \frac{11 m_c^2 \langle g_s^2GG\rangle \langle\bar{q}q\rangle^4}{15552\pi^2 T^6} \int dy\,\delta(s-\tilde{m}^2) s^2 \,{\rm exp}\left(-\frac{s}{T^2}\right) \\
\notag && + \frac{11 m_c^2  \langle g_s^2GG\rangle \langle\bar{q}q\rangle^4}{5184 \pi^2 T^2} \int dy\,\delta(s-\tilde{m}^2) \frac{1}{y^2} \,{\rm exp}\left(-\frac{s}{T^2}\right) \\
\notag && - \frac{m_c^2  \langle g_s^2GG\rangle \langle\bar{q}q\rangle^4}{2304 \pi^2 T^2} \int dy\,\delta(s-\tilde{m}^2) \frac{1}{(1-y) y} \,{\rm exp}\left(-\frac{s}{T^2}\right) \\
\notag && - \frac{m_c^2 \langle g_s^2GG\rangle \langle\bar{q}q\rangle^4}{15552 \pi^2} \int dy\,\delta(s-\tilde{m}^2) \frac{(1-y)}{y^2} \left( \frac{7s}{T^4} -\frac{4}{T^2} \right)\,{\rm exp}\left(-\frac{s}{T^2}\right) \\
\notag && + \frac{\langle g_s^2GG\rangle \langle\bar{q}q\rangle^4}{46656\pi^2} \int dy\,\delta(s-\tilde{m}^2) (1-y) y \left( \frac{29 s^3}{T^6} + \frac{63 s^2}{T^4} + \frac{78 s}{T^2} -78 \right)\,{\rm exp}\left(-\frac{s}{T^2}\right) \\
\notag && - \frac{\langle g_s^2GG\rangle \langle\bar{q}q\rangle^4}{124416 \pi^2} \int dy\,\delta(s-\tilde{m}^2) \left( \frac{19 s}{T^2} -11 \right)\,{\rm exp}\left(-\frac{s}{T^2}\right) \,,
\end{eqnarray}
\begin{eqnarray}
\notag \rho_{\mathcal{D}}(s,10) &=& -\frac{5 m_c^4 \langle g_s^2GG\rangle \langle\bar{q}q\rangle^2}{248832 \pi^6} \int dydz\,\delta(s-\bar{m}^2) \frac{(1-y-z)^3}{y^3} s \,{\rm exp}\left(-\frac{s}{T^2}\right) \,,
\end{eqnarray}
\begin{eqnarray}
\notag \rho_{\mathcal{D}}(s,12) &=& \frac{9 m_c^2  \langle\bar{q}q\rangle^4}{34992 \pi^4} \int dydz\,\delta(s-\bar{m}^2) s\,{\rm exp}\left(-\frac{s}{T^2}\right) \,.
\end{eqnarray}
The spectral density $\rho_4(s)$ of the current $\mathcal{J}_{2\alpha}$:
\begin{eqnarray}
\notag \rho_{\mathcal{A}}(s,12)_1 &=&  - \frac{11 m_c^2 \langle\bar{q}q\rangle^4}{432 \pi^2} \int dy\, \,{\rm exp}\left(-\frac{s}{T^2}\right) \\
\notag && + \frac{\langle\bar{q}q\rangle^4}{432 \pi^2} \int dy\, (1-y) y (17s-10\tilde{m}^2) \,{\rm exp}\left(-\frac{s}{T^2}\right) \,,
\end{eqnarray}
\begin{eqnarray}
\notag \rho_{\mathcal{A}}(s,12)_2 &=&  - \frac{7 g_s^2 m_c^2 \langle\bar{q}q\rangle^4}{8748 \pi^4} \int dy\, \,{\rm exp}\left(-\frac{s}{T^2}\right) \\
\notag &&  + \frac{g_s^2 \langle\bar{q}q\rangle^4}{23328 \pi^4} \int dy\, (1-y) y (49s-27\tilde{m}^2) \,{\rm exp}\left(-\frac{s}{T^2}\right) \,,
\end{eqnarray}
\begin{eqnarray}
\notag \rho_{\mathcal{A}}(s,14) &=&  -\frac{5\langle\bar{q}q\rangle^3 \langle\bar{q}g_s\sigma Gq\rangle}{72 \pi^2} \int dy\, (1-y) y \,{\rm exp}\left(-\frac{s}{T^2}\right) \\
\notag &&   + \frac{5 \langle\bar{q}q\rangle^3 \langle\bar{q}g_s\sigma Gq\rangle}{1152 \pi^2} \int dy\, (1-y) \,{\rm exp}\left(-\frac{s}{T^2}\right) \,,
\end{eqnarray}
\begin{eqnarray}
\notag \rho_{\mathcal{B}}(s,0) &=& -\frac{m_c^2}{23592960 \pi^{10}} \int dydz\,(1-y-z)^5 (s-\bar{m}^2)^6 \,{\rm exp}\left(-\frac{s}{T^2}\right) \\
\notag &&  - \frac{m_c^2}{707788800 \pi^{10}} \int dydz\, (1-y-z)^6 (s-\bar{m}^2)^5 (13s-\bar{m}^2) \,{\rm exp}\left(-\frac{s}{T^2}\right) \\
\notag && + \frac{11}{825753600 \pi^{10}} \int dydz\, y z(1-y-z)^5 (s-\bar{m}^2)^6 (8s-\bar{m}^2) \,{\rm exp}\left(-\frac{s}{T^2}\right) \,,
\end{eqnarray}
\begin{eqnarray}
\notag \rho_{\mathcal{B}}(s,4) &=& \frac{m_c^4 \langle g_s^2GG\rangle}{212336640 \pi^{10}} \int dydz\, \frac{(1-y-z)^6}{y^3} (s-\bar{m}^2)^2 (7s-\bar{m}^2) \,{\rm exp}\left(-\frac{s}{T^2}\right) \\
\notag && - \frac{m_c^2 \langle g_s^2GG\rangle}{14155776 \pi^{10}} \int dydz\, \frac{(1-y-z)^5}{y^2} (s-\bar{m}^2)^3 \left[ (1-y)(s-\bar{m}^2) - 2y s \right] \,{\rm exp}\left(-\frac{s}{T^2}\right) \\
\notag &&  - \frac{95 m_c^2 \langle g_s^2GG\rangle}{339738624 \pi^{10}} \int dydz\, \frac{(1-y-z)^4}{y} (s-\bar{m}^2)^4\,{\rm exp}\left(-\frac{s}{T^2}\right) \\
\notag && - \frac{7 m_c^2 \langle g_s^2GG\rangle}{56623104 \pi^{10}} \int dydz\,(1-y-z)^3 (s-\bar{m}^2)^4 \,{\rm exp}\left(-\frac{s}{T^2}\right) \\
\notag &&  + \frac{7 m_c^2 \langle g_s^2GG\rangle}{339738624 \pi^{10}} \int dydz\, \frac{(1-y-z)^5}{y z} (s-\bar{m}^2)^4\,{\rm exp}\left(-\frac{s}{T^2}\right) \\
\notag && - \frac{11 m_c^2 \langle g_s^2GG\rangle}{141557760 \pi^{10}} \int dydz\, \frac{z(1-y-z)^5}{y^2} (s-\bar{m}^2)^3 (5s-\bar{m}^2) \,{\rm exp}\left(-\frac{s}{T^2}\right) \\
\notag &&  - \frac{m_c^2 \langle g_s^2GG\rangle}{283115520 \pi^{10}} \int dydz\, \frac{(1-y-z)^6}{y^2} (s-\bar{m}^2)^3 (9s-\bar{m}^2) \,{\rm exp}\left(-\frac{s}{T^2}\right) \\
\notag && - \frac{m_c^2 \langle g_s^2GG\rangle}{212336640 \pi^{10}} \int dydz\, \frac{(1-y-z)^5}{y} (s-\bar{m}^2)^3 (9s-\bar{m}^2) \,{\rm exp}\left(-\frac{s}{T^2}\right) \\
\notag &&  - \frac{13m_c^2 \langle g_s^2GG\rangle}{226492416 \pi^{10}} \int dydz\, (1-y-z)^4 (s-\bar{m}^2)^3 (9s-\bar{m}^2) \,{\rm exp}\left(-\frac{s}{T^2}\right) \\
\notag && + \frac{11m_c^2 \langle g_s^2GG\rangle}{10192158720 \pi^{10}} \int dydz\, \frac{(1-y-z)^6}{y z} (s-\bar{m}^2)^3 (9s-\bar{m}^2) \,{\rm exp}\left(-\frac{s}{T^2}\right) \\
\notag && - \frac{\langle g_s^2GG\rangle}{209715200 \pi^{10}} \int dydz\,(1-y-z)^5 (s-\bar{m}^2)^4 (6s-\bar{m}^2) \,{\rm exp}\left(-\frac{s}{T^2}\right) \\
\notag && + \frac{\langle g_s^2GG\rangle}{113246208 \pi^{10}} \int dydz\, z (1-y-z)^4 (s-\bar{m}^2)^4 (46s-9\bar{m}^2) \,{\rm exp}\left(-\frac{s}{T^2}\right) \\
\notag && + \frac{\langle g_s^2GG\rangle}{94371840 \pi^{10}} \int dydz\, y z(1-y-z)^3 (s-\bar{m}^2)^4 (154s-19 \bar{m}^2) \,{\rm exp}\left(-\frac{s}{T^2}\right) \,,
\end{eqnarray}
\begin{eqnarray}
\notag \rho_{\mathcal{B}}(s,6)_1 &=&  - \frac{7 m_c^2 \langle\bar{q}q\rangle^2}{3456 \pi^6} \int dydz\, (1-y-z)^2 (s-\bar{m}^2)^3\,{\rm exp}\left(-\frac{s}{T^2}\right) \\
\notag && - \frac{5 m_c^2 \langle\bar{q}q\rangle^2}{82944 \pi^6} \int dydz\,(1-y-z)^3 (s-\bar{m}^2)^2 (7s-\bar{m}^2) \,{\rm exp}\left(-\frac{s}{T^2}\right) \\
\notag &&  + \frac{\langle\bar{q}q\rangle^2}{36864 \pi^6} \int dydz\, y z (1-y-z)^2 (s-\bar{m}^2)^3 (115s-27\bar{m}^2) \,{\rm exp}\left(-\frac{s}{T^2}\right) \,,
\end{eqnarray}
\begin{eqnarray}
\notag \rho_{\mathcal{B}}(s,6)_2 &=& - \frac{4 g_s^2 m_c^2 \langle\bar{q}q\rangle^2}{248832 \pi^8} \int dydz\, (1-y-z)^2 (s-\bar{m}^2)^3 \,{\rm exp}\left(-\frac{s}{T^2}\right) \\
\notag && - \frac{g_s^2 m_c^2 \langle\bar{q}q\rangle^2}{746496 \pi^8} \int dydz\,(1-y-z)^3 (s-\bar{m}^2)^2 (7s-\bar{m}^2) \,{\rm exp}\left(-\frac{s}{T^2}\right) \\
\notag && + \frac{11 g_s^2 \langle\bar{q}q\rangle^2}{995328 \pi^8} \int dydz\, y z (1-y-z)^2 (s-\bar{m}^2)^3 (5s-\bar{m}^2) \,{\rm exp}\left(-\frac{s}{T^2}\right) \,,
\end{eqnarray}
\begin{eqnarray}
\notag \rho_{\mathcal{B}}(s,8) &=& \frac{101 m_c^2 \langle\bar{q}q\rangle \langle\bar{q}g_s\sigma Gq\rangle}{13824 \pi^6} \int dydz\,(1-y-z) (s-\bar{m}^2)^2\,{\rm exp}\left(-\frac{s}{T^2}\right) \\
\notag &&  - \frac{37 m_c^2 \langle\bar{q}q\rangle \langle\bar{q}g_s\sigma Gq\rangle}{36864 \pi^6} \int dydz\, \frac{(1-y-z)^2}{y} (s-\bar{m}^2)^2\,{\rm exp}\left(-\frac{s}{T^2}\right) \\
\notag && + \frac{53 m_c^2 \langle\bar{q}q\rangle \langle\bar{q}g_s\sigma Gq\rangle}{221184 \pi^6} \int dydz\, (1-y-z)^2 (s-\bar{m}^2) (5s-\bar{m}^2) \,{\rm exp}\left(-\frac{s}{T^2}\right) \\
\notag && - \frac{m_c^2 \langle\bar{q}q\rangle \langle\bar{q}g_s\sigma Gq\rangle}{110592 \pi^6} \int dydz\, \frac{(1-y-z)^3}{y} (s-\bar{m}^2) (5s-\bar{m}^2) \,{\rm exp}\left(-\frac{s}{T^2}\right) \\
\notag && - \frac{\langle\bar{q}q\rangle \langle\bar{q}g_s\sigma Gq\rangle}{36864 \pi^6} \int dydz\, y z(1-y-z) (s-\bar{m}^2)^2 (303s-89\bar{m}^2) \,{\rm exp}\left(-\frac{s}{T^2}\right) \\
\notag && + \frac{\langle\bar{q}q\rangle \langle\bar{q}g_s\sigma Gq\rangle}{73728 \pi^6} \int dydz\, z (1-y-z)^2 (s-\bar{m}^2)^2 (76s-21\bar{m}^2) \,{\rm exp}\left(-\frac{s}{T^2}\right) \,,
\end{eqnarray}
\begin{eqnarray}
\notag \rho_{\mathcal{B}}(s,10)_1 &=& - \frac{1165 m_c^2 \langle\bar{q}g_s\sigma Gq\rangle^2}{1327104 \pi^6} \int dydz\, (s-\bar{m}^2)\,{\rm exp}\left(-\frac{s}{T^2}\right) \\
\notag && - \frac{137 m_c^2 \langle\bar{q}g_s\sigma Gq\rangle^2}{5308416 \pi^6} \int dydz\, \frac{(1-y-z)^2}{y z} (s-\bar{m}^2)\,{\rm exp}\left(-\frac{s}{T^2}\right) \\
\notag && + \frac{2131 m_c^2 \langle\bar{q}g_s\sigma Gq\rangle^2}{2654208 \pi^6} \int dydz\,  \frac{(1-y-z)}{y} (s-\bar{m}^2)\,{\rm exp}\left(-\frac{s}{T^2}\right) \\
\notag &&  - \frac{481 m_c^2 \langle\bar{q}g_s\sigma Gq\rangle^2}{5308416 \pi^6} \int dydz\,(1-y-z) (3s-\bar{m}^2)\,{\rm exp}\left(-\frac{s}{T^2}\right) \\
\notag && + \frac{179 m_c^2 \langle\bar{q}g_s\sigma Gq\rangle^2}{10616832 \pi^6} \int dydz\, \frac{(1-y-z)^2}{y} (3s-\bar{m}^2)\,{\rm exp}\left(-\frac{s}{T^2}\right) \\
\notag &&  - \frac{61 m_c^2 \langle\bar{q}g_s\sigma Gq\rangle^2}{7962624 \pi^6} \int dydz\, \frac{(1-y-z)^3}{y z} (3s-\bar{m}^2)\,{\rm exp}\left(-\frac{s}{T^2}\right) \\
\notag &&  - \frac{\langle\bar{q}g_s\sigma Gq\rangle^2}{1769472 \pi^6} \int dydz\, z(1-y-z) (s-\bar{m}^2) (1711s-621\bar{m}^2) \,{\rm exp}\left(-\frac{s}{T^2}\right) \\
\notag && + \frac{\langle\bar{q}g_s\sigma Gq\rangle^2}{1769472 \pi^6} \int dydz\, y z (s-\bar{m}^2) (2811s-1137\bar{m}^2) \,{\rm exp}\left(-\frac{s}{T^2}\right) \\
\notag && + \frac{\langle\bar{q}g_s\sigma Gq\rangle^2}{393216 \pi^6} \int dydz\, (1-y-z)^2 (s-\bar{m}^2) (29s-15\bar{m}^2) \,{\rm exp}\left(-\frac{s}{T^2}\right) \,,
\end{eqnarray}
\begin{eqnarray}
\notag \rho_{\mathcal{B}}(s,10)_2 &=&   \frac{5 m_c^4 \langle g_s^2GG\rangle \langle\bar{q}q\rangle^2}{497664 \pi^6} \int dydz\, \frac{(1-y-z)^3}{y^3} \,{\rm exp}\left(-\frac{s}{T^2}\right) \\
\notag && - \frac{m_c^2 \langle g_s^2GG\rangle \langle\bar{q}q\rangle^2}{20736 \pi^6} \int dydz\, \frac{(1-y-z)^2}{y^2} \left[ 14(1-y)(s-\bar{m}^2) + 7s y \right]\,{\rm exp}\left(-\frac{s}{T^2}\right) \\
\notag && - \frac{5 m_c^2 \langle g_s^2GG\rangle \langle\bar{q}q\rangle^2}{165888 \pi^6} \int dydz\, \frac{(1-y-z)^3}{y^2} (3s-\bar{m}^2)\,{\rm exp}\left(-\frac{s}{T^2}\right) \\
\notag && - \frac{m_c^2 \langle g_s^2GG\rangle \langle\bar{q}q\rangle^2}{82944 \pi^6} \int dydz\, \frac{(1-y-z)^2}{y}   (3s-\bar{m}^2)\,{\rm exp}\left(-\frac{s}{T^2}\right) \\
\notag && + \frac{17 m_c^2 \langle g_s^2GG\rangle \langle\bar{q}q\rangle^2}{1990656 \pi^6} \int dydz\, \frac{(1-y-z)^3}{y z} (3s-\bar{m}^2)\,{\rm exp}\left(-\frac{s}{T^2}\right) \\
\notag && - \frac{m_c^2 \langle g_s^2GG\rangle \langle\bar{q}q\rangle^2}{20736 \pi^6} \int dydz\, (1-y-z) (3s-\bar{m}^2)\,{\rm exp}\left(-\frac{s}{T^2}\right) \\
\notag && - \frac{199 m_c^2 \langle g_s^2GG\rangle \langle\bar{q}q\rangle^2}{331776 \pi^6} \int dydz\, (s-\bar{m}^2)\,{\rm exp}\left(-\frac{s}{T^2}\right) \\
\notag && - \frac{149 m_c^2 \langle g_s^2GG\rangle \langle\bar{q}q\rangle^2}{331776 \pi^6} \int dydz\, \frac{(1-y-z)}{y} (s-\bar{m}^2)\,{\rm exp}\left(-\frac{s}{T^2}\right) \\
\notag &&  + \frac{35 m_c^2 \langle g_s^2GG\rangle \langle\bar{q}q\rangle^2}{165888 \pi^6} \int dydz\, \frac{(1-y-z)^2}{y z} (s-\bar{m}^2)\,{\rm exp}\left(-\frac{s}{T^2}\right) \\
\notag &&  - \frac{m_c^2 \langle g_s^2GG\rangle \langle\bar{q}q\rangle^2}{165888 \pi^6} \int dydz\, \frac{z (1-y-z)^2}{y^2} (83s -81\bar{m}^2) \,{\rm exp}\left(-\frac{s}{T^2}\right) \\
\notag && +  \frac{\langle g_s^2GG\rangle \langle\bar{q}q\rangle^2}{442368 \pi^6} \int dydz\, y z (s-\bar{m}^2) (431s-159\bar{m}^2) \,{\rm exp}\left(-\frac{s}{T^2}\right) \\
\notag && + \frac{\langle g_s^2GG\rangle \langle\bar{q}q\rangle^2}{221184 \pi^6} \int dydz\, z(1-y-z) (s-\bar{m}^2) (173s-63\bar{m}^2) \,{\rm exp}\left(-\frac{s}{T^2}\right) \\
\notag &&  - \frac{\langle g_s^2GG\rangle \langle\bar{q}q\rangle^2}{884736 \pi^6} \int dydz\, (1-y-z)^2 (s-\bar{m}^2) (191s-75\bar{m}^2) \,{\rm exp}\left(-\frac{s}{T^2}\right) \,,
\end{eqnarray}
\begin{eqnarray}
\notag \rho_{\mathcal{B}}(s,12) &=&  - \frac{5 g_s^2 m_c^2 \langle\bar{q}q\rangle^4}{69984 \pi^4} \int dydz\, \,{\rm exp}\left(-\frac{s}{T^2}\right) \,,
\end{eqnarray}
\begin{eqnarray}
\notag \rho_{\mathcal{C}}(s,14) &=&  \frac{11 m_c^2 \langle\bar{q}q\rangle^3 \langle\bar{q}g_s\sigma Gq\rangle}{432 \pi^2} \int dy\,\delta(s-\tilde{m}^2) \left( \frac{s}{T^2} + 1\right)\,{\rm exp}\left(-\frac{s}{T^2}\right) \\
\notag &&  - \frac{11 m_c^2 \langle\bar{q}q\rangle^3 \langle\bar{q}g_s\sigma Gq\rangle}{1728 \pi^2} \int dy\,\delta(s-\tilde{m}^2) \frac{1}{y} \,{\rm exp}\left(-\frac{s}{T^2}\right) \\
\notag && - \frac{\langle\bar{q}q\rangle^3 \langle\bar{q}g_s\sigma Gq\rangle}{432 \pi^2} \int dy\,\delta(s-\tilde{m}^2) (1-y) y \left( \frac{7s^2}{ T^2} + 24s \right)\,{\rm exp}\left(-\frac{s}{T^2}\right) \\
\notag &&   + \frac{37 \langle\bar{q}q\rangle^3 \langle\bar{q}g_s\sigma Gq\rangle}{10368 \pi^2} \int dy\,\delta(s-\tilde{m}^2) (1-y) s\,{\rm exp}\left(-\frac{s}{T^2}\right) \,,
\end{eqnarray}
\begin{eqnarray}
\notag \rho_{\mathcal{C}}(s,16)_1 &=&  - \frac{11 m_c^2 \langle\bar{q}q\rangle^2 \langle\bar{q}g_s\sigma Gq\rangle^2}{1152\pi^2 T^6} \int dy\,\delta(s-\tilde{m}^2) s^2 \,{\rm exp}\left(-\frac{s}{T^2}\right) \\
\notag && + \frac{11 m_c^2 \langle\bar{q}q\rangle^2 \langle\bar{q}g_s\sigma Gq\rangle^2}{2304 \pi^2 T^4} \int dy\,\delta(s-\tilde{m}^2) \frac{s}{y} \,{\rm exp}\left(-\frac{s}{T^2}\right) \\
\notag && - \frac{25 m_c^2 \langle\bar{q}q\rangle^2 \langle\bar{q}g_s\sigma Gq\rangle^2}{110592 \pi^2 T^2} \int dy\,\delta(s-\tilde{m}^2) \frac{1}{(1-y) y} \,{\rm exp}\left(-\frac{s}{T^2}\right) \\
\notag &&  + \frac{\langle\bar{q}q\rangle^2 \langle\bar{q}g_s\sigma Gq\rangle^2}{1152 \pi^2} \int dy\,\delta(s-\tilde{m}^2) (1-y) y \left( \frac{7s^3}{T^6} + \frac{17 s^2}{T^4} + \frac{26 s}{T^2} -18\right)\,{\rm exp}\left(-\frac{s}{T^2}\right) \\
\notag && - \frac{\langle\bar{q}q\rangle^2 \langle\bar{q}g_s\sigma Gq\rangle^2}{13824 \pi^2} \int dy\,\delta(s-\tilde{m}^2) (1-y) \left( \frac{37s^2}{T^4} + \frac{45s}{T^2} + 16 \right)\,{\rm exp}\left(-\frac{s}{T^2}\right) \\
\notag && + \frac{\langle\bar{q}q\rangle^2 \langle\bar{q}g_s\sigma Gq\rangle^2}{1990656 \pi^2} \int dy\,\delta(s-\tilde{m}^2) \left( \frac{223s}{T^2} + 293 \right)\,{\rm exp}\left(-\frac{s}{T^2}\right) \,,
\end{eqnarray}
\begin{eqnarray}
\notag \rho_{\mathcal{C}}(s,16)_2 &=& \frac{11 m_c^4 \langle g_s^2GG\rangle \langle\bar{q}q\rangle^4}{15552 \pi^2 T^4} \int dy\,\delta(s-\tilde{m}^2) \frac{1}{y^3} \,{\rm exp}\left(-\frac{s}{T^2}\right) \\
\notag &&  - \frac{11 m_c^2 \langle g_s^2GG\rangle \langle\bar{q}q\rangle^4}{15552\pi^2 T^6} \int dy\,\delta(s-\tilde{m}^2) s^2 \,{\rm exp}\left(-\frac{s}{T^2}\right) \\
\notag &&  - \frac{11 m_c^2 \langle g_s^2GG\rangle \langle\bar{q}q\rangle^4}{5184 \pi^2 T^2} \int dy\,\delta(s-\tilde{m}^2) \frac{1}{y^2} \,{\rm exp}\left(-\frac{s}{T^2}\right) \\
\notag &&  + \frac{m_c^2 \langle g_s^2GG\rangle \langle\bar{q}q\rangle^4}{2304 \pi^2 T^2} \int dy\,\delta(s-\tilde{m}^2) \frac{1}{(1-y) y} \,{\rm exp}\left(-\frac{s}{T^2}\right) \\
\notag && - \frac{m_c^2 \langle g_s^2GG\rangle \langle\bar{q}q\rangle^4}{15552 \pi^2} \int dy\,\delta(s-\tilde{m}^2) \frac{1-y}{y^2} \left( \frac{7 s}{T^4} -\frac{4}{T^2} \right)\,{\rm exp}\left(-\frac{s}{T^2}\right) \\
\notag &&  + \frac{\langle g_s^2GG\rangle \langle\bar{q}q\rangle^4}{15552\pi^2} \int dy\,\delta(s-\tilde{m}^2) (1-y) y \left( \frac{7 s^3}{T^6} + \frac{17 s^2}{ T^4} + \frac{26s}{T^2} + 22 \right)\,{\rm exp}\left(-\frac{s}{T^2}\right) \\
\notag &&  - \frac{\langle g_s^2GG\rangle \langle\bar{q}q\rangle^4}{124416 \pi^2} \int dy\,\delta(s-\tilde{m}^2) \left( \frac{19 s}{T^2} + 11 \right)\,{\rm exp}\left(-\frac{s}{T^2}\right) \,,
\end{eqnarray}
\begin{eqnarray}
\notag \rho_{\mathcal{D}}(s,10) &=& \frac{5 m_c^4 \langle g_s^2GG\rangle \langle\bar{q}q\rangle^2}{248832 \pi^6} \int dydz\,\delta(s-\bar{m}^2) \frac{(1-y-z)^3}{y^3} s \,{\rm exp}\left(-\frac{s}{T^2}\right) \,,
\end{eqnarray}
\begin{eqnarray}
\notag \rho_{\mathcal{D}}(s,12) &=&  - \frac{5 m_c^2 \langle\bar{q}q\rangle^4}{34992 \pi^4} \int dydz\,\delta(s-\bar{m}^2) s\,{\rm exp}\left(-\frac{s}{T^2}\right) \,.
\end{eqnarray}
The spectral density $\rho_5(s)$ of the currents $\mathcal{J}_1^\pm$:
\begin{eqnarray}
\notag \rho_{\mathcal{A}}(s,12)_1 &=& - \frac{m_c^2 \langle\bar{q}q\rangle^4}{36 \pi^2} \int dy\, \,{\rm exp}\left(-\frac{s}{T^2}\right) \\
\notag && +  \frac{\langle\bar{q}q\rangle^4}{72 \pi^2} \int dy\,  (1-y) y  (3s - 2\tilde{m}^2)\,{\rm exp}\left(-\frac{s}{T^2}\right) \,,
\end{eqnarray}
\begin{eqnarray}
\notag \rho_{\mathcal{A}}(s,12)_2 &=& - \frac{g_s^2 m_c^2 \langle\bar{q}q\rangle^4}{1296 \pi^4} \int dy\,\,{\rm exp}\left(-\frac{s}{T^2}\right) \\
\notag &&+ \frac{g_s^2 \langle\bar{q}q\rangle^4}{1296 \pi^4} \int dy\,  (1-y) y (3s-2\tilde{m}^2) \,{\rm exp}\left(-\frac{s}{T^2}\right) \,,
\end{eqnarray}
\begin{eqnarray}
\notag   \rho_{\mathcal{A}}(s,14) &=& -\frac{\langle\bar{q}q\rangle^3 \langle\bar{q}g_s\sigma Gq\rangle}{12 \pi^2} \int dy\,  (1-y) y \,{\rm exp}\left(-\frac{s}{T^2}\right) \\
\notag && + \frac{\langle\bar{q}q\rangle^3 \langle\bar{q}g_s\sigma Gq\rangle}{144 \pi^2}\int dy\, (1-y)\,{\rm exp}\left(-\frac{s}{T^2}\right) \,,
\end{eqnarray}
\begin{eqnarray}
\notag \rho_{\mathcal{B}}(s,0) &=& -\frac{m_c^2}{19660800 \pi^{10}} \int dydz\,(1-y-z)^5 (s-\bar{m}^2)^6 \,{\rm exp}\left(-\frac{s}{T^2}\right) \\
\notag && + \frac{1}{68812800 \pi^{10}} \int dydz\, y z(1-y-z)^5 (s-\bar{m}^2)^6 (9s-2\bar{m}^2) \,{\rm exp}\left(-\frac{s}{T^2}\right) \,,
\end{eqnarray}
\begin{eqnarray}
\notag\rho_{\mathcal{B}}(s,4)&=&  - \frac{m_c^2 \langle g_s^2GG\rangle}{11796480 \pi^{10}} \int dydz\, \frac{(1-y-z)^5}{y^2} (s-\bar{m}^2)^3 \left[(1-y) (s-\bar{m}^2) - 2s y \right] \,{\rm exp}\left(-\frac{s}{T^2}\right) \\
\notag && - \frac{m_c^2 \langle g_s^2GG\rangle}{3145728 \pi^{10}} \int dydz\, \frac{(1-y-z)^4}{y} (s-\bar{m}^2)^4 \,{\rm exp}\left(-\frac{s}{T^2}\right) \\
\notag && - \frac{m_c^2 \langle g_s^2GG\rangle}{5898240 \pi^{10}} \int dydz\,  \frac{z(1-y-z)^5}{y^2}  (s-\bar{m}^2)^3 (3s-\bar{m}^2) \,{\rm exp}\left(-\frac{s}{T^2}\right) \\
\notag &&+ \frac{\langle g_s^2GG\rangle}{15728640 \pi^{10}} \int dydz\,  z (1-y-z)^4  (s-\bar{m}^2)^4 (7s-2\bar{m}^2) \,{\rm exp}\left(-\frac{s}{T^2}\right) \\
\notag && + \frac{\langle g_s^2GG\rangle}{3932160 \pi^{10}} \int dydz\,  y z(1-y-z)^3 (s-\bar{m}^2)^4 (7s-2\bar{m}^2) \,{\rm exp}\left(-\frac{s}{T^2}\right) \,,
\end{eqnarray}
\begin{eqnarray}
\notag \rho_{\mathcal{B}}(s,6)_1&=& - \frac{m_c^2 \langle\bar{q}q\rangle^2}{512 \pi^6} \int dydz\, (1-y-z)^2 (s-\bar{m}^2)^3\,{\rm exp}\left(-\frac{s}{T^2}\right) \\
\notag &&+ \frac{\langle\bar{q}q\rangle^2}{1024 \pi^6} \int dydz\,  y z (1-y-z)^2  (s-\bar{m}^2)^3 (3s-\bar{m}^2) \,{\rm exp}\left(-\frac{s}{T^2}\right) \,,
\end{eqnarray}
\begin{eqnarray}
\notag \rho_{\mathcal{B}}(s,6)_2 &=& - \frac{g_s^2 m_c^2 \langle\bar{q}q\rangle^2}{41472 \pi^8} \int dydz\,  (1-y-z)^2  (s-\bar{m}^2)^3 \,{\rm exp}\left(-\frac{s}{T^2}\right) \\
\notag && + \frac{g_s^2 \langle\bar{q}q\rangle^2}{41472 \pi^8} \int dydz\,  y z (1-y-z)^2 (s-\bar{m}^2)^3 (3s-\bar{m}^2) \,{\rm exp}\left(-\frac{s}{T^2}\right) \,,
\end{eqnarray}
\begin{eqnarray}
\notag \rho_{\mathcal{B}}(s,8)&=& \frac{3 m_c^2 \langle\bar{q}q\rangle \langle\bar{q}g_s\sigma Gq\rangle}{512 \pi^6} \int dydz\, (1-y-z) (s-\bar{m}^2)^2 \,{\rm exp}\left(-\frac{s}{T^2}\right) \\
\notag && - \frac{m_c^2 \langle\bar{q}q\rangle \langle\bar{q}g_s\sigma Gq\rangle}{1024 \pi^6} \int dydz\, \frac{(1-y-z)^2}{y} (s-\bar{m}^2)^2 \,{\rm exp}\left(-\frac{s}{T^2}\right) \\
\notag && -  \frac{\langle\bar{q}q\rangle \langle\bar{q}g_s\sigma Gq\rangle}{512 \pi^6} \int dydz\,  y z(1-y-z) (s-\bar{m}^2)^2 (5s-2\bar{m}^2) \,{\rm exp}\left(-\frac{s}{T^2}\right) \\
\notag &&+ \frac{\langle\bar{q}q\rangle \langle\bar{q}g_s\sigma Gq\rangle}{6144 \pi^6} \int dydz\,  z (1-y-z)^2 (s-\bar{m}^2)^2 (5s-2\bar{m}^2) \,{\rm exp}\left(-\frac{s}{T^2}\right) \,,
\end{eqnarray}
\begin{eqnarray}
\notag \rho_{\mathcal{B}}(s,10)_1&=& - \frac{143 m_c^2 \langle\bar{q}g_s\sigma Gq\rangle^2}{98304 \pi^6} \int dydz\,  (s-\bar{m}^2)\,{\rm exp}\left(-\frac{s}{T^2}\right) \\
\notag && + \frac{47 m_c^2\langle\bar{q}g_s\sigma Gq\rangle^2}{49152 \pi^6} \int dydz\, \frac{(1-y-z)}{y} (s-\bar{m}^2) \,{\rm exp}\left(-\frac{s}{T^2}\right) \\
\notag &&+ \frac{143 \langle\bar{q}g_s\sigma Gq\rangle^2}{98304 \pi^6} \int dydz\, y z (s-\bar{m}^2) (2s-\bar{m}^2) \,{\rm exp}\left(-\frac{s}{T^2}\right) \\
\notag && - \frac{\langle\bar{q}g_s\sigma Gq\rangle^2}{2048 \pi^6} \int dydz\, z(1-y-z)  (s-\bar{m}^2) (2s-\bar{m}^2) \,{\rm exp}\left(-\frac{s}{T^2}\right) \\
\notag && + \frac{\langle\bar{q}g_s\sigma Gq\rangle^2}{49152 \pi^6} \int dydz\, y(1-y-z) (s-\bar{m}^2) (2s-\bar{m}^2) \,{\rm exp}\left(-\frac{s}{T^2}\right) \,,
\end{eqnarray}
\begin{eqnarray}
\notag\rho_{\mathcal{B}}(s,10)_2&=& - \frac{41 m_c^2 \langle g_s^2GG\rangle \langle\bar{q}q\rangle^2}{124416 \pi^6} \int dydz\, \frac{(1-y-z)^2}{y^2} \left[2(1-y)(s-\bar{m}^2) -s y \right] \,{\rm exp}\left(-\frac{s}{T^2}\right) \\
\notag && - \frac{83 m_c^2 \langle g_s^2GG\rangle \langle\bar{q}q\rangle^2}{248832 \pi^6} \int dydz\, \frac{z (1-y-z)^2}{y^2} (3s-2\bar{m}^2)\,{\rm exp}\left(-\frac{s}{T^2}\right) \\
\notag &&- \frac{m_c^2 \langle g_s^2GG\rangle \langle\bar{q}q\rangle^2}{2048 \pi^6} \int dydz\, (s-\bar{m}^2)\,{\rm exp}\left(-\frac{s}{T^2}\right) \\
\notag && - \frac{m_c^2 \langle g_s^2GG\rangle \langle\bar{q}q\rangle^2}{3072\pi^6} \int dydz\,\frac{(1-y-z)}{y} (s-\bar{m}^2) \,{\rm exp}\left(-\frac{s}{T^2}\right) \\
\notag && + \frac{\langle g_s^2GG\rangle \langle\bar{q}q\rangle^2}{3072 \pi^6} \int dydz\, z(1-y-z) (s-\bar{m}^2) (2s-\bar{m}^2) \,{\rm exp}\left(-\frac{s}{T^2}\right) \\
\notag && + \frac{\langle g_s^2GG\rangle \langle\bar{q}q\rangle^2}{2048 \pi^6} \int dydz\, y z  (s-\bar{m}^2) (2s-\bar{m}^2) \,{\rm exp}\left(-\frac{s}{T^2}\right) \,,
\end{eqnarray}
\begin{eqnarray}
\notag \rho_{\mathcal{C}}(s,14) &=& \frac{m_c^2 \langle\bar{q}q\rangle^3 \langle\bar{q}g_s\sigma Gq\rangle}{36 \pi^2} \int dy\,\delta(s-\tilde{m}^2) \left(\frac{s}{ T^2} + 1 \right) \,{\rm exp}\left(-\frac{s}{T^2}\right) \\
\notag &&- \frac{m_c^2 \langle\bar{q}q\rangle^3 \langle\bar{q}g_s\sigma Gq\rangle}{144\pi^2} \int dy\,\delta(s-\tilde{m}^2) \frac{1}{y} \,{\rm exp}\left(-\frac{s}{T^2}\right) \\
\notag && - \frac{\langle\bar{q}q\rangle^3 \langle\bar{q}g_s\sigma Gq\rangle}{72 \pi^2} \int dy\,\delta(s-\tilde{m}^2) (1-y) y \left(\frac{s^2}{T^2} + 4s \right) \,{\rm exp}\left(-\frac{s}{T^2}\right) \\
\notag && + \frac{\langle\bar{q}q\rangle^3 \langle\bar{q}g_s\sigma Gq\rangle}{288 \pi^2} \int dy\,\delta(s-\tilde{m}^2) (1-y) s\,{\rm exp}\left(-\frac{s}{T^2}\right) \,,
\end{eqnarray}
\begin{eqnarray}
\notag \rho_{\mathcal{C}}(s,16)_1&=& - \frac{m_c^2 \langle\bar{q}q\rangle^2 \langle\bar{q}g_s\sigma Gq\rangle^2}{96 \pi^2 T^6} \int dy\,\delta(s-\tilde{m}^2) s^2 \,{\rm exp}\left(-\frac{s}{T^2}\right) \\
\notag &&+ \frac{m_c^2 \langle\bar{q}q\rangle^2 \langle\bar{q}g_s\sigma Gq\rangle^2}{192 \pi^2 T^4} \int dy\,\delta(s-\tilde{m}^2) \frac{s}{y} \,{\rm exp}\left(-\frac{s}{T^2}\right) \\
\notag && + \frac{\langle\bar{q}q\rangle^2 \langle\bar{q}g_s\sigma Gq\rangle^2}{192 \pi^2} \int dy\,\delta(s-\tilde{m}^2) (1-y) y \left(\frac{s^3}{T^6}+\frac{3s^2}{T^4}+\frac{6s}{T^2} + 6 \right) \,{\rm exp}\left(-\frac{s}{T^2}\right) \\
\notag && - \frac{\langle\bar{q}q\rangle^2 \langle\bar{q}g_s\sigma Gq\rangle^2}{384 \pi^2 } \int dy\,\delta(s-\tilde{m}^2) (1-y) \left(\frac{s^2}{T^4}+\frac{2s}{T^2} + 2\right) \,{\rm exp}\left(-\frac{s}{T^2}\right) \,,
\end{eqnarray}
\begin{eqnarray}
\notag \rho_{\mathcal{C}}(s,16)_2&=&  \frac{m_c^4 \langle g_s^2GG\rangle \langle\bar{q}q\rangle^4}{1296 \pi^2 T^4} \int dy\,\delta(s-\tilde{m}^2) \frac{1}{y^3} \,{\rm exp}\left(-\frac{s}{T^2}\right) \\
\notag && - \frac{m_c^2 \langle g_s^2GG\rangle \langle\bar{q}q\rangle^4}{1296 \pi^2 T^6} \int dy\,\delta(s-\tilde{m}^2) s^2 \,{\rm exp}\left(-\frac{s}{T^2}\right) \\
\notag &&- \frac{m_c^2 \langle g_s^2GG\rangle \langle\bar{q}q\rangle^4}{2592 \pi^2 T^4} \int dy\,\delta(s-\tilde{m}^2) \frac{(1-y)s}{y^2} \,{\rm exp}\left(-\frac{s}{T^2}\right) \\
\notag && - \frac{m_c^2 \langle g_s^2GG\rangle \langle\bar{q}q\rangle^4}{432 \pi^2 T^2} \int dy\,\delta(s-\tilde{m}^2) \frac{1}{y^2} \,{\rm exp}\left(-\frac{s}{T^2}\right) \\
\notag && + \frac{\langle g_s^2GG\rangle \langle\bar{q}q\rangle^4}{2592 \pi^2} \int dy\,\delta(s-\tilde{m}^2) (1-y) y  \left(\frac{s^3}{T^6}+\frac{3s^2}{T^4}+\frac{6s}{T^2} +6 \right)\,{\rm exp}\left(-\frac{s}{T^2}\right) \,.
\end{eqnarray}
The spectral density $\rho_6(s)$ of the currents $\mathcal{J}_2^\pm$:
\begin{eqnarray}
\notag \rho_{\mathcal{A}}(s,12)_1&=& \frac{m_c^2 \langle\bar{q}q\rangle^4}{36 \pi^2} \int dy\, \,{\rm exp}\left(-\frac{s}{T^2}\right) \\
\notag && + \frac{\langle\bar{q}q\rangle^4}{72 \pi^2} \int dy\,(1-y) y (3s-2\tilde{m}^2) \,{\rm exp}\left(-\frac{s}{T^2}\right) \,,
\end{eqnarray}
\begin{eqnarray}
\notag \rho_{\mathcal{A}}(s,12)_2&=& \frac{g_s^2 m_c^2 \langle\bar{q}q\rangle^4}{1296 \pi^4} \int dy\, \,{\rm exp}\left(-\frac{s}{T^2}\right) \\
\notag && + \frac{g_s^2 \langle\bar{q}q\rangle^4}{1296 \pi^4} \int dy\,(1-y) y (3s-2\tilde{m}^2)\,{\rm exp}\left(-\frac{s}{T^2}\right) \,,
\end{eqnarray}
\begin{eqnarray}
\notag \rho_{\mathcal{A}}(s,14) &=&  \frac{\langle\bar{q}q\rangle^3 \langle\bar{q}g_s\sigma Gq\rangle}{144 \pi^2} \int dy\,(1-y) \,{\rm exp}\left(-\frac{s}{T^2}\right) \\
\notag && - \frac{\langle\bar{q}q\rangle^3 \langle\bar{q}g_s\sigma Gq\rangle}{12 \pi^2} \int dy\,(1-y) y\,{\rm exp}\left(-\frac{s}{T^2}\right) \,,
\end{eqnarray}
\begin{eqnarray}
\notag \rho_{\mathcal{B}}(s,0) &=& \frac{m_c^2}{19660800 \pi^{10}} \int dydz\,(1-y-z)^5 (s-\bar{m}^2)^6\,{\rm exp}\left(-\frac{s}{T^2}\right) \\
\notag && + \frac{1}{68812800 \pi^{10}} \int dydz\, y z(1-y-z)^5 (s-\bar{m}^2)^6 (9s-2\bar{m}^2) \,{\rm exp}\left(-\frac{s}{T^2}\right) \,,
\end{eqnarray}
\begin{eqnarray}
\notag \rho_{\mathcal{B}}(s,4) &=& \frac{m_c^2 \langle g_s^2GG\rangle}{11796480 \pi^{10}} \int dydz\, \frac{(1-y-z)^5}{y^2} (s-\bar{m}^2)^3 \left[(1-y)(s-\bar{m}^2) - 2s y \right] \,{\rm exp}\left(-\frac{s}{T^2}\right) \\
\notag && + \frac{m_c^2 \langle g_s^2GG\rangle}{3145728 \pi^{10}} \int dydz\, \frac{(1-y-z)^4}{y} (s-\bar{m}^2)^4 \,{\rm exp}\left(-\frac{s}{T^2}\right) \\
\notag && - \frac{m_c^2 \langle g_s^2GG\rangle}{5898240 \pi^{10}} \int dydz\, \frac{z(1-y-z)^5}{y^2} (s-\bar{m}^2)^3 (3s-\bar{m}^2) \,{\rm exp}\left(-\frac{s}{T^2}\right) \\
\notag && + \frac{\langle g_s^2GG\rangle}{15728640 \pi^{10}} \int dydz\, z (1-y-z)^4 (s-\bar{m}^2)^4 (7s-2\bar{m}^2) \,{\rm exp}\left(-\frac{s}{T^2}\right) \\
\notag && + \frac{\langle g_s^2GG\rangle}{3932160 \pi^{10}} \int dydz\, y z(1-y-z)^3 (s-\bar{m}^2)^4 (7s-2\bar{m}^2) \,{\rm exp}\left(-\frac{s}{T^2}\right) \,,
\end{eqnarray}
\begin{eqnarray}
\notag \rho_{\mathcal{B}}(s,6)_1 &=& \frac{m_c^2 \langle\bar{q}q\rangle^2}{512 \pi^6} \int dydz\, (1-y-z)^2 (s-\bar{m}^2)^3\,{\rm exp}\left(-\frac{s}{T^2}\right) \\
\notag && + \frac{\langle\bar{q}q\rangle^2}{1024 \pi^6} \int dydz\, y z (1-y-z)^2 (s-\bar{m}^2)^3 (3s-\bar{m}^2) \,{\rm exp}\left(-\frac{s}{T^2}\right) \,,
\end{eqnarray}
\begin{eqnarray}
\notag \rho_{\mathcal{B}}(s,6)_2 &=& \frac{g_s^2 m_c^2 \langle\bar{q}q\rangle^2}{41472 \pi^8} \int dydz\, (1-y-z)^2 (s-\bar{m}^2)^3\,{\rm exp}\left(-\frac{s}{T^2}\right) \\
\notag && + \frac{g_s^2 \langle\bar{q}q\rangle^2}{41472 \pi^8} \int dydz\, y z (1-y-z)^2 (s-\bar{m}^2)^3 (3s-\bar{m}^2) \,{\rm exp}\left(-\frac{s}{T^2}\right) \,,
\end{eqnarray}
\begin{eqnarray}
\notag \rho_{\mathcal{B}}(s,8) &=& - \frac{3 m_c^2 \langle\bar{q}q\rangle \langle\bar{q}g_s\sigma Gq\rangle}{512 \pi^6} \int dydz\, (1-y-z) (s-\bar{m}^2)^2\,{\rm exp}\left(-\frac{s}{T^2}\right) \\
\notag && + \frac{m_c^2 \langle\bar{q}q\rangle \langle\bar{q}g_s\sigma Gq\rangle}{1024 \pi^6} \int dydz\,\frac{(1-y-z)^2}{y} (s-\bar{m}^2)^2\,{\rm exp}\left(-\frac{s}{T^2}\right) \\
\notag && - \frac{\langle\bar{q}q\rangle \langle\bar{q}g_s\sigma Gq\rangle}{512 \pi^6} \int dydz\, y z(1-y-z) (s-\bar{m}^2)^2 (5s-2\bar{m}^2) \,{\rm exp}\left(-\frac{s}{T^2}\right) \\
\notag && + \frac{\langle\bar{q}q\rangle \langle\bar{q}g_s\sigma Gq\rangle}{6144 \pi^6} \int dydz\, z (1-y-z)^2 (s-\bar{m}^2)^2 (5s-2\bar{m}^2) \,{\rm exp}\left(-\frac{s}{T^2}\right) \,,
\end{eqnarray}
\begin{eqnarray}
\notag \rho_{\mathcal{B}}(s,10)_1 &=& \frac{143 m_c^2 \langle\bar{q}g_s\sigma Gq\rangle^2}{98304 \pi^6} \int dydz\,(s-\bar{m}^2)\,{\rm exp}\left(-\frac{s}{T^2}\right) \\
\notag && - \frac{47 m_c^2 \langle\bar{q}g_s\sigma Gq\rangle^2}{49152 \pi^6} \int dydz\, \frac{(1-y-z)}{y}(s-\bar{m}^2)\,{\rm exp}\left(-\frac{s}{T^2}\right) \\
\notag && + \frac{143 \langle\bar{q}g_s\sigma Gq\rangle^2}{98304 \pi^6} \int dydz\, y z (s-\bar{m}^2) (2s-\bar{m}^2) \,{\rm exp}\left(-\frac{s}{T^2}\right) \\
\notag && - \frac{\langle\bar{q}g_s\sigma Gq\rangle^2}{2048 \pi^6} \int dydz\, z(1-y-z) (s-\bar{m}^2) (2s-\bar{m}^2) \,{\rm exp}\left(-\frac{s}{T^2}\right) \\
\notag && + \frac{\langle\bar{q}g_s\sigma Gq\rangle^2}{49152 \pi^6} \int dydz\, y(1-y-z) (s-\bar{m}^2) (2s-\bar{m}^2) \,{\rm exp}\left(-\frac{s}{T^2}\right) \,,
\end{eqnarray}
\begin{eqnarray}
\notag \rho_{\mathcal{B}}(s,10)_2 &=& \frac{m_c^2 \langle g_s^2GG\rangle \langle\bar{q}q\rangle^2}{3072 \pi^6} \int dydz\, \frac{(1-y-z)^2}{y^2} \left[ 2(1-y)(s-\bar{m}^2) + s y \right] \,{\rm exp}\left(-\frac{s}{T^2}\right) \\
\notag && + \frac{m_c^2 \langle g_s^2GG\rangle \langle\bar{q}q\rangle^2}{2048 \pi^6} \int dydz\, (s-\bar{m}^2) \,{\rm exp}\left(-\frac{s}{T^2}\right) \\
\notag && + \frac{m_c^2 \langle g_s^2GG\rangle \langle\bar{q}q\rangle^2}{3072 \pi^6} \int dydz\, \frac{(1-y-z)}{y} (s-\bar{m}^2) \,{\rm exp}\left(-\frac{s}{T^2}\right) \\
\notag && - \frac{m_c^2 \langle g_s^2GG\rangle \langle\bar{q}q\rangle^2}{3072 \pi^6} \int dydz\, \frac{z (1-y-z)^2}{y^2} (3s -2\bar{m}^2)\,{\rm exp}\left(-\frac{s}{T^2}\right) \\
\notag && + \frac{\langle g_s^2GG\rangle \langle\bar{q}q\rangle^2}{2048 \pi^6} \int dydz\, y z (s-\bar{m}^2) (2s-\bar{m}^2) \,{\rm exp}\left(-\frac{s}{T^2}\right) \\
\notag && + \frac{\langle g_s^2GG\rangle \langle\bar{q}q\rangle^2}{3072 \pi^6} \int dydz\, z(1-y-z) (s-\bar{m}^2) (2s-\bar{m}^2) \,{\rm exp}\left(-\frac{s}{T^2}\right) \,,
\end{eqnarray}
\begin{eqnarray}
\notag \rho_{\mathcal{C}}(s,14) &=& - \frac{m_c^2 \langle\bar{q}q\rangle^3 \langle\bar{q}g_s\sigma Gq\rangle}{36 \pi^2} \int dy\,\delta(s-\tilde{m}^2) \left( \frac{s}{T^2} + 1 \right) \,{\rm exp}\left(-\frac{s}{T^2}\right) \\
\notag && + \frac{m_c^2 \langle\bar{q}q\rangle^3 \langle\bar{q}g_s\sigma Gq\rangle}{144 \pi^2} \int dy\,\delta(s-\tilde{m}^2) \frac{1}{y} \,{\rm exp}\left(-\frac{s}{T^2}\right) \\
\notag && - \frac{\langle\bar{q}q\rangle^3 \langle\bar{q}g_s\sigma Gq\rangle}{72 \pi^2} \int dy\,\delta(s-\tilde{m}^2) (1-y) y \left( \frac{s^2}{T^2} + 4s \right) \,{\rm exp}\left(-\frac{s}{T^2}\right) \\
\notag && + \frac{\langle\bar{q}q\rangle^3 \langle\bar{q}g_s\sigma Gq\rangle}{288 \pi^2} \int dy\,\delta(s-\tilde{m}^2) (1-y) s\,{\rm exp}\left(-\frac{s}{T^2}\right) \,,
\end{eqnarray}
\begin{eqnarray}
\notag \rho_{\mathcal{C}}(s,16)_1 &=& \frac{m_c^2 \langle\bar{q}q\rangle^2 \langle\bar{q}g_s\sigma Gq\rangle^2}{96\pi^2 T^6} \int dy\,\delta(s-\tilde{m}^2) s^2 \,{\rm exp}\left(-\frac{s}{T^2}\right) \\
\notag && - \frac{m_c^2 \langle\bar{q}q\rangle^2 \langle\bar{q}g_s\sigma Gq\rangle^2}{192 \pi^2 T^4} \int dy\,\delta(s-\tilde{m}^2) \frac{s}{y} \,{\rm exp}\left(-\frac{s}{T^2}\right) \\
\notag && + \frac{\langle\bar{q}q\rangle^2 \langle\bar{q}g_s\sigma Gq\rangle^2}{192\pi^2} \int dy\,\delta(s-\tilde{m}^2) (1-y) y \left( \frac{s^3}{T^6} + \frac{3s^2}{T^4} + \frac{6s}{T^2} + 6 \right) \,{\rm exp}\left(-\frac{s}{T^2}\right) \\
\notag && + \frac{\langle\bar{q}q\rangle^2 \langle\bar{q}g_s\sigma Gq\rangle^2}{384 \pi^2} \int dy\,\delta(s-\tilde{m}^2) (1-y) \left( \frac{s^2}{T^4} - \frac{2s}{T^2} - 2 \right) \,{\rm exp}\left(-\frac{s}{T^2}\right) \,,
\end{eqnarray}
\begin{eqnarray}
\notag \rho_{\mathcal{C}}(s,16)_2 &=& - \frac{m_c^4 \langle g_s^2GG\rangle \langle\bar{q}q\rangle^4}{1296 \pi^2 T^4} \int dy\,\delta(s-\tilde{m}^2) \frac{1}{y^3} \,{\rm exp}\left(-\frac{s}{T^2}\right) \\
\notag && + \frac{m_c^2 \langle g_s^2GG\rangle \langle\bar{q}q\rangle^4}{1296\pi^2 T^6} \int dy\,\delta(s-\tilde{m}^2) s^2 \,{\rm exp}\left(-\frac{s}{T^2}\right) \\
\notag && - \frac{m_c^2 \langle g_s^2GG\rangle \langle\bar{q}q\rangle^4}{2592 \pi^2 T^4} \int dy\,\delta(s-\tilde{m}^2) \frac{(1-y)s}{y^2} \,{\rm exp}\left(-\frac{s}{T^2}\right) \\
\notag &&+ \frac{m_c^2 \langle g_s^2GG\rangle \langle\bar{q}q\rangle^4}{432 \pi^2 T^2} \int dy\,\delta(s-\tilde{m}^2) \frac{1}{y^2} \,{\rm exp}\left(-\frac{s}{T^2}\right) \\
\notag && + \frac{\langle g_s^2GG\rangle \langle\bar{q}q\rangle^4}{2592 \pi^2} \int dy\,\delta(s-\tilde{m}^2) (1-y) y \left( \frac{s^3}{T^6} + \frac{3s^2}{T^4} + \frac{6s}{T^2} + 6 \right)\,{\rm exp}\left(-\frac{s}{T^2}\right) \,.
\end{eqnarray}
The spectral density $\rho_7(s)$ of the currents $\mathcal{J}_{1\alpha}^\pm$:
\begin{eqnarray}
\notag \rho_{\mathcal{A}}(s,12)_1 &=& \frac{m_c^2 \langle\bar{q}q\rangle^4}{36 \pi^2} \int dy\, \,{\rm exp}\left(-\frac{s}{T^2}\right) \\
\notag && + \frac{\langle\bar{q}q\rangle^4}{72 \pi^2} \int dy\, (1-y) y (2s-\tilde{m}^2)\,{\rm exp}\left(-\frac{s}{T^2}\right) \,,
\end{eqnarray}
\begin{eqnarray}
\notag \rho_{\mathcal{A}}(s,12)_2 &=& \frac{g_s^2 m_c^2 \langle\bar{q}q\rangle^4}{1296 \pi^4} \int dy\, \,{\rm exp}\left(-\frac{s}{T^2}\right) \\
\notag && + \frac{g_s^2 \langle\bar{q}q\rangle^4}{1296 \pi^4} \int dy\, (1-y) y (2s-\tilde{m}^2)\,{\rm exp}\left(-\frac{s}{T^2}\right) \,,
\end{eqnarray}
\begin{eqnarray}
\notag \rho_{\mathcal{A}}(s,14) &=& -\frac{\langle\bar{q}q\rangle^3 \langle\bar{q}g_s\sigma Gq\rangle}{24 \pi^2} \int dy\, (1-y) y \,{\rm exp}\left(-\frac{s}{T^2}\right) \\
\notag && + \frac{\langle\bar{q}q\rangle^3 \langle\bar{q}g_s\sigma Gq\rangle}{288 \pi^2} \int dy\, (1-y) \,{\rm exp}\left(-\frac{s}{T^2}\right) \,,
\end{eqnarray}
\begin{eqnarray}
\notag \rho_{\mathcal{B}}(s,0) &=& \frac{m_c^2}{19660800 \pi^{10}} \int dydz\,(1-y-z)^5 (s-\bar{m}^2)^6 \,{\rm exp}\left(-\frac{s}{T^2}\right) \\
\notag && + \frac{1}{68812800 \pi^{10}} \int dydz\, y z(1-y-z)^5 (s-\bar{m}^2)^6 (8s-\bar{m}^2) \,{\rm exp}\left(-\frac{s}{T^2}\right) \,,
\end{eqnarray}
\begin{eqnarray}
\notag \rho_{\mathcal{B}}(s,4) &=& \frac{m_c^2 \langle g_s^2GG\rangle}{11796480 \pi^{10}} \int dydz\, \frac{(1-y-z)^5}{y^2} (s-\bar{m}^2)^3 \left[ (1-y) (s-\bar{m}^2) - 2s y \right] \,{\rm exp}\left(-\frac{s}{T^2}\right) \\
\notag &&  + \frac{m_c^2 \langle g_s^2GG\rangle}{3145728 \pi^{10}} \int dydz\, \frac{(1-y-z)^4}{y} (s-\bar{m}^2)^4\,{\rm exp}\left(-\frac{s}{T^2}\right) \\
\notag && - \frac{m_c^2 \langle g_s^2GG\rangle}{11796480 \pi^{10}} \int dydz\, \frac{z(1-y-z)^5}{y^2} (s-\bar{m}^2)^3 (5s-\bar{m}^2) \,{\rm exp}\left(-\frac{s}{T^2}\right) \\
\notag &&  + \frac{\langle g_s^2GG\rangle}{15728640 \pi^{10}} \int dydz\, z (1-y-z)^4 (s-\bar{m}^2)^4 (6s-\bar{m}^2) \,{\rm exp}\left(-\frac{s}{T^2}\right) \\
\notag && + \frac{\langle g_s^2GG\rangle}{3932160 \pi^{10}} \int dydz\, y z(1-y-z)^3 (s-\bar{m}^2)^4 (6s-\bar{m}^2) \,{\rm exp}\left(-\frac{s}{T^2}\right) \,,
\end{eqnarray}
\begin{eqnarray}
\notag \rho_{\mathcal{B}}(s,6)_1 &=& \frac{m_c^2 \langle\bar{q}q\rangle^2}{512 \pi^6} \int dydz\, (1-y-z)^2 (s-\bar{m}^2)^3\,{\rm exp}\left(-\frac{s}{T^2}\right) \\
\notag && + \frac{\langle\bar{q}q\rangle^2}{2048 \pi^6} \int dydz\, y z (1-y-z)^2 (s-\bar{m}^2)^3 (5s-\bar{m}^2) \,{\rm exp}\left(-\frac{s}{T^2}\right) \,,
\end{eqnarray}
\begin{eqnarray}
\notag \rho_{\mathcal{B}}(s,6)_2 &=& \frac{g_s^2 m_c^2 \langle\bar{q}q\rangle^2}{41472 \pi^8} \int dydz\, (1-y-z)^2 (s-\bar{m}^2)^3\,{\rm exp}\left(-\frac{s}{T^2}\right) \\
\notag && + \frac{g_s^2 \langle\bar{q}q\rangle^2}{82944 \pi^8} \int dydz\, y z (1-y-z)^2 (s-\bar{m}^2)^3 (5s-\bar{m}^2) \,{\rm exp}\left(-\frac{s}{T^2}\right) \,,
\end{eqnarray}
\begin{eqnarray}
\notag \rho_{\mathcal{B}}(s,8) &=& - \frac{3 m_c^2 \langle\bar{q}q\rangle \langle\bar{q}g_s\sigma Gq\rangle}{512 \pi^6} \int dydz\, (1-y-z) (s-\bar{m}^2)^2\,{\rm exp}\left(-\frac{s}{T^2}\right) \\
\notag &&  + \frac{m_c^2 \langle\bar{q}q\rangle \langle\bar{q}g_s\sigma Gq\rangle}{1024 \pi^6} \int dydz\, \frac{(1-y-z)^2}{y} (s-\bar{m}^2)^2\,{\rm exp}\left(-\frac{s}{T^2}\right) \\
\notag && - \frac{\langle\bar{q}q\rangle \langle\bar{q}g_s\sigma Gq\rangle}{512 \pi^6} \int dydz\, y z(1-y-z) (s-\bar{m}^2)^2 (4s-\bar{m}^2) \,{\rm exp}\left(-\frac{s}{T^2}\right) \\
\notag && + \frac{\langle\bar{q}q\rangle \langle\bar{q}g_s\sigma Gq\rangle}{6144 \pi^6} \int dydz\, z (1-y-z)^2 (s-\bar{m}^2)^2 (4s-\bar{m}^2) \,{\rm exp}\left(-\frac{s}{T^2}\right) \,,
\end{eqnarray}
\begin{eqnarray}
\notag \rho_{\mathcal{B}}(s,10)_1 &=&  \frac{143 m_c^2 \langle\bar{q}g_s\sigma Gq\rangle^2}{98304 \pi^6} \int dydz\,(s-\bar{m}^2) \,{\rm exp}\left(-\frac{s}{T^2}\right) \\
\notag && - \frac{47 m_c^2 \langle\bar{q}g_s\sigma Gq\rangle^2}{49152 \pi^6} \int dydz\, \frac{(1-y-z)}{y}(s-\bar{m}^2)\,{\rm exp}\left(-\frac{s}{T^2}\right) \\
\notag && + \frac{143 \langle\bar{q}g_s\sigma Gq\rangle^2}{196608 \pi^6} \int dydz\, y z (s-\bar{m}^2) (3s-\bar{m}^2) \,{\rm exp}\left(-\frac{s}{T^2}\right) \\
\notag && - \frac{23 \langle\bar{q}g_s\sigma Gq\rangle^2}{98304 \pi^6} \int dydz\, z(1-y-z) (s-\bar{m}^2) (3s-\bar{m}^2) \,{\rm exp}\left(-\frac{s}{T^2}\right) \,,
\end{eqnarray}
\begin{eqnarray}
\notag \rho_{\mathcal{B}}(s,10)_2 &=&  \frac{m_c^2 \langle g_s^2GG\rangle \langle\bar{q}q\rangle^2}{3072 \pi^6} \int dydz\, \frac{(1-y-z)^2}{y^2}\left[2(1-y) (s-\bar{m}^2) - s y \right] \,{\rm exp}\left(-\frac{s}{T^2}\right) \\
\notag && + \frac{m_c^2 \langle g_s^2GG\rangle \langle\bar{q}q\rangle^2}{2048 \pi^6} \int dydz\, (s-\bar{m}^2)\,{\rm exp}\left(-\frac{s}{T^2}\right) \\
\notag && + \frac{m_c^2 \langle g_s^2GG\rangle \langle\bar{q}q\rangle^2}{3072 \pi^6} \int dydz\,  \frac{(1-y-z)}{y}(s-\bar{m}^2)\,{\rm exp}\left(-\frac{s}{T^2}\right) \\
\notag && - \frac{m_c^2 \langle g_s^2GG\rangle \langle\bar{q}q\rangle^2}{9216 \pi^6} \int dydz\, \frac{z (1-y-z)^2}{y^2} (7s -3\bar{m}^2)\,{\rm exp}\left(-\frac{s}{T^2}\right) \\
\notag && + \frac{\langle g_s^2GG\rangle \langle\bar{q}q\rangle^2}{4096 \pi^6} \int dydz\, y z (s-\bar{m}^2) (3s-\bar{m}^2) \,{\rm exp}\left(-\frac{s}{T^2}\right) \\
\notag && + \frac{\langle g_s^2GG\rangle \langle\bar{q}q\rangle^2}{6144 \pi^6} \int dydz\, z(1-y-z) (s-\bar{m}^2) (3s-\bar{m}^2) \,{\rm exp}\left(-\frac{s}{T^2}\right) \,,
\end{eqnarray}
\begin{eqnarray}
\notag \rho_{\mathcal{C}}(s,14) &=& - \frac{m_c^2 \langle\bar{q}q\rangle^3 \langle\bar{q}g_s\sigma Gq\rangle}{36 \pi^2} \int dy\,\delta(s-\tilde{m}^2) \left( \frac{s}{T^2} + 1 \right) \,{\rm exp}\left(-\frac{s}{T^2}\right) \\
\notag && +\frac{m_c^2 \langle\bar{q}q\rangle^3 \langle\bar{q}g_s\sigma Gq\rangle}{144 \pi^2} \int dy\,\delta(s-\tilde{m}^2) \frac{1}{y} \,{\rm exp}\left(-\frac{s}{T^2}\right) \\
\notag && - \frac{\langle\bar{q}q\rangle^3 \langle\bar{q}g_s\sigma Gq\rangle}{72 \pi^2} \int dy\,\delta(s-\tilde{m}^2) (1-y) y \left( \frac{s^2}{T^2} + 3s \right) \,{\rm exp}\left(-\frac{s}{T^2}\right) \\
\notag && + \frac{\langle\bar{q}q\rangle^3 \langle\bar{q}g_s\sigma Gq\rangle}{288 \pi^2} \int dy\,\delta(s-\tilde{m}^2) (1-y) s\,{\rm exp}\left(-\frac{s}{T^2}\right) \,,
\end{eqnarray}
\begin{eqnarray}
\notag \rho_{\mathcal{C}}(s,16)_1 &=& \frac{m_c^2 \langle\bar{q}q\rangle^2 \langle\bar{q}g_s\sigma Gq\rangle^2}{96\pi^2 T^6} \int dy\,\delta(s-\tilde{m}^2) s^2 \,{\rm exp}\left(-\frac{s}{T^2}\right) \\
\notag && - \frac{m_c^2 \langle\bar{q}q\rangle^2 \langle\bar{q}g_s\sigma Gq\rangle^2}{192 \pi^2 T^4} \int dy\,\delta(s-\tilde{m}^2) \frac{s}{y} \,{\rm exp}\left(-\frac{s}{T^2}\right) \\
\notag && + \frac{\langle\bar{q}q\rangle^2 \langle\bar{q}g_s\sigma Gq\rangle^2}{192 \pi^2} \int dy\,\delta(s-\tilde{m}^2) (1-y) y \left( \frac{s^3}{T^6} + \frac{2s^2}{T^4} + \frac{2s}{T^2} \right)\,{\rm exp}\left(-\frac{s}{T^2}\right) \\
\notag && - \frac{\langle\bar{q}q\rangle^2 \langle\bar{q}g_s\sigma Gq\rangle^2}{384 \pi^2} \int dy\,\delta(s-\tilde{m}^2) (1-y) \left( \frac{s^2}{T^4} + \frac{s}{T^2} \right)\,{\rm exp}\left(-\frac{s}{T^2}\right) \,,
\end{eqnarray}
\begin{eqnarray}
\notag \rho_{\mathcal{C}}(s,16)_2 &=& - \frac{m_c^4 \langle g_s^2GG\rangle \langle\bar{q}q\rangle^4}{1296 \pi^2 T^4} \int dy\,\delta(s-\tilde{m}^2) \frac{1}{y^3} \,{\rm exp}\left(-\frac{s}{T^2}\right) \\
\notag && + \frac{m_c^2 \langle g_s^2GG\rangle \langle\bar{q}q\rangle^4}{1296 \pi^2 T^6} \int dy\,\delta(s-\tilde{m}^2) s^2 \,{\rm exp}\left(-\frac{s}{T^2}\right) \\
\notag && + \frac{m_c^2 \langle g_s^2GG\rangle \langle\bar{q}q\rangle^4}{432 \pi^2 T^2} \int dy\,\delta(s-\tilde{m}^2) \frac{1}{y^2} \,{\rm exp}\left(-\frac{s}{T^2}\right) \\
\notag && - \frac{m_c^2 \langle g_s^2GG\rangle \langle\bar{q}q\rangle^4}{2592 \pi^2 } \int dy\,\delta(s-\tilde{m}^2) \frac{1-y}{y^2} \left( \frac{s}{T^4} +\frac{1}{T^2} \right)\,{\rm exp}\left(-\frac{s}{T^2}\right) \\
\notag && - \frac{\langle g_s^2GG\rangle \langle\bar{q}q\rangle^4}{7776 \pi^2 } \int dy\,\delta(s-\tilde{m}^2) (1-y) y \left(\frac{s^3}{T^6} - \frac{6s}{T^2} \right)\,{\rm exp}\left(-\frac{s}{T^2}\right) \,.
\end{eqnarray}
The spectral density $\rho_8(s)$ of the currents $\mathcal{J}_{2\alpha}^\pm$:
\begin{eqnarray}
\notag \rho_{\mathcal{A}}(s,12)_1&=& - \frac{m_c^2 \langle\bar{q}q\rangle^4}{36 \pi^2} \int dy\, \,{\rm exp}\left(-\frac{s}{T^2}\right) \\
\notag && + \frac{\langle\bar{q}q\rangle^4}{72 \pi^2} \int dy\, (1-y) y (2s-\tilde{m}^2)\,{\rm exp}\left(-\frac{s}{T^2}\right) \,,
\end{eqnarray}
\begin{eqnarray}
\notag \rho_{\mathcal{A}}(s,12)_2&=& - \frac{g_s^2 m_c^2 \langle\bar{q}q\rangle^4}{1296 \pi^4} \int dy\, \,{\rm exp}\left(-\frac{s}{T^2}\right) \\
\notag && + \frac{g_s^2 \langle\bar{q}q\rangle^4}{1296 \pi^4} \int dy\, (1-y) y (2s-\tilde{m}^2)\,{\rm exp}\left(-\frac{s}{T^2}\right) \,,
\end{eqnarray}
\begin{eqnarray}
\notag \rho_{\mathcal{A}}(s,14) &=& -\frac{\langle\bar{q}q\rangle^3 \langle\bar{q}g_s\sigma Gq\rangle}{24 \pi^2} \int dy\, (1-y) y \,{\rm exp}\left(-\frac{s}{T^2}\right) \\
\notag && + \frac{\langle\bar{q}q\rangle^3 \langle\bar{q}g_s\sigma Gq\rangle}{288 \pi^2} \int dy\, (1-y) \,{\rm exp}\left(-\frac{s}{T^2}\right) \,,
\end{eqnarray}
\begin{eqnarray}
\notag \rho_{\mathcal{B}}(s,0)&=& -\frac{m_c^2}{19660800 \pi^{10}} \int dydz\,(1-y-z)^5 (s-\bar{m}^2)^6 \,{\rm exp}\left(-\frac{s}{T^2}\right) \\
\notag && + \frac{1}{68812800 \pi^{10}} \int dydz\, y z(1-y-z)^5 (s-\bar{m}^2)^6 (8s-\bar{m}^2) \,{\rm exp}\left(-\frac{s}{T^2}\right) \,,
\end{eqnarray}
\begin{eqnarray}
\notag \rho_{\mathcal{B}}(s,4) &=& - \frac{m_c^2 \langle g_s^2GG\rangle}{11796480 \pi^{10}} \int dydz\, \frac{(1-y-z)^5}{y^2}) (s-\bar{m}^2)^3 \left[(1-y) (s-\bar{m}^2) - 2s y \right]\,{\rm exp}\left(-\frac{s}{T^2}\right) \\
\notag && - \frac{m_c^2 \langle g_s^2GG\rangle}{3145728 \pi^{10}} \int dydz\, \frac{(1-y-z)^4}{y} (s-\bar{m}^2)^4\,{\rm exp}\left(-\frac{s}{T^2}\right) \\
\notag && - \frac{m_c^2 \langle g_s^2GG\rangle}{11796480 \pi^{10}} \int dydz\, \frac{z(1-y-z)^5}{y^2} (s-\bar{m}^2)^3 (5s-\bar{m}^2) \,{\rm exp}\left(-\frac{s}{T^2}\right) \\
\notag && + \frac{\langle g_s^2GG\rangle}{15728640 \pi^{10}} \int dydz\, z (1-y-z)^4 (s-\bar{m}^2)^4 (6s-\bar{m}^2) \,{\rm exp}\left(-\frac{s}{T^2}\right) \\
\notag && + \frac{\langle g_s^2GG\rangle}{3932160 \pi^{10}} \int dydz\, y z(1-y-z)^3 (s-\bar{m}^2)^4 (6s-\bar{m}^2) \,{\rm exp}\left(-\frac{s}{T^2}\right) \,,
\end{eqnarray}
\begin{eqnarray}
\notag \rho_{\mathcal{B}}(s,6)_1 &=& - \frac{m_c^2 \langle\bar{q}q\rangle^2}{512 \pi^6} \int dydz\, (1-y-z)^2 (s-\bar{m}^2)^3 \,{\rm exp}\left(-\frac{s}{T^2}\right) \\
\notag && + \frac{\langle\bar{q}q\rangle^2}{2048 \pi^6} \int dydz\, y z (1-y-z)^2 (s-\bar{m}^2)^3 (5s-\bar{m}^2) \,{\rm exp}\left(-\frac{s}{T^2}\right) \,,
\end{eqnarray}
\begin{eqnarray}
\notag \rho_{\mathcal{B}}(s,6)_2 &=& - \frac{g_s^2 m_c^2 \langle\bar{q}q\rangle^2}{41472 \pi^8} \int dydz\, (1-y-z)^2 (s-\bar{m}^2)^3\,{\rm exp}\left(-\frac{s}{T^2}\right) \\
\notag && + \frac{g_s^2 \langle\bar{q}q\rangle^2}{82944 \pi^8} \int dydz\, y z (1-y-z)^2 (s-\bar{m}^2)^3 (5s-\bar{m}^2) \,{\rm exp}\left(-\frac{s}{T^2}\right) \,,
\end{eqnarray}
\begin{eqnarray}
\notag \rho_{\mathcal{B}}(s,8) &=&  \frac{3 m_c^2 \langle\bar{q}q\rangle \langle\bar{q}g_s\sigma Gq\rangle}{512 \pi^6} \int dydz\, (1-y-z) (s-\bar{m}^2)^2\,{\rm exp}\left(-\frac{s}{T^2}\right) \\
\notag && - \frac{m_c^2 \langle\bar{q}q\rangle \langle\bar{q}g_s\sigma Gq\rangle}{1024 \pi^6} \int dydz\, \frac{(1-y-z)^2}{y} (s-\bar{m}^2)^2\,{\rm exp}\left(-\frac{s}{T^2}\right) \\
\notag && - \frac{\langle\bar{q}q\rangle \langle\bar{q}g_s\sigma Gq\rangle}{512 \pi^6} \int dydz\, y z(1-y-z) (s-\bar{m}^2)^2 (4s-\bar{m}^2) \,{\rm exp}\left(-\frac{s}{T^2}\right) \\
\notag && + \frac{\langle\bar{q}q\rangle \langle\bar{q}g_s\sigma Gq\rangle}{6144 \pi^6} \int dydz\, z (1-y-z)^2 (s-\bar{m}^2)^2 (4s-\bar{m}^2) \,{\rm exp}\left(-\frac{s}{T^2}\right) \,,
\end{eqnarray}
\begin{eqnarray}
\notag \rho_{\mathcal{B}}(s,10)_1&=& -\frac{143 m_c^2 \langle\bar{q}g_s\sigma Gq\rangle^2}{98304 \pi^6} \int dydz\,(s-\bar{m}^2)\,{\rm exp}\left(-\frac{s}{T^2}\right) \\
\notag && + \frac{47 m_c^2 \langle\bar{q}g_s\sigma Gq\rangle^2}{49152 \pi^6} \int dydz\, \frac{(1-y-z)}{y}(s-\bar{m}^2)\,{\rm exp}\left(-\frac{s}{T^2}\right) \\
\notag && + \frac{143 \langle\bar{q}g_s\sigma Gq\rangle^2}{196608 \pi^6} \int dydz\, y z (s-\bar{m}^2) (3s-\bar{m}^2) \,{\rm exp}\left(-\frac{s}{T^2}\right) \\
\notag && - \frac{23 \langle\bar{q}g_s\sigma Gq\rangle^2}{98304 \pi^6} \int dydz\, z(1-y-z) (s-\bar{m}^2) (3s-\bar{m}^2) \,{\rm exp}\left(-\frac{s}{T^2}\right) \,,
\end{eqnarray}
\begin{eqnarray}
\notag \rho_{\mathcal{B}}(s,10)_2&=& - \frac{m_c^2 \langle g_s^2GG\rangle \langle\bar{q}q\rangle^2}{3072 \pi^6} \int dydz\, \frac{(1-y-z)^2}{y^2} \left[ 2(1-y)(s-\bar{m}^2) - s y \right] \,{\rm exp}\left(-\frac{s}{T^2}\right) \\
\notag && - \frac{m_c^2 \langle g_s^2GG\rangle \langle\bar{q}q\rangle^2}{2048 \pi^6} \int dydz\, (s-\bar{m}^2) \,{\rm exp}\left(-\frac{s}{T^2}\right) \\
\notag && - \frac{m_c^2 \langle g_s^2GG\rangle \langle\bar{q}q\rangle^2}{3072 \pi^6} \int dydz\, \frac{(1-y-z)}{y}(s-\bar{m}^2)\,{\rm exp}\left(-\frac{s}{T^2}\right) \\
\notag && - \frac{m_c^2 \langle g_s^2GG\rangle \langle\bar{q}q\rangle^2}{3072 \pi^6} \int dydz\, \frac{z (1-y-z)^2}{y^2} (2s-\bar{m}^2)\,{\rm exp}\left(-\frac{s}{T^2}\right) \\
\notag && + \frac{\langle g_s^2GG\rangle \langle\bar{q}q\rangle^2}{4096 \pi^6} \int dydz\, y z (s-\bar{m}^2) (3s-\bar{m}^2) \,{\rm exp}\left(-\frac{s}{T^2}\right) \\
\notag && + \frac{\langle g_s^2GG\rangle \langle\bar{q}q\rangle^2}{6144 \pi^6} \int dydz\, z(1-y-z) (s-\bar{m}^2) (3s-\bar{m}^2) \,{\rm exp}\left(-\frac{s}{T^2}\right) \,,
\end{eqnarray}
\begin{eqnarray}
\notag \rho_{\mathcal{C}}(s,14) &=& \frac{m_c^2 \langle\bar{q}q\rangle^3 \langle\bar{q}g_s\sigma Gq\rangle}{36 \pi^2} \int dy\,\delta(s-\tilde{m}^2) \left( \frac{s}{T^2} + 1 \right)\,{\rm exp}\left(-\frac{s}{T^2}\right) \\
\notag && - \frac{m_c^2 \langle\bar{q}q\rangle^3 \langle\bar{q}g_s\sigma Gq\rangle}{144 \pi^2} \int dy\,\delta(s-\tilde{m}^2) \frac{1}{y} \,{\rm exp}\left(-\frac{s}{T^2}\right) \\
\notag && - \frac{\langle\bar{q}q\rangle^3 \langle\bar{q}g_s\sigma Gq\rangle}{216 \pi^2} \int dy\,\delta(s-\tilde{m}^2) (1-y) y \left( \frac{3s^2}{T^2} + 221 s \right)\,{\rm exp}\left(-\frac{s}{T^2}\right) \\
\notag && + \frac{\langle\bar{q}q\rangle^3 \langle\bar{q}g_s\sigma Gq\rangle}{288 \pi^2} \int dy\,\delta(s-\tilde{m}^2) (1-y) s\,{\rm exp}\left(-\frac{s}{T^2}\right) \,,
\end{eqnarray}
\begin{eqnarray}
\notag \rho_{\mathcal{C}}(s,16)_1 &=& - \frac{m_c^2 \langle\bar{q}q\rangle^2 \langle\bar{q}g_s\sigma Gq\rangle^2}{96 \pi^2 T^6} \int dy\,\delta(s-\tilde{m}^2) s^2 \,{\rm exp}\left(-\frac{s}{T^2}\right) \\
\notag && + \frac{m_c^2 \langle\bar{q}q\rangle^2 \langle\bar{q}g_s\sigma Gq\rangle^2}{192 \pi^2 T^4} \int dy\,\delta(s-\tilde{m}^2) \frac{s}{y} \,{\rm exp}\left(-\frac{s}{T^2}\right) \\
\notag && + \frac{\langle\bar{q}q\rangle^2 \langle\bar{q}g_s\sigma Gq\rangle^2}{576 \pi^2} \int dy\,\delta(s-\tilde{m}^2) (1-y) y \left( \frac{s^3}{ T^6}+\frac{6s^2}{T^4} + \frac{6s}{T^2} \right)\,{\rm exp}\left(-\frac{s}{T^2}\right) \\
\notag && - \frac{\langle\bar{q}q\rangle^2 \langle\bar{q}g_s\sigma Gq\rangle^2}{384 \pi^2} \int dy\,\delta(s-\tilde{m}^2) (1-y) \left( \frac{s^2}{ T^4} + \frac{s}{T^2} \right)\,{\rm exp}\left(-\frac{s}{T^2}\right) \,,
\end{eqnarray}
\begin{eqnarray}
\notag \rho_{\mathcal{C}}(s,16)_2 &=& \frac{m_c^4 \langle g_s^2GG\rangle \langle\bar{q}q\rangle^4}{1296 \pi^2 T^4} \int dy\,\delta(s-\tilde{m}^2) \frac{1}{y^3} \,{\rm exp}\left(-\frac{s}{T^2}\right) \\
\notag && - \frac{m_c^2 \langle g_s^2GG\rangle \langle\bar{q}q\rangle^4}{1296 \pi^2 T^6} \int dy\,\delta(s-\tilde{m}^2) s^2 \,{\rm exp}\left(-\frac{s}{T^2}\right) \\
\notag && - \frac{m_c^2 \langle g_s^2GG\rangle \langle\bar{q}q\rangle^4}{432 \pi^2 T^2} \int dy\,\delta(s-\tilde{m}^2) \frac{1}{y^2} \,{\rm exp}\left(-\frac{s}{T^2}\right) \\
\notag && - \frac{m_c^2 \langle g_s^2GG\rangle \langle\bar{q}q\rangle^4}{2592 \pi^2} \int dy\,\delta(s-\tilde{m}^2) \frac{1-y}{y^2} \left( \frac{s}{T^4} -\frac{1}{T^2} \right)\,{\rm exp}\left(-\frac{s}{T^2}\right) \\
\notag && + \frac{\langle g_s^2GG\rangle \langle\bar{q}q\rangle^4}{2592 \pi^2} \int dy\,\delta(s-\tilde{m}^2) (1-y) y \left( \frac{s^3}{T^6} + \frac{2s^2}{T^4} + \frac{2s}{T^2} \right)\,{\rm exp}\left(-\frac{s}{T^2}\right) \,.
\end{eqnarray}
\end{document}